\documentclass[usenatbib]{mnras}
\usepackage{newtxtext,newtxmath}
\usepackage[T1]{fontenc}
\usepackage{ae,aecompl}
\usepackage{times}
\usepackage{graphicx}
\usepackage{amsmath}
\usepackage{lscape}
\usepackage{multirow}
\usepackage{subcaption}
\usepackage{rotating}
\usepackage{adjustbox} 
\usepackage{natbib}
\usepackage{upgreek}
\usepackage[utf8x]{inputenc}
\newcommand{\kms}{\,km\,s$^{-1}$\,}
\newcommand{\msol}{\,M$_{\odot}$\,}
\newcommand{\sfr}{\,M$_{\odot}$\,yr$^{-1}$\,}
\newcommand{\sfrd}{\,M$_{\odot}$\,yr$^{-1}$\,kpc$^{-2}$\,}

\usepackage[dvipsnames]{xcolor}

\title[The rocky road to quiescence]{The rocky road to quiescence: compaction and quenching of quasar host galaxies at z\,$\sim$\,2 }

\author[H.\,R.~Stacey et al.]{H.\,R.~Stacey,$^{\! 1,2,3}$\thanks{E-mail: stacey@mpa-garching.mpg.de}
J.\,P.~McKean,$^{\! 1,2}$
D.\,M.~Powell,$^{\! 3}$
S.~Vegetti,$^{\! 3}$
F.~Rizzo,$^{\! 3}$
C.~Spingola,$^{\! 4,5}$
\newauthor 
M.\,W.~Auger,$^{\! 6,7}$
R.\,J.~Ivison$^{8}$
and
P.\,P.~van~der~Werf$^{\,9}$ \smallskip \\ 
$^{1}$ASTRON, Netherlands Institute for Radio Astronomy, Oude Hoogeveensedijk 4, 7991 PD, Dwingeloo, The Netherlands \\
$^{2}$Kapteyn Astronomical Institute, University of Groningen, PO Box 800, 9700 AV Groningen, The Netherlands \\
$^{3}$Max Planck Institute for Astrophysics, Karl-Schwarzschild Str. 1, D-85748 Garching bei M\"unchen, Germany \\
$^{4}$INAF $-$ Istituto di Radioastronomia, Via Gobetti 101, I$-$40129, Bologna, Italy \\
$^{5}$Dipartimento di Fisica e Astronomia, Universit\`a degli Studi di Bologna, Via Gobetti 93/2, I$-$40129 Bologna, Italy \\
$^{6}$Institute of Astronomy, University of Cambridge, Madingley Road, Cambridge CB3 0HA, UK \\ 
$^{7}$Kavli Institute for Cosmology, University of Cambridge, Madingley Road, Cambridge CB3 0HA, UK \\
$^{8}$European Southern Observatory, Karl-Schwarzschild-Str. 2, D-85748 Garching bei M\"unchen, Germany \\
$^{9}$Leiden Observatory, Leiden University, PO Box 9513, NL-2300 RA Leiden, The Netherlands \\
}
\date{Accepted 2020 November 02. Received 2020 October 28; in original form 2019 December 13}

\pubyear{2020}

\begin{document}
\label{firstpage}
\pagerange{\pageref{firstpage}--\pageref{lastpage}}
\maketitle

\begin{abstract}
We resolve the host galaxies of seven gravitationally lensed quasars at redshift 1.5 to 2.8 using observations with the Atacama Large (sub-)Millimetre Array. Using a visibility-plane lens modelling technique, we create pixellated reconstructions of the dust morphology, and CO line morphology and kinematics. We find that the quasar hosts in our sample can be distinguished into two types: 1) galaxies characterised by clumpy, extended dust distributions ($R_{\rm eff}\sim2$~kpc) and mean star formation rate surface densities comparable to sub-mm-selected dusty star-forming galaxies ($\Sigma_{\rm SFR}\sim3$~\sfrd); 2) galaxies that have sizes in dust emission similar to coeval passive galaxies and compact starbursts ($R_{\rm eff}\sim0.5$~kpc), with high mean star formation rate surface densities ($\Sigma_{\rm SFR}=400$--4500~\sfrd) that may be Eddington-limited or super-Eddington. The small sizes of some quasar hosts suggests that we observe them at a stage in their transformation into compact spheroids via dissipative contraction, where a high density of dynamically unstable gas leads to efficient star formation and black hole accretion. For the one system where we probe the bulk of the gas reservoir, we find a gas fraction of just $0.06\pm0.04$ and a depletion timescale of $50\pm40$~Myr, suggesting it is transitioning into quiescence. In general, we expect that the extreme level of star formation in the compact quasar host galaxies will rapidly exhaust their gas reservoirs and could quench with or without help from active galactic nuclei feedback.
\end{abstract}




\section{Introduction}

In the last two decades, large-area sub-millimetre surveys have revealed a population of high-redshift galaxies with extreme levels of star formation, which were largely undetected in optical surveys as their ultraviolet (UV) emission is obscured by dust (\citealt{Blain:2002}; \citealt{Casey:2014}, \citealt{Hodge:2020} for reviews). These dusty star-forming galaxies (DSFGs) are expected to be precursors to locally observed massive elliptical galaxies, which are characterised by dense, old stellar populations with dispersion-dominated dynamics \citep{Hopkins:2008}. A key aspect of the study of galaxy evolution is understanding how these galaxies formed such high stellar densities and grew concurrently with their central supermassive black holes \citep{Magorrian:1998}. Recent near-infrared surveys have revealed that the population of compact quiescent galaxies start to appear at $z\!\sim\!2$ and rapidly increase in number density before apparently declining at $z\!\sim\!1$ \citep{Trujillo:2006,vanDokkum:2008,vanDokkum:2015}. These galaxies are around four times smaller in size than $z\!\sim\!0$ massive ellipticals and are thought to form the centre of these galaxies, which later grow in size (but little in mass) due to a series of gas-poor, minor mergers \citep{Naab:2007,Naab:2009}.

The characteristics and rapid formation of compact quiescent galaxies can be reproduced if a very high density of gas is concentrated within a region of $\sim$\,1~kpc to generate a brief, intense starburst \citep{Valentino:2019}. However, the mechanisms that cause such rapid morphological change and quenching of star formation are currently unclear. Compaction may be due to a net loss of angular momentum that results from dynamical instabilities induced by gas-rich mergers or tidal interactions \citep{Mihos:1996,Hopkins:2008}, or by non-axisymmetric structures caused by rapid accretion and clumpy star formation \citep{Dekel:2009,Dekel:2014,Zolotov:2015}. Alternatively, these galaxies could have formed secularly, in-situ at earlier epochs without the need for mergers or rapid evolution \citep{Damjanov:2011,Carollo:2013,Williams:2014,Wellons:2015}.

Star formation terminates as a result of depletion or cessation of the supply of molecular gas. In compact quiescent galaxies, this may happen as a result of compaction, where star formation is self-quenched by radiation pressure from massive stars and supernovae-driven winds \citep{Murray:2005,Andrews:2011,Diamond-Stanic:2012}. Eddington-limited `maximum' starbursts have been discovered in DSFGs at high redshift \citep{Riechers:2013,Oteo:2016,Canameras:2017,Spilker:2019} as well as in nearby ultra-luminous infrared galaxies (ULIRGs; \citealt{Barcos-Munoz:2017}). These can be sufficiently vigorous to drive large-scale outflows and quench their hosts \citep{Canameras:2017,Spilker:2018}, from the inside-out, on timescales of 10s~Myr \citep{Spilker:2019}.

Hydro-dynamical simulations and semi-analytic models of galaxy formation find that feedback from active galactic nuclei (AGN) is necessary, in addition to stellar feedback, to reproduce observed galaxy stellar populations and luminosity functions \citep{DiMatteo:2005,Sijacki:2007,Somerville:2008,Schaye:2015}. AGN feedback occurs in the form of jets or radiative winds, which can suppress star formation by mechanically coupling to the molecular gas and/or by preventing (re-)accretion from the circumgalactic medium (\citealt{Fabian:2012}, for review).

Due to computational limitations, current cosmological simulations involve only phenomenological implementations of feedback, which are calibrated to match observational data (see \citealt{Somerville:2015}). Observations find circumstantial evidence that AGN could play a role in the evolution of their hosts, in that the compaction phase seems to coincide with the appearance of an AGN \citep{Barro:2013,Kocevski:2017}. However, studies have found conflicting results as to whether AGN have any effect on their host galaxies. While some earlier studies found evidence of suppressed star formation in AGN hosts (e.g. \citealt{Page:2012}), more recent studies of statistical samples have found no correlation between star formation and black hole accretion \citep{Harrison:2012,Rosario:2013,Harris:2016,Stanley:2017,Pitchford:2016,Kirkpatrick:2019,Schulze:2019}. This could be because the effects of AGN feedback can only be detected later, or because black hole accretion is stochastic \citep{Gabor:2014,Hickox:2014}. Even in simulations, star-formation rates averaged over $\sim$\,100~Myr can show no clear correlation with black hole accretion \citep{Harrison:2017}, suggesting the effect of AGN may not be obvious in studies of the global properties of quasar hosts.

These key tests of the evolutionary sequence require investigations of the size, structure and dynamical properties of individual quasar hosts during the cosmic peak of galaxy growth ($z\!\sim\!2$). This demands high spatial resolution (100s~pc), which is most efficiently achieved by observing objects that are gravitationally lensed (e.g. \citealt{Swinbank:2010,ALMA:2015}). By observing galaxies that are strongly gravitationally lensed, it is possible to recover the properties of the source with greater angular resolution and sensitivity. \cite{Stacey:2018a} conducted a survey of gravitationally lensed quasar systems with the {\it Herschel Space Observatory} to measure the level of obscured star formation in the quasar host galaxies. This paper presents observations with the Atacama Large (sub-)Millimetre Array (ALMA) of a sub-sample of seven optically luminous quasars from the parent sample of \citeauthor{Stacey:2018a} to resolve their host galaxy emission. Using a pixellated lens modelling technique applied to the interferometric data, we reconstruct dust and gas in the host galaxies and derive their intrinsic properties. In Section~\ref{section:obs} we describe the targets, observations and data reduction process. In Section~\ref{section:modelling} we describe our lens modelling and source reconstruction technique. Section~\ref{section:results} reports the results of the structure of dust and gas, gas dynamics and star formation properties of the individual objects. We compare the morphological and star formation properties of the quasar host galaxies with a sample of DSFGs. In Section~\ref{section:discussion}, we discuss the implications of our results in the context of evolutionary models and possible avenues to test our conjectures. Section~\ref{section:conclusion} presents a summary of our findings and avenues for future work.

Throughout, we assume the \cite{Planck:2016} instance of a flat $\Lambda$CDM cosmology with $H_{0}= 67.8~$km\,s$^{-1}$ Mpc$^{-1}$, $\Omega_{\rm M}=0.31$ and $\Omega_{\Lambda}=0.69$.

\section{Sample and observations}
\label{section:obs}

\begin{table*}
    \caption{Summary of the targets and ALMA observations. We give the phase centre right ascension and declination (in degrees, J2000), lens and source redshift (from optical spectroscopy -- improved redshift estimates for these systems are presented in Table~\ref{table:obs2}), central frequency of the observation, total on-source integration, the FWHM of the naturally weighted beam, and project code for the seven targets in this work. The redshift of the lens is not known for HS~0810+2554, H1413+117 or WFI~J2026$-$4536, but this has no bearing on our results.}
    \centering
    \begin{tabular}{ l c c c c c c c c } \hline
                     & RA & Dec & $z_{\rm l}$ & $z_{\rm s}$ & $\nu_{\rm obs}$ & $t_{s}$ & FWHM & Project code \\
                     & (deg)  & (deg) &  &  & (GHz) & (min) & (arcsec) &  \\ \hline
       HS~0810+2254  & 123.38053 & $+$25.75068 & -- & 1.51 & 145~GHz & 32 & $0.17\times0.13$  & 2017.1.01368.S \\
       RX~J0911+0551 & 137.86458 & $+$05.84833 & 0.70 & 2.79 & 145~GHz & 123 & $0.40\times0.35$ & 2017.1.01081.S \\
       SDSS~J0924+0219 & 141.23258 & $+$02.32347 & 0.39 & 1.52 & 358~GHz & 44 & $0.28\times0.23$ & 2018.1.01591.S \\
       PG~1115+080 & 169.57083 & $+$07.76603  & 0.31 & 1.74 & 346~GHz & 27 & $0.32\times0.21$ & 2018.1.01591.S \\
       H1413+117 & 213.94271 & $+$11.49539 & -- & 2.56 & 285~GHz & 10 & $0.24\times0.21$ & 2012.1.00175.S \\
       WFI~J2026$-$4536 & 306.54346 & $-$45.60753 & -- & 2.22 & 350~GHz & 28 & $0.14\times0.14$ & 2018.1.01591.S \\
       WFI~J2033$-$4723 & 308.42533 & $-$47.39528 & 0.66 & 1.66 & 341~GHz & 28 & $0.31\times0.28$ & 2018.1.01591.S \\ \hline 
    \end{tabular}
    \label{table:obs1}
\end{table*}

In this section, we first summarise the sample, the observations with ALMA and the data reduction processes, before giving a detailed description of the properties of each target studied here.

\subsection{Summary of the sample, observations and data reduction}

The targets in this work are seven four-image gravitationally lensed quasar (Type 1 AGN) systems. Four targets, SDSS~J0924+0219, PG~1115+080, WFI~J2033$-$4723 and WFI~J2026$-$4536 are from our own programme (PI: McKean) and were not selected on the basis of their FIR/sub-mm properties, but for the configuration of their lensed quasar images in optical data for reasons that are not relevant to this work. We also include here archival data of HS~0810+2554 (PI: Chartas), RX~J0911+0551 (PI: Leung) and H1413+117 (PI: van~der~Werf), which are known to have bright sub-mm emission \citep{Barvainis:2002}. Details of the observations are given in Table~\ref{table:obs1}. From previous modelling of their spectral energy distributions (SEDs; \citealt{Stacey:2018a}; see Figs.~\ref{fig:seds1} and \ref{fig:seds1}) we expect the sub-mm emission to be entirely dominated by thermal dust. As these latter systems were primarily selected in optical imaging and subsequently selected on the basis of their FIR luminosity, their selection is largely insensitive to dust temperature. We compile all accessible data with sufficiently high angular resolution to resolve the size and structure of the emission from cold dust at sub-mm/mm wavelengths. As a result, this sample is heterogeneous, and the observations probe different CO line transitions (see Table~\ref{table:obs2}).

The raw data were calibrated using the ALMA pipeline in the Common Astronomy Software Applications package (CASA; \citealt{McMullin:2007}) to produce calibrated visibilities. The data were inspected to confirm the quality of the pipeline calibration and determine whether further flagging was required. RX~J0911+0551, PG~1115+080, H1413+117 and WFI~J2026$-$4536 were self-calibrated using the line-free spectral windows with solution intervals of each scan length. Self-calibration was attempted for SDSS~J0924+0219 and WFI~J2033$-$4723, but was not successful, probably because of the lower signal-to-noise ratio of the surface brightness. We did not attempt to self-calibrate the data for HS~0810+2554 as the continuum emission is extremely weak (see below).

The targets were imaged with natural weighting of the visibilities and deconvolved using CLEAN \citep{Hogbom:1974}. The deconvolved images are shown in Figs.~\ref{fig:alma_images} and \ref{fig:1115_images}. The continuum emission is consistent with the level of thermal dust emission expected from SED-fitting \citep{Stacey:2018a}. We detect the host galaxies of the quasars that, in some cases, forms Einstein rings and gravitational arcs. We also find compact dust emission that, in all cases, approximately coincides with the quasar positions seen in optical data. No emission is detected from any of the lensing galaxies.

All data sets, except in the case of PG~1115+080, included observations of an emission line of CO. The spectral line data were prepared by fitting a linear model for the continuum to the line-free spectral windows and subtracting this from the visibilities. This produced a visibility data set that included only the line emission. The resulting line profiles, fit with single Gaussians, are shown in Fig.~\ref{fig:line_profiles}. In some cases, the line peaks are offset from the systemic velocity (the rest-frame of the galaxy), which is possibly due to an uncertainty in the inferred redshift from optical spectroscopy: redshifts of quasars determined from optical spectroscopy can trace outflows of ionised gas, leading to incorrect inference on the systemic velocity. A summary of the spectral line observations is presented in Table~\ref{table:obs2}.

The spectral line moment maps (velocity-integrated flux density, velocity field, velocity dispersion) are shown in Figs.~\ref{fig:0810_images} to \ref{fig:2033_images}. The velocity field and velocity dispersion are made by masking channel pixels below a signal-to-noise ratio threshold (stated within the figure captions).

\subsection{HS~0810+2554}

HS~0810+2554 is a quasar at $z_{\rm s}=1.51$ that is lensed into a characteristic fold-configuration \citep{Reimers:2002,Hewett:2010}. The redshift of the foreground galaxy is unknown, but is estimated to be 0.9 based on the lens population distribution \citep{Mosquera:2011}. This quasar is faint at radio wavelengths, but VLBI investigations find evidence of compact, low-luminosity radio-jet structure that dominates the radio emission \citep{Hartley:2019}. HS~0810+2554 was observed with ALMA at 145 GHz, which also targeted the CO (3--2) emission line.

We resolve thermal dust emission from this system with ALMA at 145~GHz that has a peak surface brightness of 6$\sigma$ (where $\sigma$ is the rms noise per beam). The achieved rms noise level is 13~$\upmu$Jy~beam$^{-1}$ with natural weighting. Emission is detected around the close images, where the magnification is high and there is a hint of emission from the counter images. The total flux density of 0.3~mJy is consistent with the level of thermal dust emission expected at the observing frequency, based on SED fitting (see Fig.~\ref{fig:seds1}). The weak synchrotron emission seen at cm wavelengths is not expected to be detectable at the higher frequencies investigated here.

The CO~(3--2) line profile for HS~0810+2554 is shown in Fig.~\ref{fig:line_profiles}, which has a FWHM of about $380\pm10$~\kms based on a single Gaussian fit. We find that the peak of the line emission is shifted $-465\pm6$~\kms from the assumed optical systemic velocity, likely due to an uncertain redshift estimate. ALMA imaging of the line emission shows that the molecular gas is extended and lensed into an Einstein ring (see Fig.~\ref{fig:0810_images}).

\subsection{RX~J0911+0551}

RX~J0911+0551 (RX~J0911.4+0551) is a quasar at $z_{\rm s}=2.79$ that is lensed by a foreground galaxy at $z_{\rm l}=0.77$ \citep{Burud:1998,Kneib:2000}. The lens system has a characteristic cusp configuration with three close images. The environment of this lens system is quite complex: the primary lens has a satellite galaxy within the Einstein radius, and the lens galaxy is part of a cluster that contributes a high level of tidal shear \citep{Kneib:2000}. Investigations at radio wavelengths have not been able to determine whether there is radio jet emission \citep{Jackson:2015}. However, its radio luminosity is consistent with the expectations for star formation based on the radio--infrared correlation \citep{Stacey:2018a}. For our analysis, we use ALMA imaging at 145~GHz that also targeted the CO (5--4) emission line from RX~J0911+0551.

We detect thermal dust continuum for this system with a total flux density at 145~GHz of 1.4~mJy, consistent with the expectations from SED fitting (see Figs.~\ref{fig:seds1} and \ref{fig:seds1}). We achieve an rms noise of 7~$\upmu$Jy~beam$^{-1}$ with natural weighting of the visibilties. Imaging of the dust shows compact emission that is resolved around the triplet images (see Fig.~\ref{fig:0911_images}). From the lens configuration, it is clear that the galaxy is crossing the cusp of the lens caustic.

The CO~(5--4) line profile for RX~J0911+0551 is presented in Fig.~\ref{fig:line_profiles}. The spatially integrated line profile has a FWHM of $133\pm3$~\kms\ based on a single Gaussian fit, which is similar to the $120\pm14$~\kms\ reported for the CO~(7--6) line by \cite{Tuan-Anh:2017}. Imaging of the CO~(5--4) line emission shows compact structure that is similar to the dust continuum (see Fig.~\ref{fig:0911_images}). The imaging is visually similar to the CO~(11--10), CO~(10--9), CO~(7--6) and CO~(1--0) emission reported by \cite{Tuan-Anh:2017} and \cite{Sharon:2016}: none of which are extended into rings or arcs, suggesting the emission is similarly compact.

\subsection{SDSS~J0924+0219}

SDSS~J0924+0219 is a radio-quiet quasar at $z_{\rm s}\!=\!1.525$ that is lensed by a galaxy at $z_{\rm l}=0.39$ \citep{Inada:2003,Hewett:2010}. This lens system was discovered at optical wavelengths and has a characteristic fold-configuration. Observations with ALMA at 358 GHz, which also targeted the CO (8--7) line, were taken as part of our own observing programme.

The continuum emission from SDSS~J0924+0219 is detected with a total flux density of 9.5~mJy at 358~GHz, consistent with the expectations for thermal dust emission from SED fitting (see Fig.~\ref{fig:seds1}). We achieve an rms noise of 28~$\upmu$Jy\,beam$^{-1}$ with natural weighting of the visibilities. The continuum imaging shows an Einstein ring of clumpy dust with compact emission at the locations of the quasar images. The extent of the dust seems consistent with a `red' ring seen in optical/infrared imaging \citep{Eigenbrod:2006}. The same study also finds lensed arcs from a `blue' component, of which there is a suggestion in the ALMA imaging where the Southern arc appears to split. These different optical components could be evidence of an ongoing merger or interaction (e.g. \citealt{Rybak:2015b}).

The CO~(8--7) line profile is presented in Fig.~\ref{fig:line_profiles}. The line profile is found from a Gaussian fit to have a FWHM of $176\pm11$~\kms\ that is shifted by $36\pm5$~\kms\ from the optical systemic velocity. Imaging of the CO (see Fig.~\ref{fig:0924_images}) shows resolved emission at the position of the quasar images and evidence for a velocity gradient in the two close images. The velocity gradient is similar to that reported by \cite{Badole:2020} for CO~(5--4), where the CO emission is more extended. The line profile has a similar FWHM, but the shape of the CO~(8--7) is more peaked than for the CO~(5--4) and the surface brightness sensitivity is lower.

\subsection{PG~1115+080}

PG~1115+080 is a quasar at $z_{\rm s}=1.74$ that is lensed by a galaxy at $z_{\rm l}=0.31$ \citep{Weymann:1980,Hewett:2010}. Investigations at radio wavelengths have not been able to determine whether the quasar has radio jet emission \citep{Jackson:2015}, however its radio luminosity is consistent with the expectations for star formation based on the radio--infrared correlation \citep{Stacey:2018a}. PG~1115+080 was targeted as part of our own programme with ALMA at 346~GHz.

\begin{figure}
    \centering
    \includegraphics[width=0.4\textwidth]{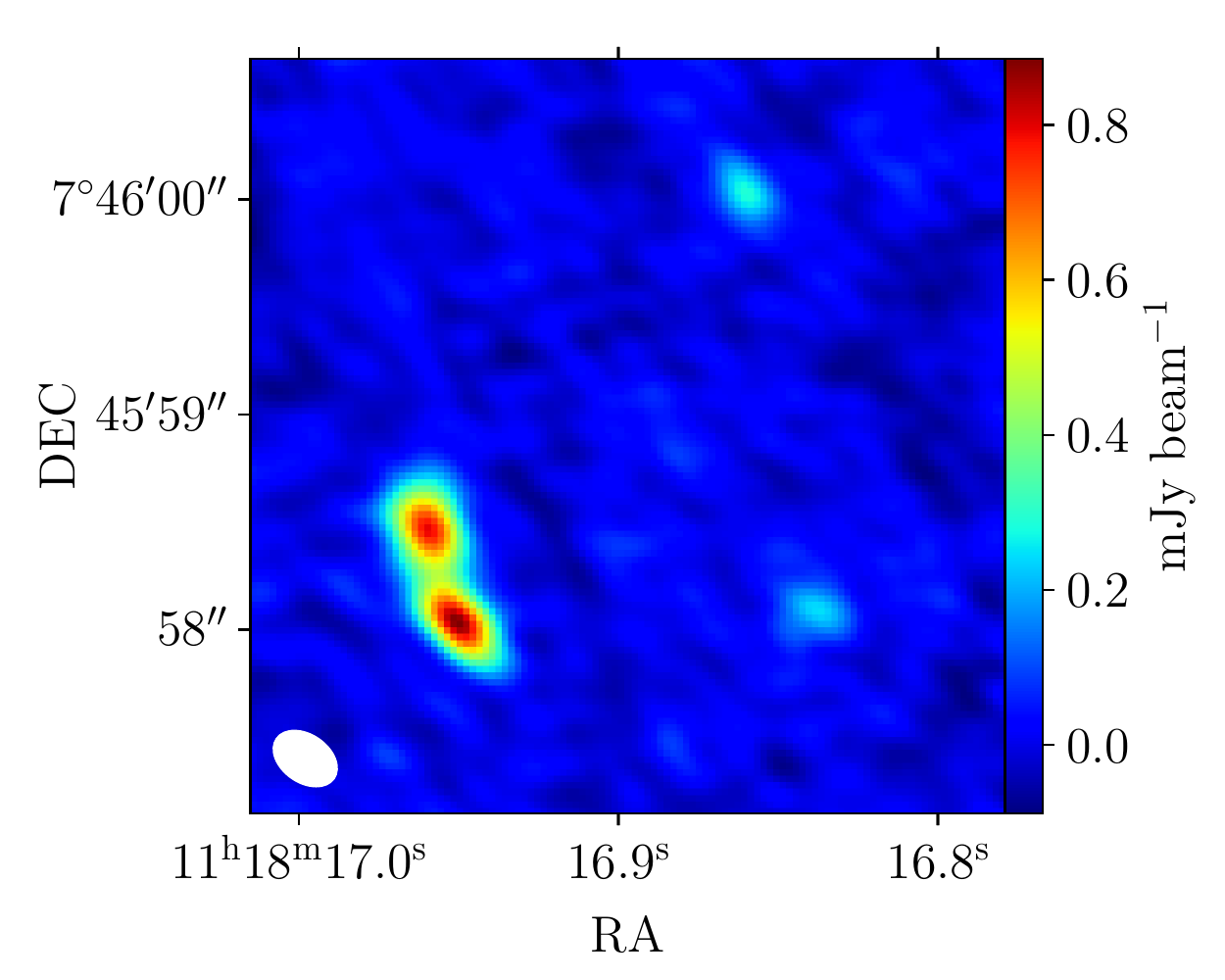}
    \caption{ALMA 346~GHz continuum images of PG~1115+080 with natural weighting of the visibility data. The synthesised beam FWHM is shown in the bottom left-hand corner.}
    \label{fig:1115_images}
\end{figure}

We detect continuum emission with a total flux density of 2.6~mJy, consistent with the expectations for thermal dust emission from SED fitting (see Fig.~\ref{fig:seds1}). The imaging of the dust emission for PG~1115+080 is presented in Fig.~\ref{fig:1115_images} and shows compact structure at the position of the quasar images, without any evidence of an extended Einstein ring. Given that the total flux density is close to what is expected from this system, we do not believe that significant emission from an extended component has been resolved out with these data. However, the emission is marginally radially resolved around the two close images. Unlike in the cases of the other quasar hosts in this sample, the unobscured star formation seen in the optical and near-infrared appears more extended than in the obscured star formation we observe here \citep{Sluse:2012,Chen:2019}.

\subsection{H1413+117}
\label{section:1413_obs}

H1413+117 (the Cloverleaf) is a quasar at $z_{\rm s}=2.56$ that is lensed by a galaxy at unknown redshift \citep{Magain:1988,Riechers:2011b}. Due to its extreme luminosity at far-infrared (FIR) wavelengths, the lens system is one of the best studied quasar-starbursts. While it is not extremely radio-luminous, this system does have a radio-excess relative to its infrared luminosity \citep{Stacey:2018a} and has evidence of structure at radio wavelengths that may be from a low-luminosity radio jet (Stacey et al. in prep). We use for our analysis archival data for H1413+117 that was taken at 285~GHz with ALMA, which also targeted the CO (9--8) emission line.

Thermal dust emission is detected in the ALMA imaging with a total flux density of 26~mJy and an rms noise level of 93~$\upmu$Jy~beam$^{-1}$, using a natural weighting of the visibilities. The CO~(9--8) line emission is detected with a spatially integrated peak flux density of 83~mJy and a FWHM of $394\pm8$~\kms, from a Gaussian profile fit (see Fig.~\ref{fig:line_profiles}); this is the brightest line emission detected from our sample.

We resolve the four images of the quasar host galaxy in continuum and the CO\,(9--8) molecular line emission, with high surface brightness peaks at the position of the quasar images and an extended, almost complete Einstein ring that connects them (see Fig.~\ref{fig:1413_images}). The structure of the emission appears similar to previous observations of dust and gas in this system \citep{Alloin:1997,Ferkinhoff:2015}. We also identify an additional source with a flux density of $\simeq3.5$~mJy within the field at a distance $\sim6$~arcsec North of the lens. This is a potential source of confusion for the SED fitting, where the photometric measurements are typically made at a low angular resolution (e.g. IRAS or {\it Herschel}/SPIRE). This additional source may also explain the scatter in the photometric measurements from H1413+117 (e.g. see \citealt{Stacey:2018a}; Fig.~\ref{fig:seds1}). From a cursory inspection of the available ALMA archival data at a frequency of $\sim100$~GHz (project code 2015.1.01309.S), we find that the additional source is detected with a similar flux density ratio as at 290~GHz with respect to H1413+117, suggesting that these galaxies have similar redshifts.

\subsection{WFI~J2026$-$4536}

WFI~J2026$-$4536 is a quasar at $z=2.22$ \citep{Morgan:2004} that is lensed by a galaxy at unknown redshift into four images in a typical fold configuration. There are no radio observations of this system available to characterise its radio properties at 1.4~GHz, but the lack of detection in the Sydney University Molonglo Sky Survey (SUMSS) Source Catalog \citep{Mauch:2003} suggests a radio flux density of $<12$~mJy at 1.4~GHz, assuming a typical synchrotron spectral index of $-0.7$. WFI~J2026$-$4536 was observed with ALMA at 350~GHz as part of our ALMA programme, which also observed the CO~(10--9) spectral line transition.

The ALMA observations detect continuum emission with a flux density of 28.3~mJy and achieve an rms noise level of 30~$\upmu$Jy~beam$^{-1}$ with natural weighting. This flux density is in good agreement with thermal dust emission, based on SED fitting (Fig.~\ref{fig:seds1}). The CLEANed image shows bright emission from the location of the quasar images and extended emission at lower surface brightness. 

The CO~(10--9) line emission is also spatially resolved with a similar morphology to the dust emission. The spatially integrated CO line profile has a peak flux density of 49~mJy and shows a hint of a double-horned feature characteristic of disc kinematics. The FWHM of the line is $252\pm7$~\kms\ and is offset by $\sim1000$~\kms\ from the systemic velocity assumed from optical spectroscopy.

\subsection{WFI~J2033$-$4723}

WFI~J2033$-$4723 is a quasar at $z_{\rm s}=1.66$ that is lensed by a galaxy at $z_{\rm l}=0.66$ \citep{Morgan:2004} into four images in a typical fold configuration. The lack of detection in the SUMSS suggests a radio flux density of $<12$~mJy at 1.4~GHz. No other radio observations are available for this lens system. The target was observed with ALMA at 341~GHz as part of our programme, which also observed the CO~(8--7) emission line from the lensed quasar.

\begin{figure*}
    \begin{subfigure}[t]{\textwidth}
    \centering
    \includegraphics[width=0.87\textwidth]{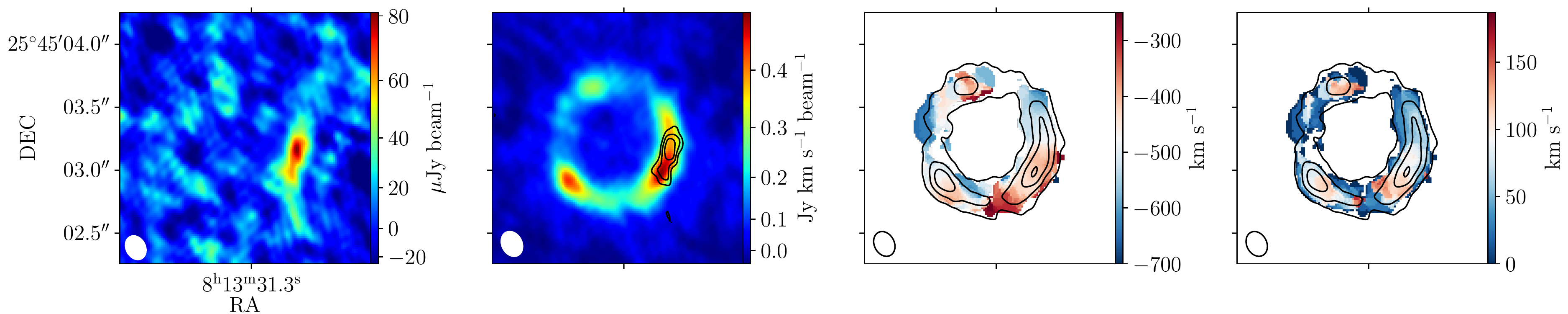}
    \caption{145~GHz continuum and CO~(3--2) spectral line images of HS~0810+2554. The moment images were made by masking channel pixels below 4$\sigma$.}
    \label{fig:0810_images}
    \end{subfigure}
    
    \begin{subfigure}[t]{\textwidth}
    \centering
    \vspace*{0.15cm}
    \includegraphics[width=0.87\textwidth]{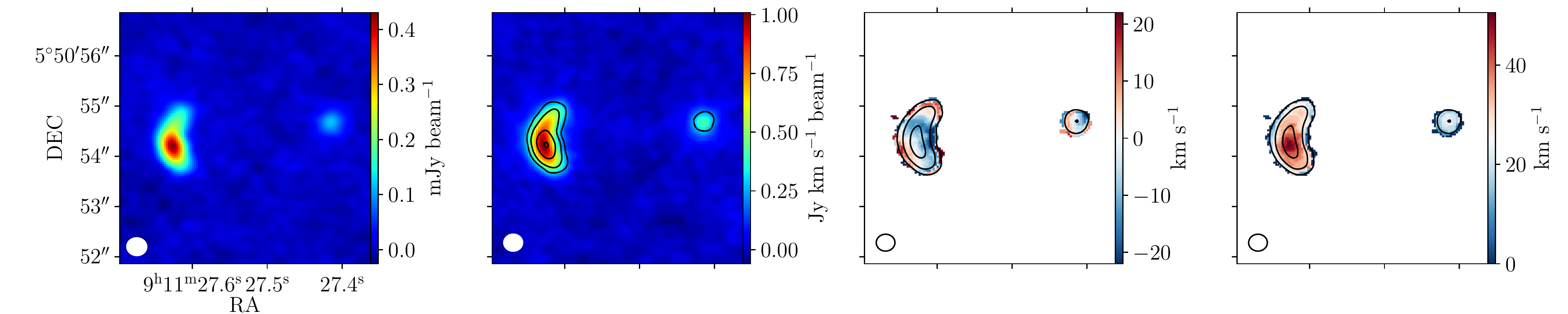}
    \caption{145~GHz continuum and CO~(5--4) spectral line images of RX~J0911+0551. The moment images were made by masking channel pixels below 5$\sigma$.}
    \label{fig:0911_images}
    \end{subfigure}
  
    \begin{subfigure}[t]{\textwidth}
    \centering
    \vspace*{0.15cm}
    \includegraphics[width=0.87\textwidth]{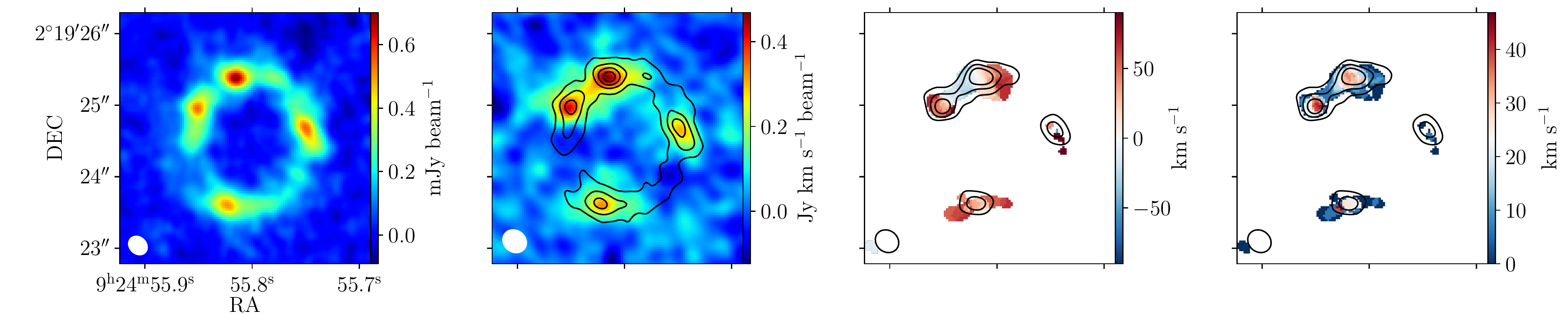}
    \caption{360~GHz continuum and CO~(8--7) spectral line images of SDSS0924+0219. The moment images were made by masking channel pixels below 4$\sigma$.}
    \label{fig:0924_images}
    \end{subfigure}
    
    \begin{subfigure}[t]{\textwidth}
    \centering
    \vspace*{0.15cm}
    \includegraphics[width=0.87\textwidth]{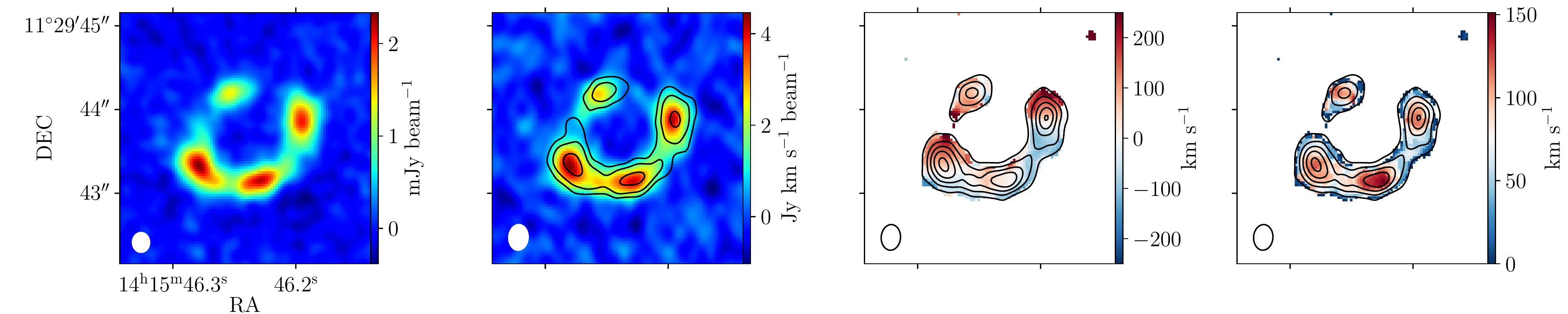}
    \caption{285~GHz continuum and CO~(9--8) spectral line images of H1413+117. The moment images were made by masking channel pixels below 4$\sigma$.}
    \label{fig:1413_images}
    \end{subfigure}
    
    \begin{subfigure}[t]{\textwidth}
    \centering
    \vspace*{0.15cm}
    \includegraphics[width=0.87\textwidth]{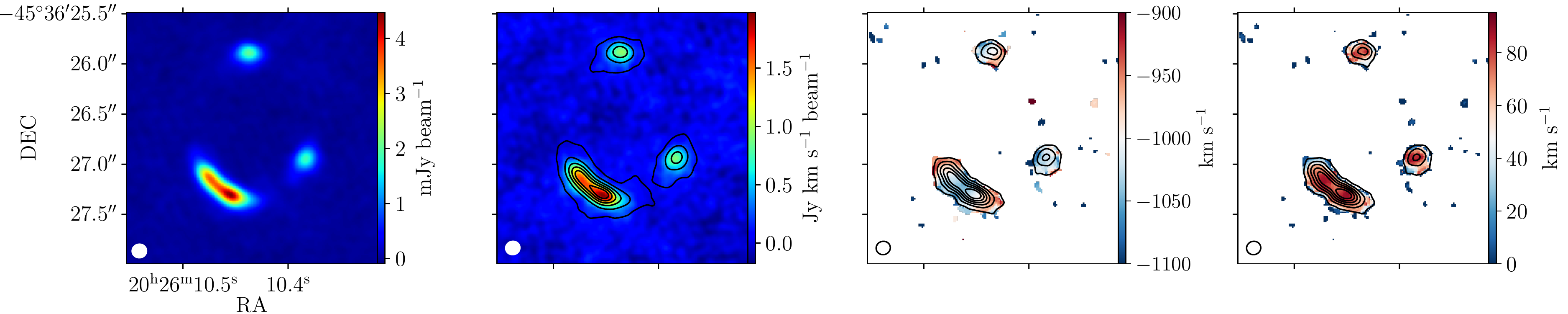}
    \caption{350~GHz continuum and CO~(10--9) spectral line images of WFI~J2026$-$4536. The moment images were made by masking channel pixels below 4$\sigma$.}
    \label{fig:2026_images}
    \end{subfigure}
    
    \begin{subfigure}[t]{\textwidth}
    \centering
    \vspace*{0.15cm}
    \includegraphics[width=0.87\textwidth]{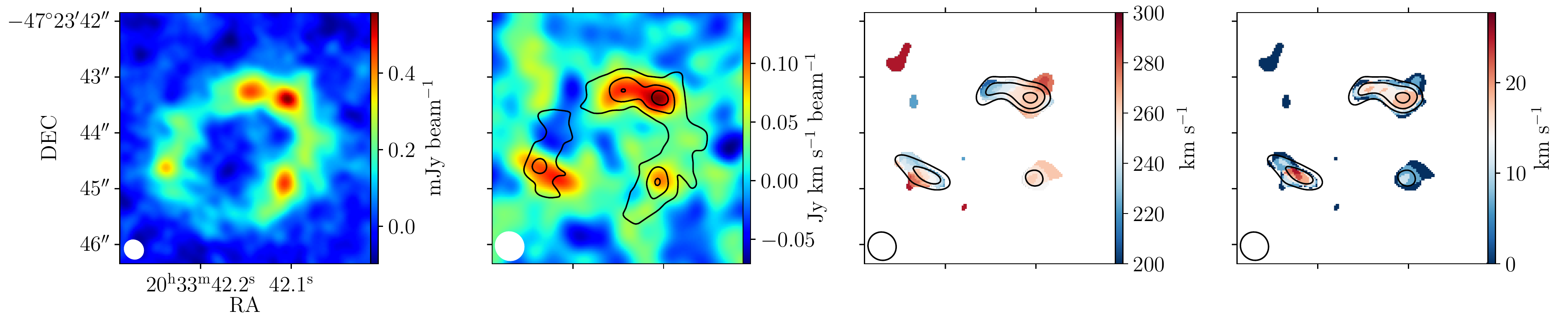}
    \caption{340~GHz continuum and CO~(8--7) spectral line images of WFI~J2033$-$4723. The moment images were made by masking channel pixels below 3$\sigma$.}
    \label{fig:2033_images}
    \end{subfigure}
    \caption{Deconvolved ALMA images for the six systems with spectral line data, using natural weighting of the visibilities. Left to right: the continuum; moment 0 (velocity-integrated line intensity); moment 1 (line velocity); moment 2 (velocity dispersion). The beam FWHM is shown in the lower-left corner.}
    \label{fig:alma_images}
\end{figure*}

The ALMA observations detect thermal dust emission with a total flux density of 10.4~mJy and achieve an rms noise of 30~$\upmu$Jy~beam$^{-1}$ with natural weighting of the visibilities. Imaging of WFI~J2033$-$4723 shows a ring of extended, clumpy dust emission and compact emission from the approximate location of the quasar images (see Fig.~\ref{fig:2033_images}). Arcs of faint emission from the host galaxy have also been observed at optical wavelengths, similar to the extended dust emission observed here \citep{Rusu:2020}. The observed flux-density at 341~GHz is in good agreement with the expectations of SED fitting to a thermal dust model (see Fig.~\ref{fig:seds1}).

\begin{figure*}
    \includegraphics[width=0.33\textwidth]{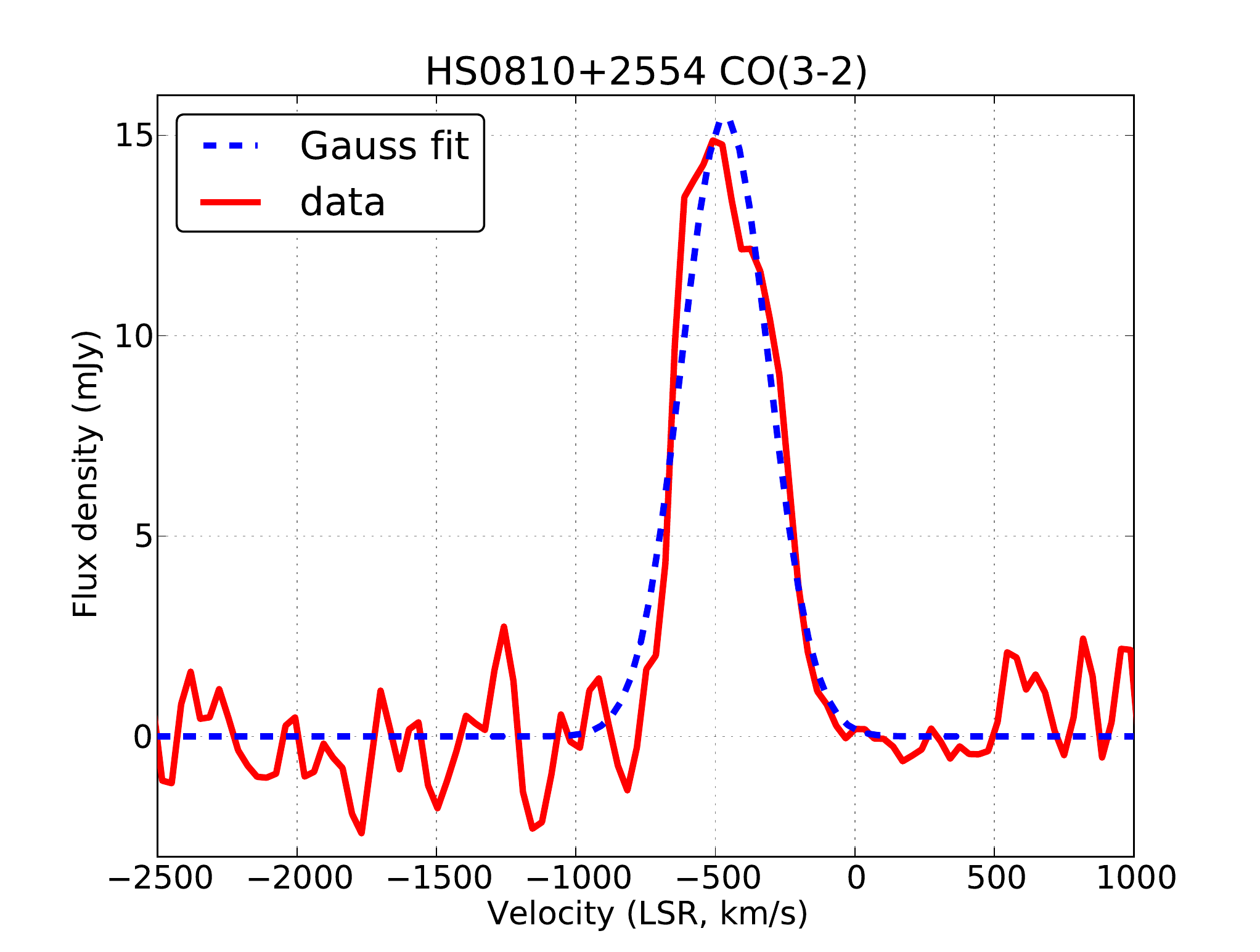}
    \includegraphics[width=0.33\textwidth]{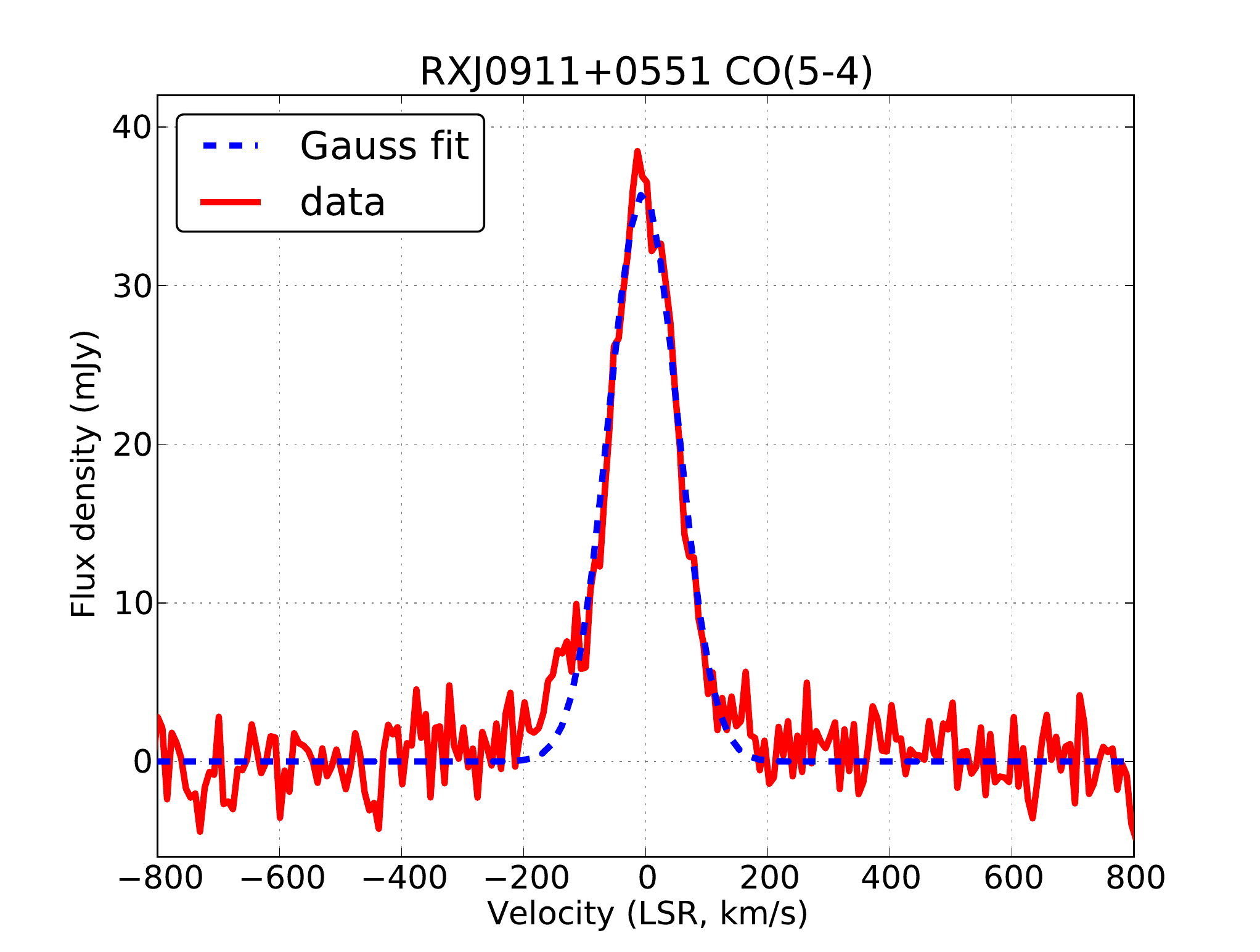}
    \includegraphics[width=0.33\textwidth]{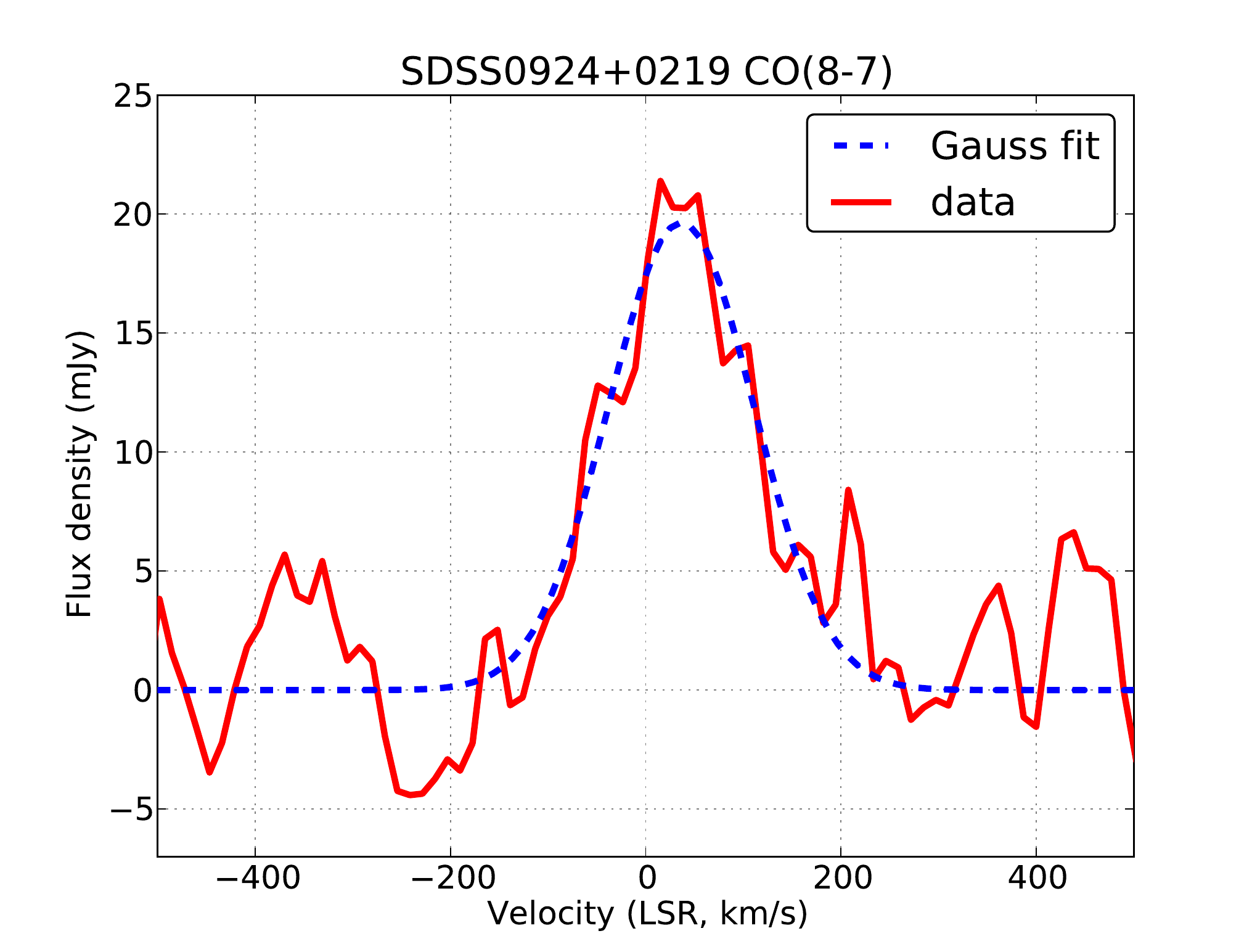}
    \includegraphics[width=0.33\textwidth]{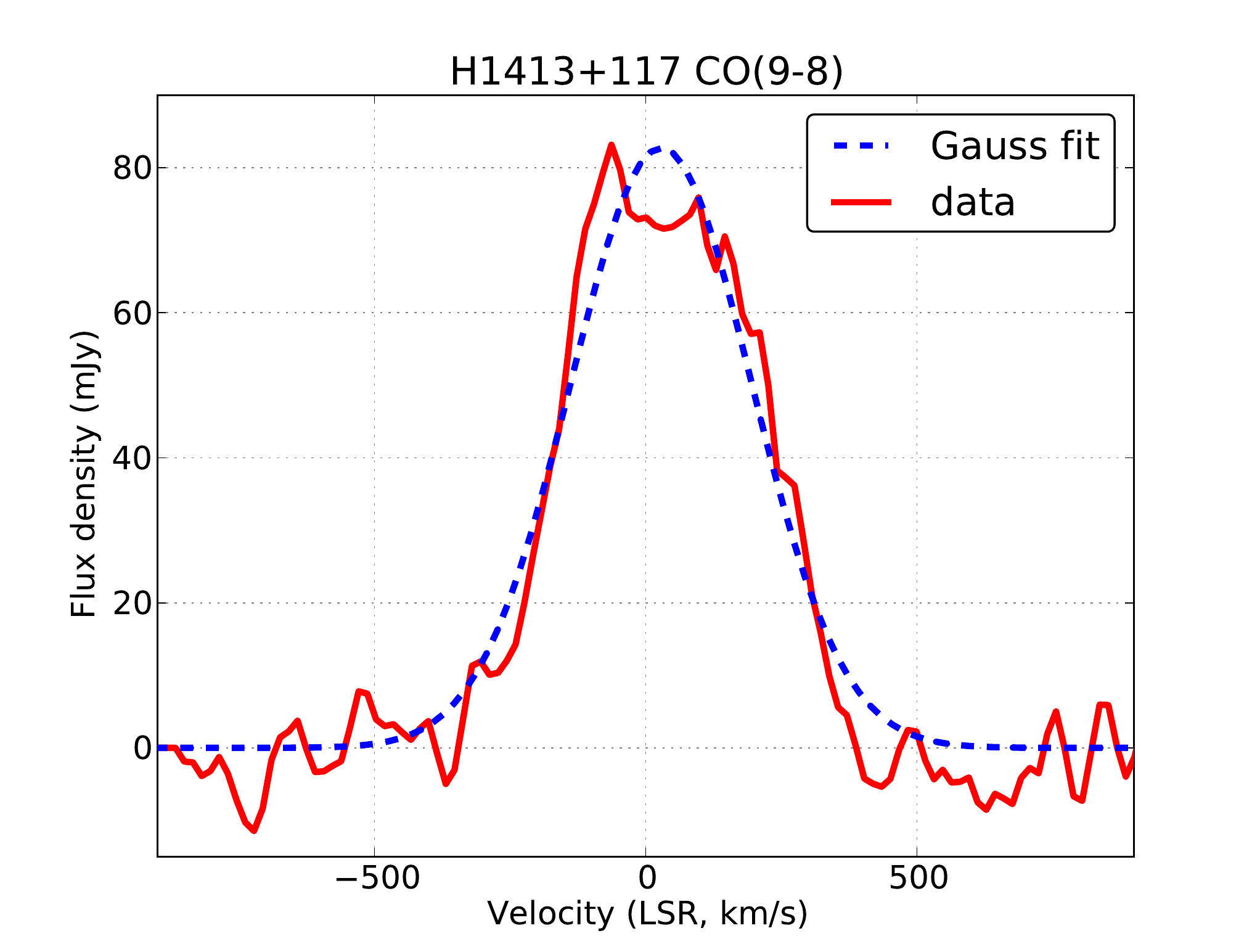}
    \includegraphics[width=0.33\textwidth]{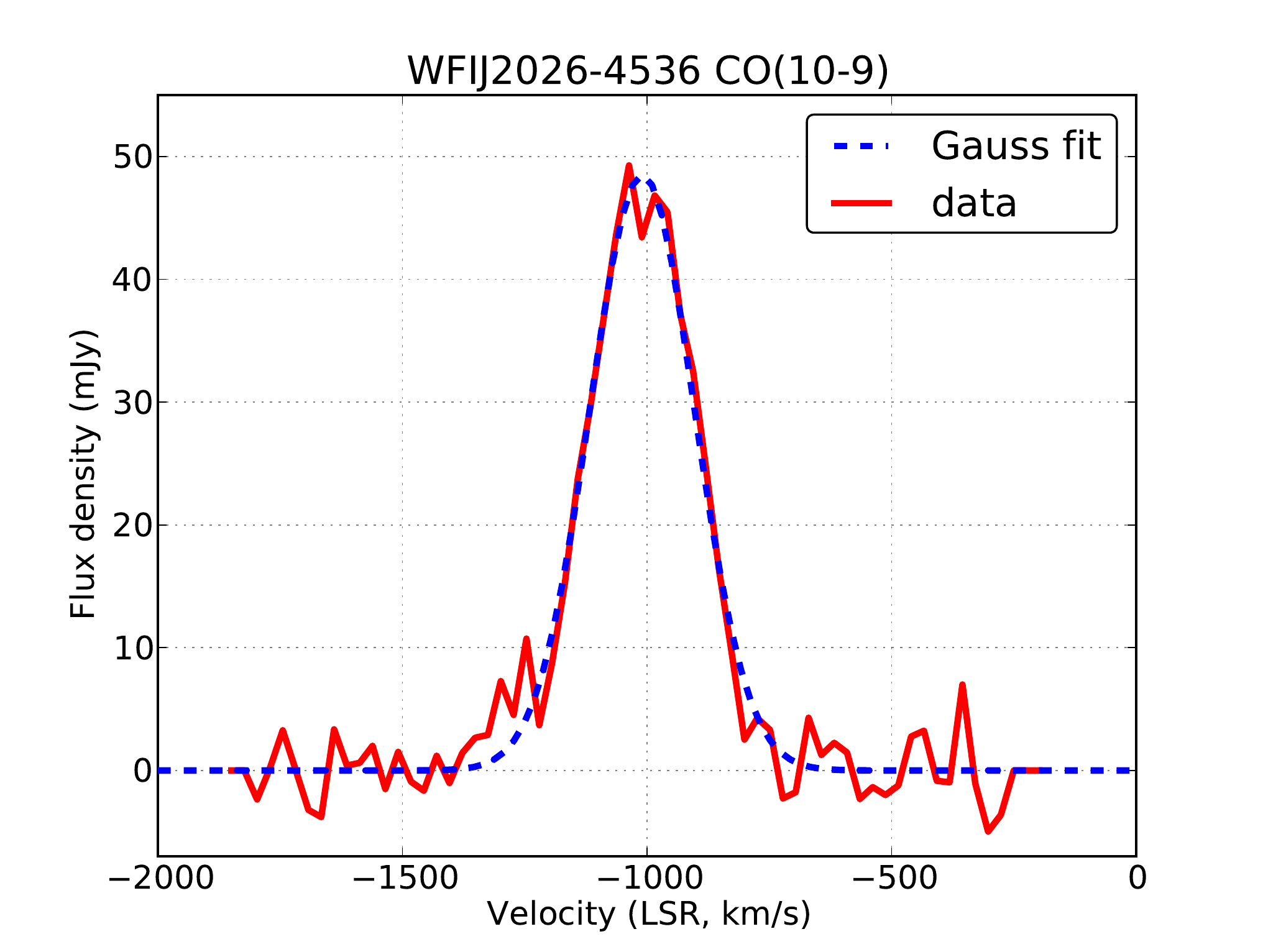}
    \includegraphics[width=0.33\textwidth]{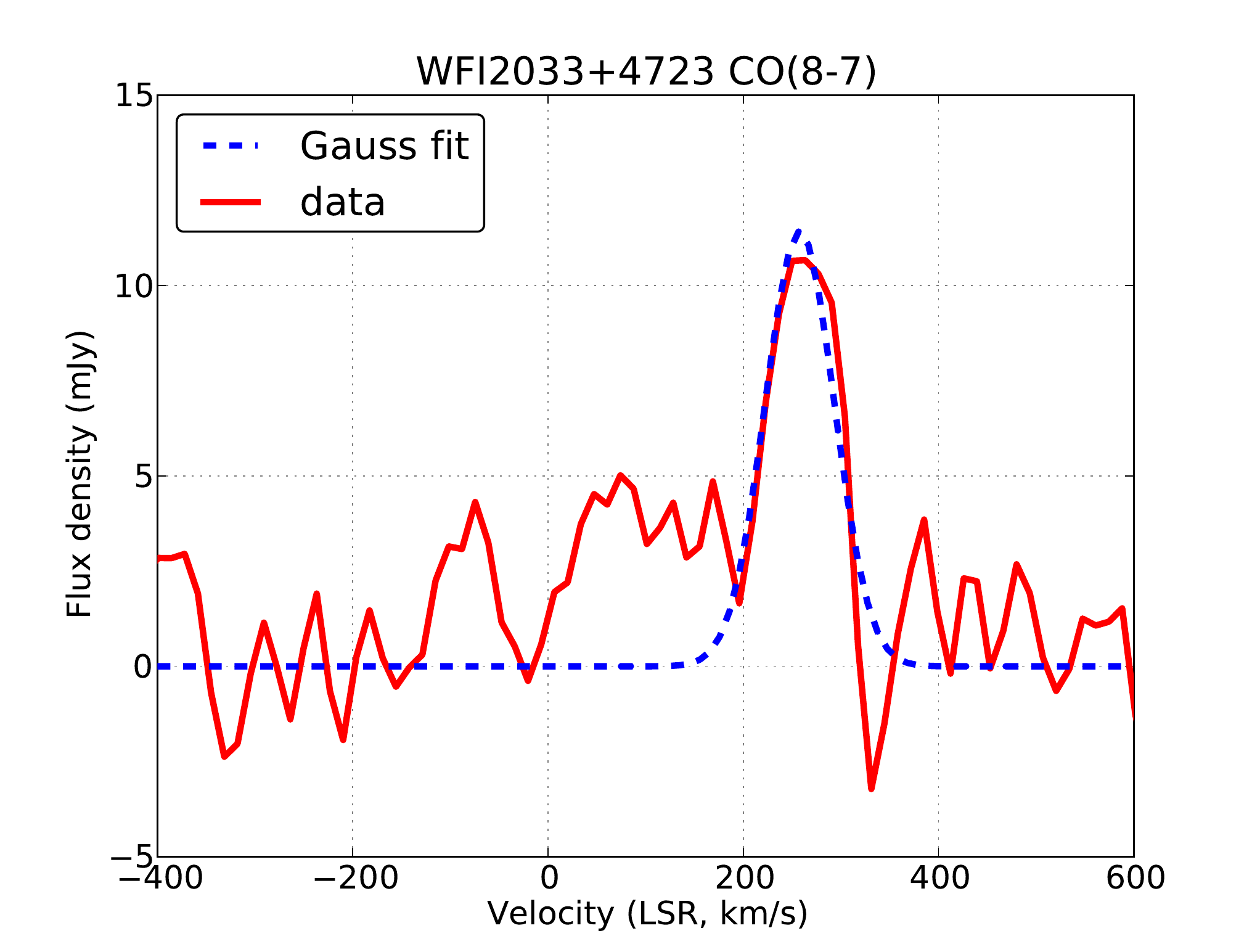}
    \caption{Line profiles for the six objects with CO observations. The red line shows the data; the blue dotted line is a Gaussian fit to the data. The systemic velocity is relative to the best redshift from the literature, using the radio definition of velocity.}
    \label{fig:line_profiles}
\end{figure*}

We detect the CO~(8--7) line emission with a spatially integrated peak flux density of 11~mJy. The FWHM of the line profile is quite narrow, at $83\pm10$~\kms, and is offset by $258\pm4$~\kms\ from the optical systemic velocity (see Fig.~\ref{fig:line_profiles}). The line profile shows a broad blue-shifted component to the main peak, but it is not clear whether this is a real feature or due to the low signal-to-noise ratio of the data. Imaging of the CO line shows resolved, compact emission that is coincident with the quasar positions (see Fig.~\ref{fig:2033_images}).

\begin{table*}
\caption{Summary of the continuum and line measurements. $\nu_{\rm line}$ is the rest frequency of the CO line. No CO line was observed for PG~1115+080. We derive the FWHM and redshift of the CO line emission based on a single Gaussian fit to the observed line profile (see Fig.~\ref{fig:line_profiles}). Here, we give flux densities and luminosities {\it uncorrected} for the lensing magnification: values corrected for the lensing magnification are presented in Tables~\ref{table:intrinsic_dust} and \ref{table:intrinsic_CO}.}
\adjustbox{width=\textwidth}{
\def\arraystretch{1.1}
    \begin{tabular}{l c c c c c c c c } \hline
        & $S_{\rm cont}$ & line & FWHM & $I_{\rm CO}$ & $L_{\rm CO}$ & $L^{'}_{\rm CO}$ & $z_{\rm CO}$ \\
        & (mJy) &  & (\kms) & (Jy~km~s$^{-1}$) & (L$_{\odot}$) & (K~km~s$^{-1}$~pc$^{2}$) &  \\ \hline
    HS~0810+2554 & $0.30\pm0.04$ & CO~(3--2) & $370\pm10$ & $6.0\pm0.2$ & $(4.9\pm0.2)\times10^{8}$ & $(3.0\pm0.1)\times10^{11}$ & $1.50849\pm0.00005$ \\
    RX~J0911+0551 & $1.35\pm0.02$ & CO~(5--4) & $133\pm3$ & $4.9\pm0.1$ & $(4.4\pm0.1)\times10^{8}$ & $(7.2\pm0.1)\times10^{10}$ & $2.79607\pm0.00002$ \\
    SDSS~J0924+0219 & $9.5\pm0.2$ & CO~(8--7) & $180\pm10$ & $4.1\pm0.3$ & $(8.9\pm0.6)\times10^{8}$ & $(3.5\pm0.3)\times10^{10}$ & $1.52495\pm0.00004$ \\
    PG~1115+080 & $2.6\pm0.1$ & -- & -- & -- & -- & --\\
    H1413+117 & $26.0\pm0.6$ & CO~(9--8) &  $394\pm8$ & $32.0\pm1.6$ & $(5.5\pm0.3)\times10^{9}$ & $(1.5\pm0.1)\times10^{11}$ & $2.55784\pm0.00004$  \\
    WFI~J2026$-$4536 & $28.31\pm0.02$ & CO~(10--9) &  $252\pm7$ & $15.2\pm0.4$ & $(1.84\pm0.05)\times10^9$ & $(3.8\pm0.1)\times10^{10}$ & $2.21217\pm0.00003$ \\
    WFI~J2033$-$4723 & $10.4\pm0.3$ & CO~(8--7) & $83\pm10$ & $1.0\pm0.1$ & $(2.1\pm0.2)\times10^{8}$ & $(8.2\pm0.8)\times10^{9}$ & $1.6629\pm0.0002$ \\
 \hline
    \end{tabular}}
    \label{table:obs2}
\end{table*}

\section{Lens Modelling}
\label{section:modelling}  

Our lens modelling analysis is adapted from the grid-based modelling technique of \cite{Vegetti:2009}. The technique simultaneously optimises for the parameters of the lens galaxy mass distribution and the source surface brightness to produce a pixellated reconstruction of the source. This method is appropriate for optical or infrared images. However, interferometers do not measure the sky surface brightness distribution directly, but measure complex `visibilities' in the Fourier ($uv$) plane. Imaging interferometric data is a non-linear process, and the resulting image fidelity can depend heavily on the sparsity of the $uv$-plane coverage and choice of deconvolution method (e.g. CLEAN, \citealt{Hogbom:1974}). 

These systematic uncertainties can be overcome by performing model fitting on the visibility data directly. This approach has been extended to lens modelling, with parametric source models \citep[e.g.][]{Hezaveh:2013,Bussmann:2013,Bussmann:2015} and also pixellated sources \citep[e.g.][]{Rybak:2015a,Rybak:2015b,Hezaveh:2016,Dye:2018}. However, these frameworks are limited by the size of the data and require averaging in time and/or frequency in order to be computationally tractable. Recently, \cite{Powell:2020} introduced several algorithmic improvements to overcome this limitation and model large interferometric data sets directly. We employ this method for optimising the visibility-plane posterior probability and source-plane regularisation.

Under the assumption of a smooth mass distribution, we parameterise the primary lens potential as a singular power-law ellipsoid with external shear, i.e. $\rho(r)\propto r^{-\zeta}$ \citep{Kormann:1994}. For all objects, with the exception of HS~0810+2554, we perform the model optimisation on the continuum data. For HS~0810+2554, we perform the lens modelling with the integrated CO~(3--2) spectral line, which provides more extended emission and a higher signal-to-noise ratio than the continuum.

The lens redshifts are not known for HS~0810+2554, H1413+117 or WFI~J2026$-$4536. However, the lens redshift only scales the angular size and mass of the perturber, so the inferred properties of the source are not affected.

Robust lens models have been produced for PG~1115+080 and WFI~J2033$-$4723, based on higher quality optical imaging of the quasars and extended arcs as part of the H$_0$ Lenses in COSMOGRAIL's Wellspring (H0LiCOW) project \citep{Chen:2019,Rusu:2020}. Therefore, in these cases we keep the lens slope, ellipticity and position angle fixed to literature values, and re-optimise only the lens galaxy position (which is not observed in our data) and shear parameters (for which we do not consider second-order corrections). In all other cases, we assume isothermal mass distributions (i.e. $\zeta=2$) and optimise for the lens galaxy position, ellipticity parameters and external shear parameters.

For WFI~J2033$−$4723, our lens model also includes a luminous satellite galaxy close to the Einstein radius of the primary lens and a massive galaxy 4~arcsec from the primary lens, both of which we parameterise as a single isothermal sphere (SIS). We find the source size changes by $\geq$10~percent with the addition of these galaxies, so it is necessary to explicitly include them in our model. The position and mass of these galaxies are fixed to the values found by \cite{Rusu:2020}. The lensing galaxy of RX~J0911+0551 also has a small satellite, however this is well within the Einstein radius and we find it does not have a significant effect on our lens modelling.

The uncertainties on the lens model parameters were obtained from the posterior distributions derived using MultiNest \citep{Feroz:2013}, with the source regularisation constant fixed to the maximum a posteriori value. Flat prior ranges were assumed for each free parameter, typically $\pm20$~percent of the optimised values. For WFI~J2026$-$4536, it was necessary to keep the lens position fixed to the maximum a posteriori (consistent within 20~mas of the optical position) for stability during the nested sampling. The likelihood-weighted posterior probability distributions for the lens parameters are given in Table~\ref{table:lens_models}.

Reconstructed velocity cubes were generated from the spectral line data by optimising for the source regularisation, using the best fit lens model. For WFI~J2033$-$4723 and SDSS~J0924+0219, we use only three velocity channels and restrict the $uv$ data to within a maximum baseline of 300~m to improve the surface brightness sensitivity. 

\section{Results}
\label{section:results}

In this section, we present the results of the source reconstruction and an analysis of their physical properties.

\subsection{Source reconstructions}
\label{section:reconst}

The maximum a posteriori lens models are shown in Figs.~\ref{fig:reconst_cont} and \ref{fig:reconst_line} of the Appendix. The residual maps are a dirty image produced after the model has been subtracted from the visibility data, normalised to the rms noise in the data. We find our lens models fit the data well and the residual surface brightness features are below the 4~$\sigma$ level.

\begin{figure*}
    
    \begin{subfigure}[t]{\textwidth}
    \centering
    \includegraphics[width=0.84\textwidth]{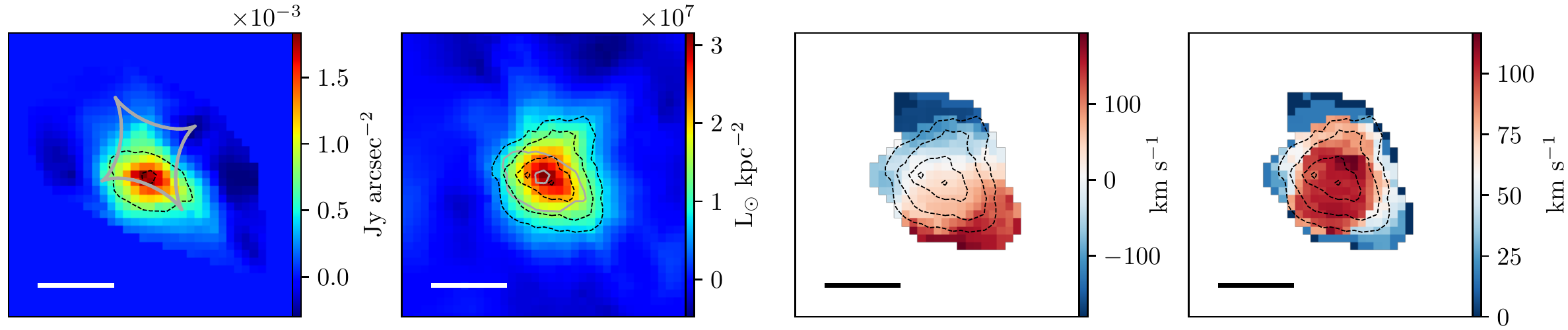}
    \caption{Reconstructed dust and CO~(3--2) moments for HS~0810+2554. We show signal-to-noise ratio contours of multiples of 5 for the line intensity.}
    \label{fig:0810_source}
    \end{subfigure}
    
    \begin{subfigure}[t]{\textwidth}
    \centering
    \includegraphics[width=0.84\textwidth]{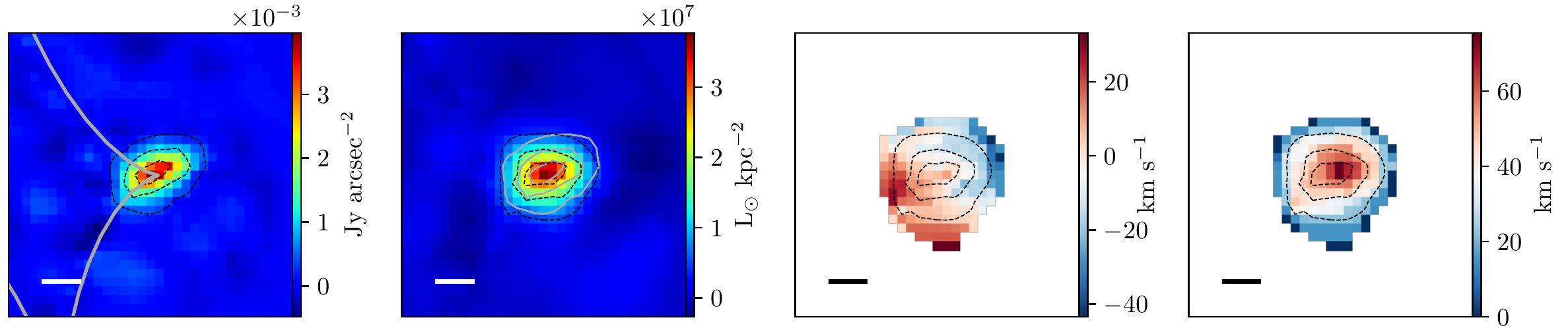}
    \caption{Reconstructed dust and CO~(5--4) moments for RX~J0911+0551. We show signal-to-noise ratio contours of 3 and multiples of 5.}
    \label{fig:0911_source}
    \end{subfigure}

    \begin{subfigure}[t]{\textwidth}
    \centering
    \includegraphics[width=0.84\textwidth]{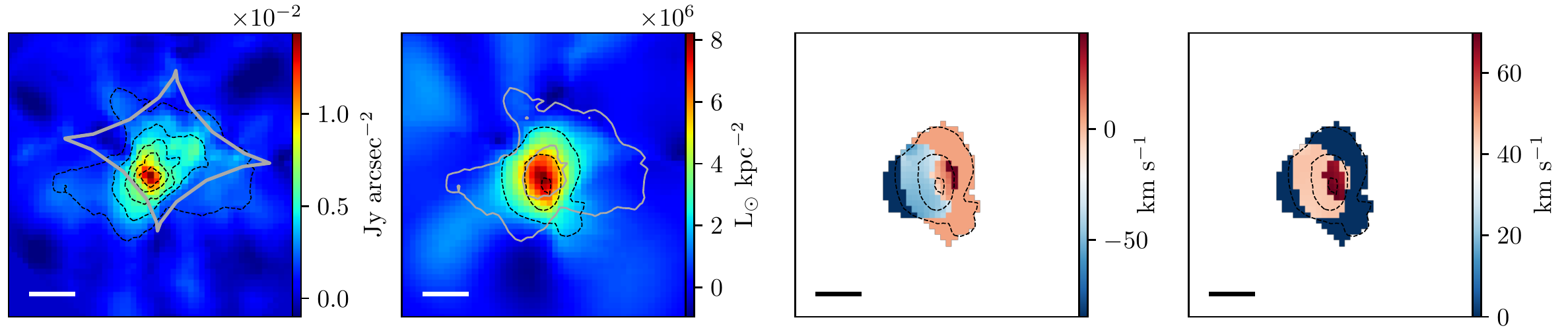}
    \caption{Reconstructed dust and CO~(8--7) moments for SDSS~J0924+0219. We show signal-to-noise ratio contours in multiples of 3.}
    \label{fig:0924_source}
    \end{subfigure}

    \begin{subfigure}[t]{\textwidth}
        \centering
    \includegraphics[width=0.84\textwidth]{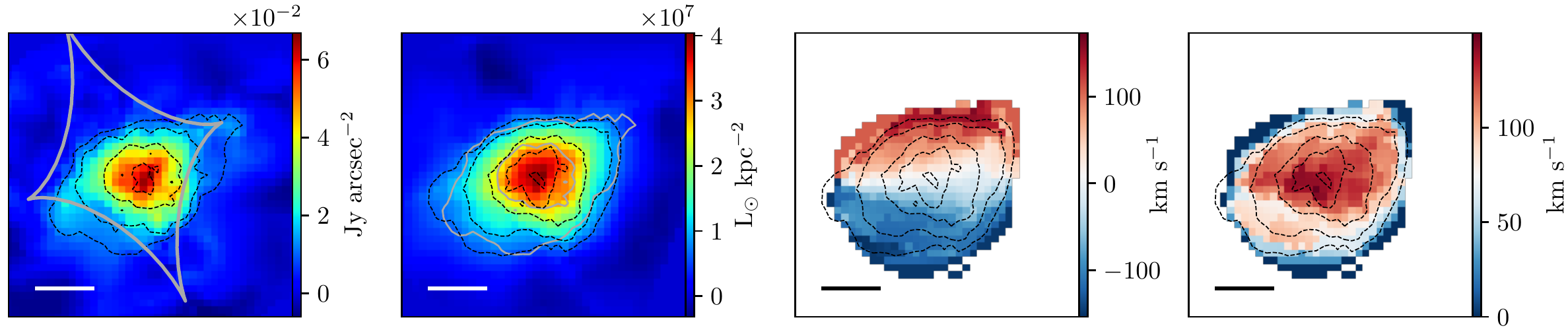}
    \caption{Reconstructed dust and CO~(9--8) moments for H1413+117. We show signal-to-noise ratio contours of 3 and multiples of 5.}
    \label{fig:1413_source}
    \end{subfigure}
    
    \begin{subfigure}[t]{\textwidth}
        \centering
    \includegraphics[width=0.84\textwidth]{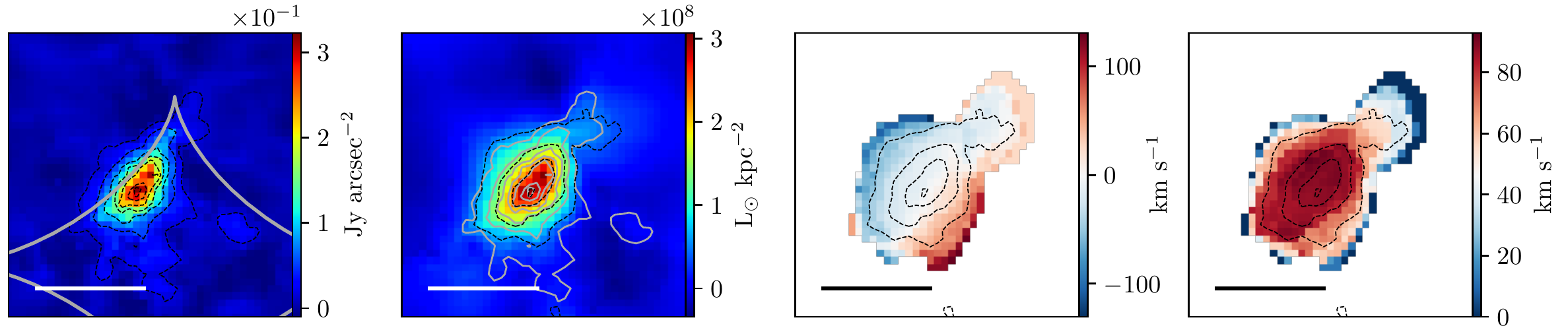}
    \caption{Reconstructed dust and CO~(10--9) moments for WFI~J2026$-$4536. We show signal-to-noise ratio contours of 3 and multiples of 10.}
    \label{fig:2026_source}
    \end{subfigure}

    \begin{subfigure}[t]{\textwidth}
        \centering
    \includegraphics[width=0.84\textwidth]{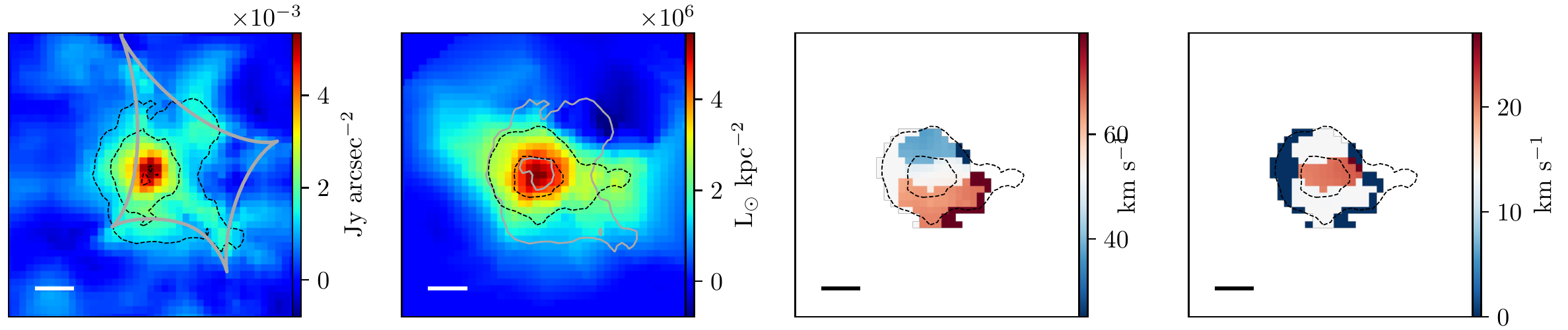}
    \caption{Reconstructed dust and CO~(8--7) moments for WFI~J2033$-$4723. We show signal-to-noise ratio contours of 3 and multiples of 5.}
    \label{fig:2033_source}
    \end{subfigure}
    
    \caption{Reconstructed dust emission and moments maps for the six lensed quasar systems with spectral line data. The dust emission is shown in units of Jy~arcsec$^{-2}$ with the lensing caustics shown in grey. Dashed contours show signal-to-noise ratio contours, starting at 3. 
    Continuum contours are shown in grey over the moment 0 image. The bar shows 1~kpc at the redshift of the source. The line reconstructions are generated by masking pixels in each channel below $3\sigma$. The CO velocity is corrected to the redshift derived from a Gaussian fit to the CO line profile.}
    \label{fig:moments}
\end{figure*}

\begin{figure}
    \includegraphics[width=0.47\textwidth]{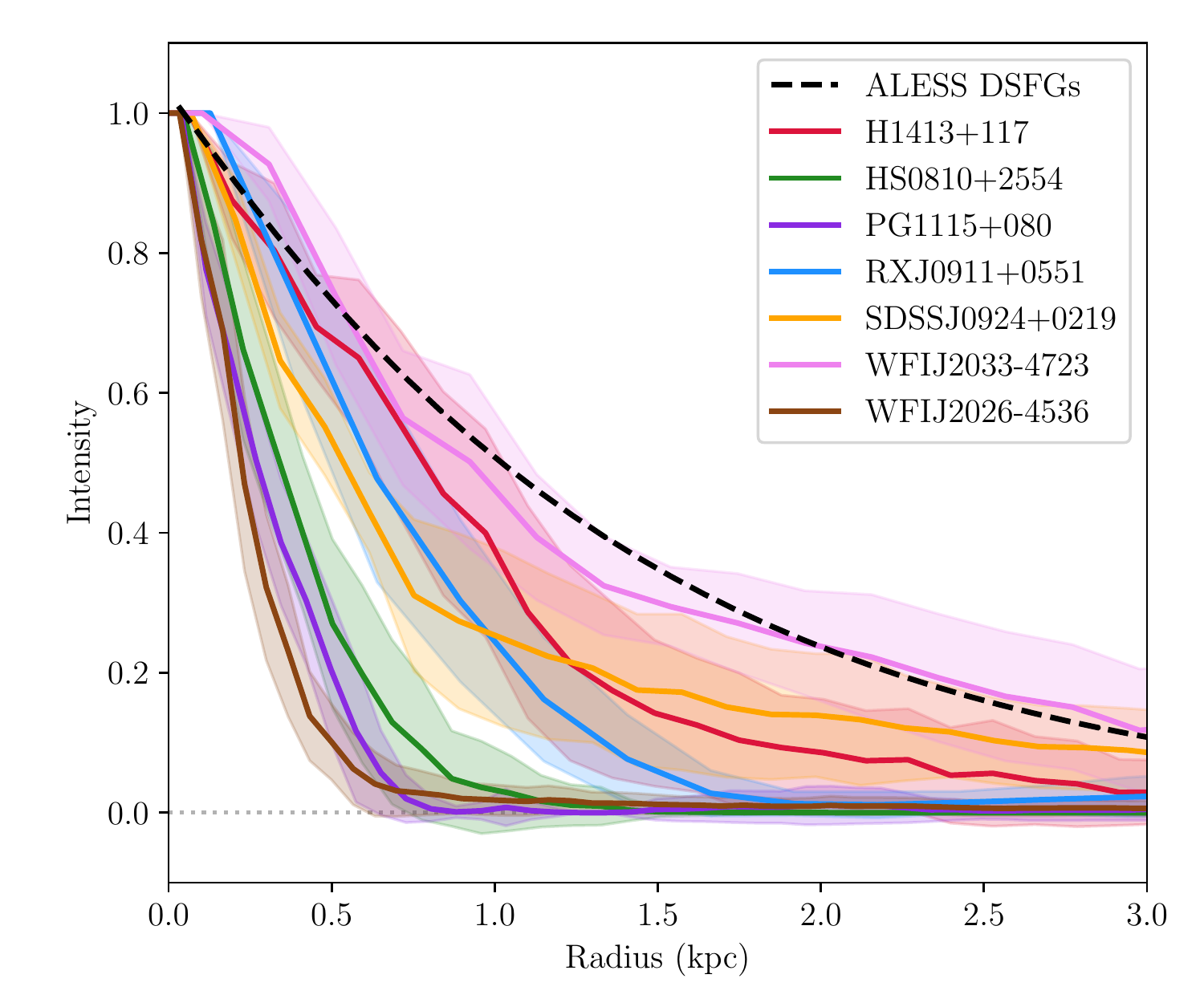}
    \caption{Normalised, azimuthally averaged surface brightness profiles of the reconstructed dust emission from the quasar hosts, accounting for ellipticity. The shaded regions show the respective standard deviation of the dust emission. The radius of HS~0810+2554 and PG~1115+080 are scaled to the size of the minor axis, based on the axis ratio from the S\'ersic model fits (see Section~\ref{section:reconst}). The black dotted line shows the mean profile of (unlensed) ALESS DSFGs from S\'ersic fits \citep{Hodge:2016,Hodge:2019}. }
    \label{fig:profiles}
\end{figure}

The reconstructed sources and moment maps are shown in Fig.~\ref{fig:moments}. For SDSS~J0924+0219 and WFI~J2033$-$4723, the dust emission is extended and resolved into clumpy features on kpc-scales. For HS~0810+2554, RX~J0911+0551, PG~1115+080, H1413+117 and WFI~J2026$-$4536, the reconstructed dust emission appears to be very compact and does not show evidence for clumpy features on the same scales. We caution that, for HS~0810+2554, there may be extended dust emission that is not detected here due to the low signal-to-noise ratio of the data. The apparent smoothness of the compact systems may be due to angular resolution limitations, however, we note that other unlensed dusty galaxies at $z\sim2$ are found to have smooth dust distributions (e.g. \citealt{Falgarone:2017}). The radially averaged surface brightness distributions of the reconstructed continuum sources are shown in Fig.~\ref{fig:profiles}. 

The robustness of the reconstructed surface brightness distribution of the lensed source depends on the quality of the data. As our source is non-parametric, our fitting down to the level of the noise can allow correlated noise features in the dirty image to be absorbed into the source reconstruction. Our investigations find that the uncertainties in the source surface brightness resulting from noise artefacts are significantly larger than the uncertainties due to the lens model parameters (Rizzo et al. in prep). To estimate these uncertainties we assume our maximum a posteriori source model and create mock data with 100 different realisations of the noise at the level measured in the real data\footnote{We find that 100 realisations is sufficient to produce a smooth distribution that can be approximated by a Gaussian function.}. We create reconstructions of these mock data sets and measure the mean and standard deviation of the surface brightness in each pixel (in each channel, for spectral line data). This allows us to discriminate features in the reconstructed source from noise artefacts. 

To estimate the uncertainty on the source size, we fit a S\'ersic profile \citep{Sersic:1963} to the source from each noise realisation of continuum and integrated line emission using a basin-hopping optimisation algorithm within the {\sc SciPy} package \citep{Wales:1998,Jones:2001}. For the integrated line emission of SDSS~J0924+0219 and WFI~J2033$-$4723, we fix the index of the S\'ersic profile to that found for the continuum as the fitting was found to be unstable due to the low signal-to-noise ratio of the data. We assume the standard deviation of the inferred S\'ersic parameters as their uncertainty. We estimate the magnification of each lens system by generating a model for the lensed emission for each noise realisation, where source pixels are masked below a signal-to-noise ratio of 4. This value is chosen so as not to include any noise features that may be present in the individual realisations that may bias the magnification to lower values. We take the mean and standard deviation of the magnifications as the inferred value and its uncertainty (given in Tables~\ref{table:intrinsic_dust} and \ref{table:intrinsic_CO}). 

In order to test the robustness of our inferred source sizes, we created mock images of Gaussian sources lensed by the maximum a-priori lens model for each object. The Gaussian sources were located at the position of peak surface brightness of the maximum a-priori source and given a range of effective radii. Mock visibilities were created with the same $uv$-coverage, baseline-based rms noise and image peak surface brightness as the real observations. We modelled and analysed these mock data sets with the same methodology as the real data, optimising for the source regularisation parameter. We found that the inferred source sizes were inflated by a constant value that can be related to the strength of the source regularisation. This inflation has little effect where the source is extended on kpc-scales, but becomes significant where the source is very compact. We use these simulations to infer a correction to the sizes derived from the S\'ersic fits.

From our simulations, we confirmed that we can recover the source size from the real observations. However, for PG~1115+080 and HS~0810+2554 continuum, we determine that the lensed emission is only marginally resolved in the tangential direction. This results in a source that is constrained only by the angular resolution of the data in one dimension. Therefore, for these two cases, we assume the minor axis of the fitted S\'ersic profile as the source size. The inferred magnifications from the real data are generally consistent with those inferred from the mock data sets, allowing for minor differences due to source structure. However, the inferred magnifications of PG~1115+080 and HS~0810+2554 (continuum) were underestimated due to the effect of source inflation, so we infer a correction using the mock data sets.

A consequence of the lack of radial resolution is that there is a degeneracy between the normalised convergence ($\kappa_0$) and the slope of the lens mass density profile ($\zeta$). The effect of this degeneracy is a geometric scaling of the reconstructed source. In most cases we assume an isothermal profile for our lens models (i.e. $\zeta\equiv2$), however the empirical scatter in the total mass slope of early-type galaxies (0.16; \citealt{Auger:2010}) implies an additional uncertainty in the source size. We tested this by modelling PG~1115+080 with a profile fixed to a value $\pm1\sigma$ from the median found by \cite{Auger:2010}. We find that the size and normalisation of a S\'ersic fit to the reconstructed source changes by $\leq10$~percent. Therefore, where we assume only an isothermal profile is fit to the data, we propagate an additional $\pm10$~percent error into the uncertainty in our estimated sizes to account for the empirical scatter in $\zeta$.

All the derived parameters of the continuum and line emission for each source can be found in Tables~\ref{table:intrinsic_dust} and \ref{table:intrinsic_CO}, respectively. Note that the given magnifications are {\it mean} magnifications, as the magnification changes across the source and between velocity channels. 

\begin{figure}
    \centering
    \includegraphics[width=0.35\textwidth]{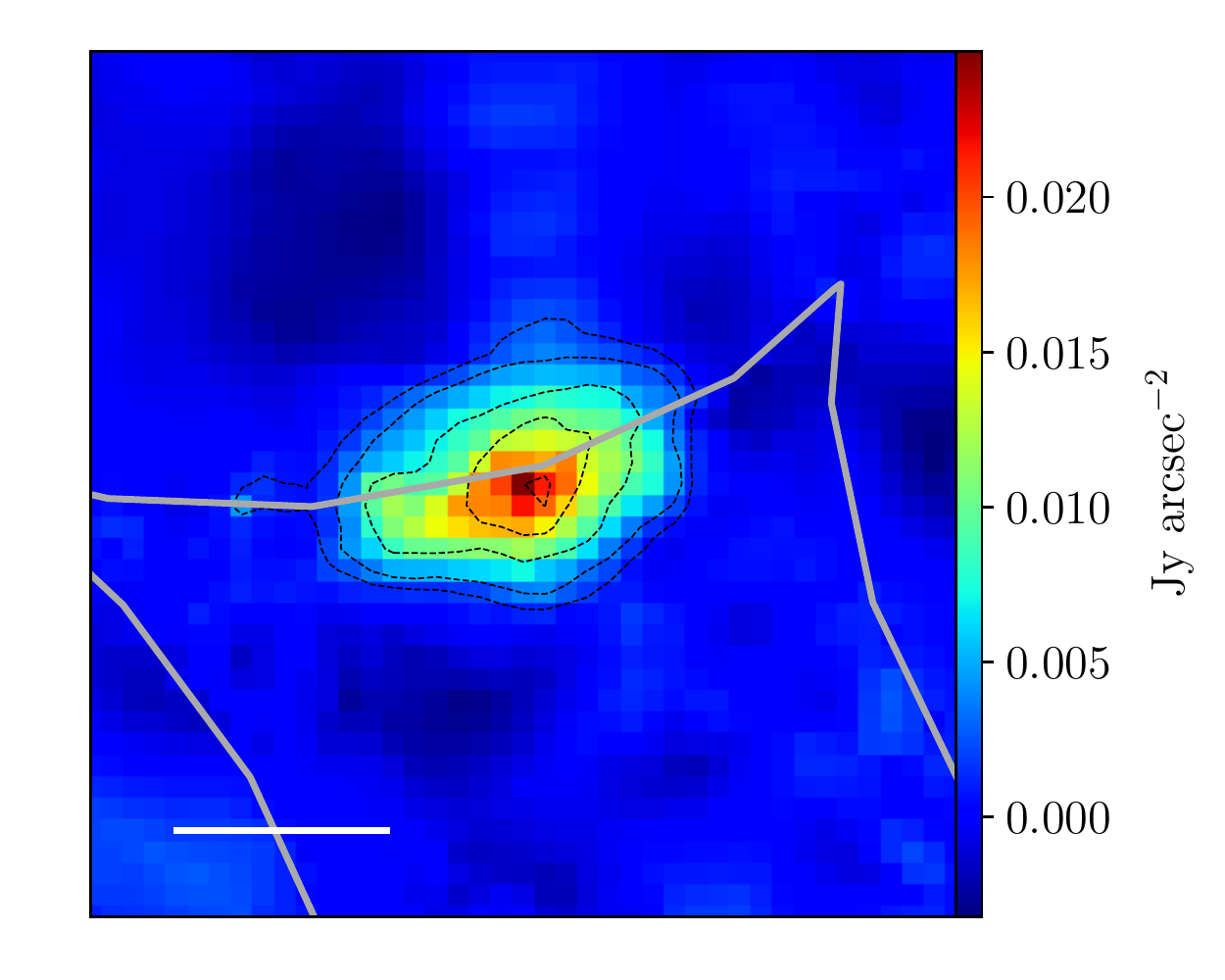}
    \caption{Reconstructed dust emission of PG~1115+080, shown in units of Jy\,arcsec$^{-2}$ with the lensing caustics shown in grey. The white bar shows 1~kpc at the redshift of the source.}
    \label{fig:1115_source}
\end{figure}


\subsection{Dust properties}
\label{section:sed_fitting}

We use the new sub-mm data to refine the spectral energy distribution (SED) models for our sample. The SED fitting follows a similar procedure described by \cite{Stacey:2018a} and uses ancillary data listed therein. \cite{Stacey:2018a} assumed the case of optically thin dust emission. However, there is growing evidence that an optically thin dust model under-predicts the dust temperature of high-redshift galaxies because the dust remains optically thick into the far-infrared (\citealt{Riechers:2013}, \citealt{Cortzen:2020}). This difference has little effect on the SED shape (hence, integrated luminosity) but shifts the peak of the dust emission to longer wavelengths, such that warmer dust temperatures can appear colder (by a few degrees, on average). Therefore, we fit a dust model described by
\begin{equation}
    S_{\nu}\propto (1-e^{-\tau(\nu)}) \frac{\nu^{3}}{e^{h \nu /k T_{\rm d}} - 1},
\end{equation}
where $\tau(\nu)=(\nu/\nu_0)^{\beta}$ and $\nu_0$ is the frequency at which the dust opacity is unity, which we assume to be rest-frame 3~THz (100~$\upmu$m) as predicted by theory \citep{Draine:2006} and supported by observational studies \citep{Riechers:2013}.

We leave both the dust temperature and emissivity (which governs the steepness of the Rayleigh-Jeans slope of the spectrum) as free parameters. Depending on the available ancillary data, we also fit a power-law component described by
\begin{equation}
    S_{\nu}\propto\nu^{\alpha},
\end{equation} where $\alpha$ is the spectral index, to account for optically thin radio synchrotron emission.

We apply a Markov Chain Monte Carlo (MCMC) analysis using the Python implementation {\sc{emcee}} \citep{Foreman-Mackey:2013} to infer the marginalised posterior distributions for each free parameter and the integrated FIR luminosity. The SED models are shown in Fig.~\ref{fig:seds1} of the Appendix. The parameters from the SED modelling, corrected for the lensing magnification, are listed in Table~\ref{table:intrinsic_dust}.

The dust temperature we infer from our SED fitting is an {\it effective} dust temperature, which should not be considered a true measure of dust temperature, but weighted by multiple components of dust emission. This may include a cold dust component from cirrus dust of the interstellar medium ($\sim20$~K) and a warmer dust component associated with star-forming regions ($\sim50$~K). Indeed, \cite{Swinbank:2014} find DSFGs require at least three dust components to account for the full mid-infrared (MIR) to FIR SED. AGN host galaxies may also have contributions from AGN-heated dust emission. For HS~0810+2554 and H1413+117, we account for a warm dust component, which we assume is associated with the AGN. For HS~0810+2554, $\beta$ is kept fixed in the SED modelling as we do not have sufficient data points to fit both $\beta$ and two dust components. We fix $\beta$ to a value of 2, as 1.5 cannot produce a satisfactory fit to the data\footnote{For comparison, the median and distribution of the parent sample is $\beta=2.0^{+0.4}_{-0.5}$, based on optically thin dust models \citep{Stacey:2018a}.}. With the exception of H1413+117 and HS~0810+2554, we do not have data at rest-frame wavelengths $\sim10$ to 100~$\upmu$m to consider additional dust emission components. 

Note that, while we leave $\beta$ as a free parameter, its value is not necessarily physically significant as, for poorly sampled SEDs, the strong temperature--$\beta$ degeneracy absorbs the observational noise in the data \citep{Juvela:2012}. We allow this as a free parameter to better estimate the true uncertainty on the fitted temperature and luminosity.

We follow the \cite{Helou:1988} definition of FIR luminosity ($L_{\rm FIR}$), as the integrated spectra from 40 to 120~$\upmu$m. We convert this to total infrared luminosity (8 to 1000~$\upmu$m) using a colour correction factor of 1.91 (i.e. $L_{\rm IR} = 1.91\times L_{\rm FIR}$) given by \cite{Dale:2001} to account for spectral features in the MIR, assuming our fitted cold dust component is associated only with obscured star formation. We convert this to a star formation rate (SFR; \sfr) assuming a Salpeter initial mass function with the conversion factor of \cite{Kennicutt:1998},
\begin{equation} 
SFR = \frac{L_{\rm IR}}{5.8\times10^{9}} ,
\label{eq:sfr}
\end{equation} where $L_{\rm IR}$ is in units of L$_{\odot}$\footnote{Note that there is evidence that starbursts have top-heavy IMFs \citep{Zhang:2018}. Our assumption of a Salpeter IMF may overestimate the star formation rate by overestimating the number of low-mass stars; thus, a different choice of IMF would result in an overall re-scaling of the star formation rates. However, as we are mostly concerned with {\it relative} star formation rates in this work, our choice of IMF does not significantly impact our results.}. Using these inferred star formation rates and magnifications, we transform the dust emission to units of star formation rate surface density. 

We follow \cite{Dunne:2000} to estimate the dust mass from our SED fits using
\begin{equation}
    M_{\rm dust} = \frac{D_{\rm L}^{2}\ S_{850}^{\rm obs}}{(1+z)\  \kappa_{850}\ B_{850} (T_{\rm d})} , 
\end{equation} where $S_{850}^{\rm obs}$ is the observed flux density at rest-frame wavelength 850~$\upmu$m (350~GHz), $D_{\rm L}$ is the luminosity distance, $\kappa_{850}$ is the dust mass opacity (0.077~m$^{2}$~kg$^{-1}$, from \citealt{Dunne:2000}), and $B_{850}(T)$ is the value of a black body of temperature $T$ at 850~$\upmu$m. 

It has often been found that a dust mass estimated from single temperature SED model may underestimate the total dust content, as most of the mass is in cold cirrus dust and not the dust that contributes most to the effective (luminosity-weighted) temperature (e.g. \citealt{Scoville:2014}). However, there is also evidence that multi-component dust models could lead to overly large dust masses \citep{Cortzen:2020} and it has not been established to what extent the approach is appropriate for quasar host galaxies. Note that, if the temperature of dust in the diffuse ISM is much lower than the effective dust temperature for quasar hosts, the dust masses derived for of our sample could be underestimated. We attempted fitting two-component models to our SEDs but, as most are sparsely sampled, the uncertainties on the mass of the cold component was very large and encompasses the masses derived with a single temperature fit. Thus, we report the dust masses from a single-temperature fit, but with this caveat.

The dust temperatures, emissivities, intrinsic luminosities, star formation rates, star formation rate surface densities and dust masses for the sample are given in Table~\ref{table:intrinsic_dust}. For H1413+117, the nearby field source we identify in our ALMA imaging (Section~\ref{section:1413_obs}) may cause modest confusion in SED fitting, which leads to overestimates of the inferred star formation properties of the source of interest. Photometric measurements at multiple frequencies will be required to accurately de-blend this emission from the total infrared luminosity. Such an analysis is beyond the scope of this work, however, the consistent relative flux density of the objects at 100 and 290~GHz suggests that the neighbour contributes $\sim20$~percent to the photometry of H1413+117. Therefore, our measurements based on infrared luminosity are likely to be overestimated by a similar percentage. We estimate the effect of this by scaling all the photometric measurements above 300~GHz by a factor of 0.8. The physical properties based on the re-scaled data are also shown in Table~\ref{table:intrinsic_dust}.

\subsection{Molecular gas properties}

Fig.~\ref{fig:size_temp} shows the intrinsic effective radius of the dust and CO emission against effective dust temperature and total infrared luminosity (8--1000~$\upmu$m). The sample in this work is quite heterogeneous and different rotational transitions of CO were observed. These CO lines correspond to different physical conditions in the ISM. At one extreme, CO~(10--9), CO~(9--8) and CO~(8--7) trace warm gas (150--200~K) in regions of intense star formation or that are associated with AGN. For the four objects where we observe one of these lines (SDSS~J0924+0219, H1413+117, WFI~J2026$-$4536 and WFI~J2033$-$4723), the molecular gas is a similar size or more compact than the dust continuum. For HS~0810+2554 and RX~J0911+0551, in which we observe CO~(5--4) and CO~(3--2), which are usually associated with star-forming regions, we find the gas to be more extended than the dust. Different relative sizes of the emission regions of CO line transitions have been found for other DSFGs 
(e.g. \citealt{Apostolovski:2019}). While little can be inferred about the relative contributions to the energy budget from observations of a single line transition, the observed sizes are consistent with high levels of star formation and with the expectations for a radially decreasing gas column density and temperature \citep{Weiss:2007}.

We convert our line flux densities to luminosities ($L_{\rm CO}$ and $L^{\prime}_{\rm CO}$) with relations as given by \cite{Solomon:2005}. We use the source-plane reconstructions to derive the intrinsic line properties to account for differential magnification across the velocity channels. As some line velocity components may be more strongly magnified than others, the integrated line emission measured in the lens-plane is not necessarily the same as the intrinsic line emission multiplied by the mean magnification (i.e. $I^{\rm lensed}_{\rm CO} \neq \bar{\mu}\ \times I_{\rm CO}$). The intrinsic line luminosities (corrected for lensing magnification) are given in Table~\ref{table:intrinsic_CO}. The reconstructed line emission in units of luminosity surface density, as well as the velocity field and velocity dispersion, are shown for each source in Figs.~\ref{fig:0810_source} to \ref{fig:2033_source}. 

In all cases, the reconstructed line velocity structure shows evidence of rotation around the peak of the continuum emission, suggesting the gas is in disc. From the reconstructed line emission, we estimate the enclosed dynamical mass of the galaxies, assuming
\begin{equation}
    M_{\rm dyn} = \frac{R_{\rm 2 eff} V_{\max}^{2}}{G} ,
\end{equation} where $R_{\rm 2 eff}$ is twice the effective radius of the CO emission based on the S\'ersic model fit described in Section~\ref{section:reconst}. Here, $\sigma$ is the maximum rotational velocity, assuming $V_{\rm max} = V_{\rm obs} / \sin i$, where $i$ is the inclination angle. We derive $\sin i$ from the S\'ersic model fit according to
\begin{equation}
    \sin i = \sqrt{1-\left( \frac{b}{a} \right) ^{2}},
\end{equation} where $b$ is the minor axis and $a$ is the major axis. Where we do not have a reliable constraint on the axis ratio due to a lack of radial resolution (i.e. RX~J0911+0551, PG~1115+080 and WFI~J2033$-$4723) we assume an inclination angle of $45\pm15$~deg, corresponding to axis ratios of 0.5 to 0.9. In this case, we do not consider axis ratios below 0.5 as they are smaller than the apparent axis ratio. This method assumes that the circular velocity accurately traces the gas rotation, ignoring the effect of a non-spherical gravitational potential and turbulent pressure. Simulations suggest that this assumption could result in an underestimate of the dynamical mass at very small or large radii, but is accurate at the intermediate radii ($\sim$\,1~kpc) we probe here \citep{Wellons:2019}. The dynamical masses for the quasar hosts are given in Table~\ref{table:intrinsic_CO}. Note that the dynamical mass inferred from the CO lines investigated here (particularly the high-excitation lines) may not probe the total mass of the galaxy \citep{Casey:2018}.

In addition to the availability of gas, the level of star formation in a galaxy depends on the local balance between turbulent pressure and self-gravity. In dynamically unstable systems, self-gravity dominates over turbulent pressure allowing gas to fragment and efficiently form stars. As the systems in this work are known to have high star formation rates and, in some cases, have clumpy non-axisymmetric dust structures, we expect the gas discs to be globally dynamically unstable. As shown in Figs.~\ref{fig:moments} to \ref{fig:1115_source}, the reconstructed velocity maps appear to show radially increasing rotational velocity and high central dispersions. However, these 2-dimensional maps may be strongly affected by beam-smearing, which can misguide interpretation of the dynamics of the systems (velocity dispersion in particular, e.g. \citealt*{Lelli:2010}). Also note that for SDSS~J0924+0219 and WFI~J2033$-$4723 we use only three velocity channels for the reconstructions. We defer a 3D analysis to future work, combining lens and source kinematic modelling as introduced by \cite{Rizzo:2018} (see also \citealt{Rizzo:2020}).

\subsection{Comparison with dusty star-forming galaxies}

We make a comparison between our lensed quasar hosts and high angular resolution ALMA observations of DSFGs. These samples include galaxies from the LABOCA ECDFS Submm Survey (ALESS; \citealt{Hodge:2016}), strongly lensed DSFGs selected with {\it Herschel} surveys, and several DSFGs at $z>4$. The choice of these samples is motivated by the different selection biases that dominate in FIR/sub-mm surveys: we choose these samples to represent a diverse range of properties observed for DSFGs and combat the effects of selection bias.

DSFGs in the ALESS sample cover a similar redshift range ($z\sim2$, including one at $z=3.4$) and infrared luminosities to our lensed quasar sample. We perform SED fitting for the ALESS sample with photometry from \cite{Swinbank:2014} and \cite{Hodge:2019}. For consistency, we use the same methodology as for the quasar hosts to obtain effective dust temperatures and infrared luminosities with an optically thick dust model. Following this methodology, we only include the 11 ALESS galaxies where there are sufficient detections at FIR--sub-mm wavelengths to constrain the dust temperature.

We note that sub-mm source selection may preferentially identify sources with lower dust temperatures at a given redshift, due to the effect of negative $k$-correction \citep{Chapin:2011}. This does not strongly influence the selection of the quasar hosts (which were mostly selected on the basis of their optical properties and lens configuration, as detailed in Section~\ref{section:obs}) but may influence the selection of the coeval DSFGs. To account for this selection bias, we include a sample of DSFGs at $z>4$ and several lensed DSFGs. The $z>4$ sample includes one of the ALESS DSFGs \citep{Hodge:2019} and four other objects that are spatially resolved, and have spectroscopic redshifts and multiple detections at FIR--sub-mm wavelengths. These include GN20 \citep{Hodge:2015}, HFLS3 \citep{Riechers:2013}, AzTEC-3 \citep{Riechers:2014}, the two components of SGP38326 \citep{Oteo:2016} and the two components of ADFS-27 \citep{Riechers:2017}. A summary of the DSFGs is given in Table~\ref{table:dsfgs} of the Appendix. 

The lensed DSFGs were selected in FIR surveys with {\it Herschel}/SPIRE \citep{Negrello:2017}. We perform the same analysis for the lensed DSFGs as for the lensed quasar sample; further details of the lens modelling and source reconstructions will be presented in a follow-up paper (Stacey et al. in prep). The selection of lensed DSFGs is expected to be strongly biased towards compact systems, where the flux density is boosted by high magnification \citep{Serjeant:2012,Hezaveh:2012}. We therefore expect these lensed systems to be examples of more compact DSFGs.

Fig.~\ref{fig:size_sigma} (left) shows the continuum source size (effective radius) of the quasar hosts in this work and the $z\sim2$ DSFGs. This shows that more compact dust emission is associated with higher effective dust temperatures. A Kendall rank test \citep{Kendall:1945} for correlation between dust size and temperature yields a coefficient of $\tau=-0.37$ with a significance of $p=0.04$ (where $p=0.05$ is often taken as significant). Such a relationship is expected as a natural consequence of the Stefan-Boltzmann law, which relates the luminosity of a black body to its temperature and physical size. If the luminosity and temperature of the thermal dust emission we observe from the quasar hosts is related only to star formation, they should follow this relation.

Fig.~\ref{fig:dustlir} shows the intrinsic infrared luminosity (and equivalent star formation rate) against effective dust temperature for our sample and the DSFGs. The relationship expected for the modified Stefan-Boltzmann law from \cite{Yan:2016} is also shown for different source sizes, assuming a dust model with $\beta=1.5$ and $\nu_0=3$~THz. The DSFGs and quasar hosts can be seen to broadly follow this relationship with some scatter, as expected due to source structure as well as variations in dust emissivity and opacity. Fig.~\ref{fig:dustlir} shows that two lensed quasars have sizes in dust emission larger than implied by their dust temperature, which could indicate the sub-mm emission has a significant contribution from AGN-heating. However, we note that for these systems (SDSS~J0924+0219 and WFI~J2033$-$4723) the dust distribution is resolved into multiple clumps and a single S\'ersic component is likely a poor descriptor of their surface brightness distribution (Figs.~\ref{fig:0924_source} and \ref{fig:2033_source}). This is also the case for some lensed DSFGs. This blending effect was also suggested by \cite{Yan:2016}, who found some DSFGs have sizes in dust emission larger than expected from the Stefan-Boltzmann relation.

Fig.~\ref{fig:dustlir} suggests that the dust emission of quasar hosts generally have smaller sizes and higher temperatures than DSFGs of similar infrared luminosity, and smaller sizes and higher luminosities than DSFGs of similar temperature. The $z\sim2$ sub-mm-selected DSFGs are a factor of $\sim3$ larger than quasar hosts, the $z>4$ DSFGs have intermediate sizes between the $z\sim2$ DSFGs and quasar hosts, and lensed DSFGs span a broad range of sizes in dust emission ($0.5<R_{\rm eff}<3.8$) the smaller of which are similar to lensed quasars. 

\subsection{Intensity of star formation}
\label{section:eddington}

\begin{figure*}
    \includegraphics[width=\textwidth]{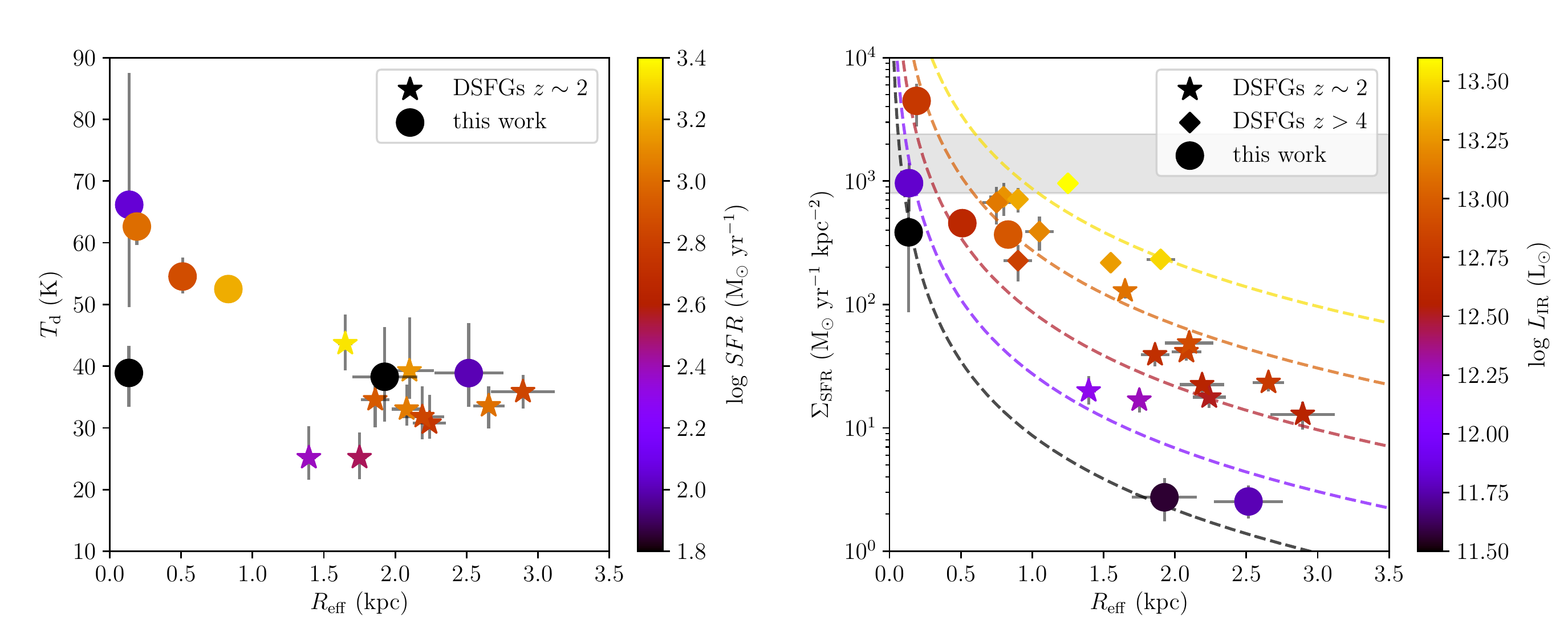}
    \caption{Left: Effective radius ($R_{\rm eff}$) against effective dust temperature ($T_{\rm d}$), coloured by star formation rate. Right: Effective radius against mean star formation rate surface density ($\Sigma_{\rm SFR}$), where the grey region shows our estimated Eddington limit and its uncertainty. $\Sigma_{\rm SFR}$ and the dashed curves show the tracks for $\log L_{\rm IR}$ values of 11.5, 12, 12.5, 13, 13.5. Circles show the quasar hosts in this work (with H1413+117 corrected for SED blending); stars show coeval DSFGs at $z\sim2$ \protect\citep{Hodge:2016,Hodge:2019}; triangles show DSFGs at $z>4$ \protect\citep{Hodge:2015,Hodge:2016,Riechers:2013,Riechers:2014,Riechers:2017}. Four of the seven quasar hosts have star formation rate surface densities close to the Eddington limit, similar to DSFGs selected at $z>4$.}
    \label{fig:size_sigma}
    \vspace{0.3cm}
    \includegraphics[width=0.85\textwidth]{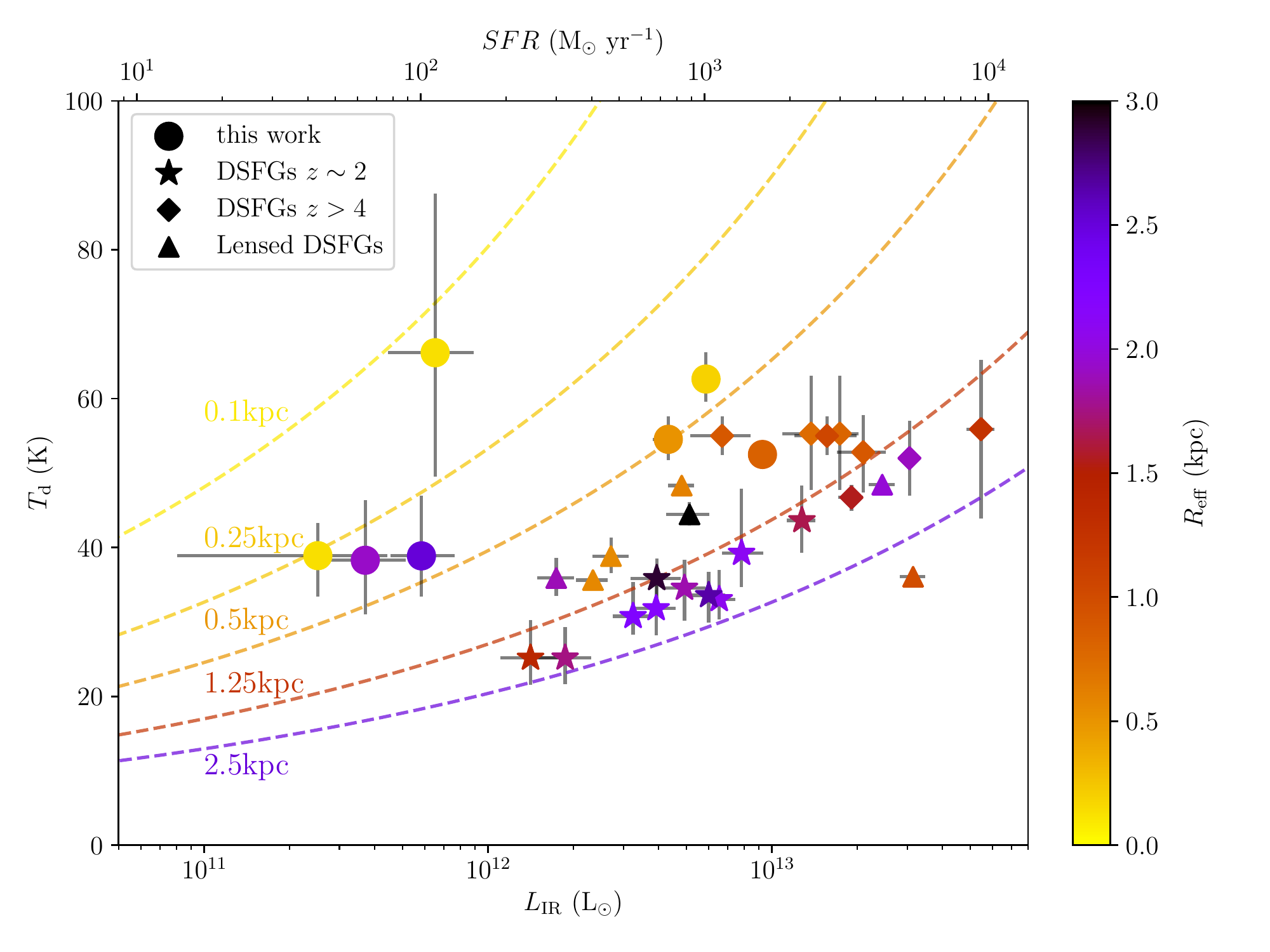}
    \caption{Effective dust temperature against infrared luminosity, coloured by effective radius. The dashed lines show the modified Stefan-Boltzmann relation from \protect\cite{Yan:2016} for sources of indicated physical sizes, assuming a dust model with $\beta=1.5$ and $\nu_0=3$~THz. Sources from this work are shown with circles. DSFGs at $z\sim2$ are shown with stars; DSFGs at $z>4$ are shown with diamonds; lensed DSFGs are shown with triangles. The quasar hosts in this sample have higher effective dust temperature and are generally more compact than dusty star-forming galaxies of similar infrared luminosity.}
    \label{fig:dustlir}
\end{figure*}

We estimate the galaxy-averaged star formation rate surface density ($\Sigma_{\rm SFR}$; \sfrd), for our sample and the DSFGs, assuming
\begin{equation}
    \Sigma_{\rm SFR} = 0.5 \times \frac{SFR}{\pi R_{\rm eff}^{2}} ,
\label{eq:sfrd}
\end{equation} where $R_{\rm eff}$ is the dust effective radius based on the S\'ersic model fit. Fig.~\ref{fig:size_sigma} (right) shows the galaxy-averaged star formation rate surface density against effective radius, where the dotted tracks show the analytic relationship for log$L_{\rm IR}$ of 11.5, 12, 12.5, 13 and 13.5, following Eq.~\ref{eq:sfrd}.

We estimate the optically thick Eddington flux limit ($F_{\rm Edd}$) assuming the relation from \cite{Andrews:2011} for warm starbursts,
\begin{equation}
F_{\rm Edd} \sim  10^{13}\ {\rm L_{\odot}\ kpc^{-2}}\ f_{\rm gas}^{-1/2}\ f_{\rm dg,150}^{-1} ,
\end{equation} where $f_{\rm gas}$ is the gas fraction and $f_{\rm dg, 150}$ is the dust-to-gas ratio multiplied by 150. Using a typical gas fraction of $0.5\pm0.2$ \citep{Spilker:2016} and a dust-to-gas ratio of $0.010\pm0.005$ (roughly solar metallicity), we derive an estimate of the Eddington-limited star formation rate surface density of $1600\pm800$~\sfrd. As can be seen in Fig.~\ref{fig:size_sigma}, the mean star formation rate density of five quasar hosts are within a factor of 2 to 3 of the estimated Eddington limit. As this is a mean star formation rate density, the proximity to the Eddington limit may imply that at least some star formation is Eddington limited. The peak star formation rate surface densities range from 25 to $>4500$~\sfrd, suggesting super-Eddington star formation in the most extreme case (WFI~J2026$-$4536).

HS~0810+2554 is the only object where we observe CO~(3--2), which is a common proxy for molecular hydrogen (H$_2$; \citealt{Greve:2014}). We assume $\alpha_{\rm CO}=0.8$~M$_{\odot}$~(K~km~s$^{-1}$~pc$^{2}$)$^{-1}$, as is often assumed for dusty starbursts, to convert $L^{\prime}_{\rm CO}$ to total gas mass ($M_{\rm gas}$). For this object we find $M_{\rm gas}=(2.1\pm0.1)\times10^9$~\msol\ and $\Sigma_{\rm gas}=4.5\pm0.7\times10^{9}$~\msol\,kpc$^{-2}$ (derived in the same manner as Eq.~\ref{eq:sfrd}). This implies a very low gas fraction of just $0.06\pm0.04$ compared to its dynamical mass, and a dust-to-gas ratio is $0.006\pm0.002$ (slightly below solar metallicity). The implied gas depletion timescale (i.e. $t_{\rm dep} \equiv M_{\rm gas}/SFR$) is $50\pm40$~Myr.

\begin{figure}
    \includegraphics[width=0.47\textwidth]{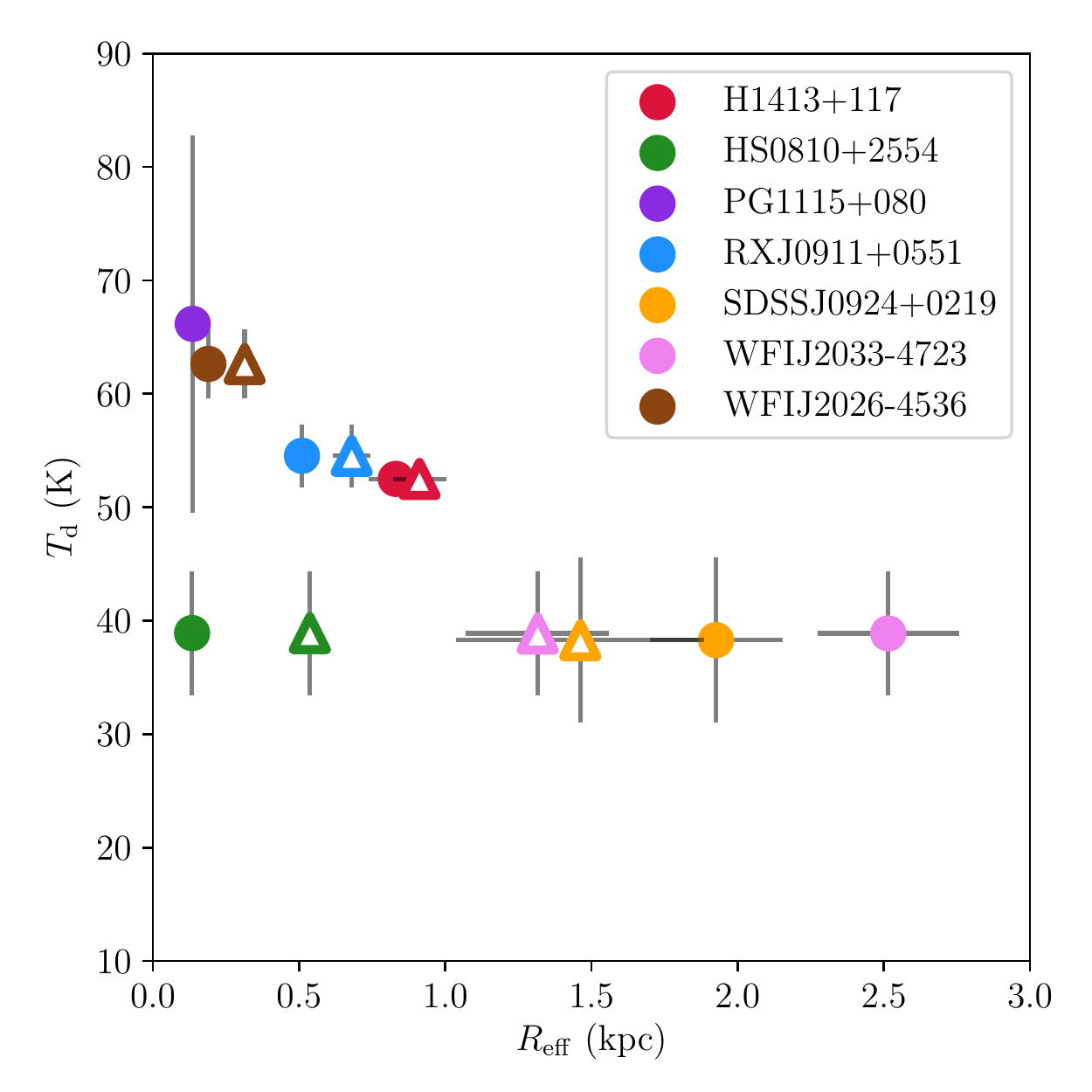}
    \caption{The effective radius ($R_{\rm eff}$) of the dust continuum (solid circles) and CO line emission (open triangles) against effective dust temperature ($T_{\rm d}$) for the seven lensed quasar hosts investigated here.}
    \label{fig:size_temp}
\end{figure}

\begin{landscape}
\begin{table}
\caption{Parameters of the lens models for the seven lensed quasars. We give the median and percentile ranges of the likelihood-weighted posterior probability distribution found with MultiNest, where the maximum a posteriori model parameters are given in square brackets. For WFI~2033$-$4723, parameters for the secondary lenses are also given. Parameters denoted by `$\equiv$' are fixed as detailed in Section~\ref{section:modelling}. The power-law index, $\zeta$, is 2.16 for PG~1115+080 \citep{Chen:2019}, 1.95 for WFI~2033$-$4723 \citep{Rusu:2020} and 2.0 for for all other lenses. $\delta x$ and $\delta y$ positions are relative to the phase centre positions given in Table~\ref{table:obs1}; $\kappa_0$ is the normalised convergence; $e$ is the ellipticity of the lens; $\gamma$ is the reduced external shear. In all cases, position angles are measured East of North.}
\adjustbox{width=1.33\textwidth}{
    \begin{tabular}{p{2.4cm} c c c c c c c} \hline 
                     & $\delta x$ & $\delta y$ & $\kappa_0$ & $e$ & $\theta_{e}$ & $\gamma$ & $\theta_{\gamma}$ \\
                     & (")  & (") & (") & & ($^\circ$) & & ($^\circ$) \\ \hline
    HS~0810+2554     & $0.277\pm0.008$ [0.278] & $0.839\pm0.013$ [0.841] & $0.470\pm0.005$ [0.470] & $0.26\pm0.03$ [0.28] & $32.6\pm2.7$ [29.9] & $0.005\pm0.003$ [0.009] & $98\pm50$ [112] \\
    RX~J0911+0551    & $-0.972\pm0.031$ [$-$0.965] & $2.422\pm0.015$ [2.424] & $1.099\pm0.021$ [1.099] & $0.25\pm0.08$ [0.27] & $-66\pm10$ [$-$67] & $0.350\pm0.024$ [0.382] & $9.7\pm1.4$ [10.3] \\
    SDSS~J0924+0219   & $0.257\pm0.011$ [0.255] & $0.037\pm0.008$ [0.044] & $0.888\pm0.008$ [0.892] & $0.36\pm0.05$ [0.39] & $-96.1\pm3.6$ [$-$95.5] & $0.020\pm0.015$ [0.022] & $-4\pm41$ [2] \\
    PG~1115+080      & $1.700\pm0.013$ [1.701] & $0.967\pm0.010$ [0.965] & $1.036\pm0.013$ [1.030] & $\equiv0.892$ & $\equiv96.9$ & $0.153\pm0.007$ [0.149] & $58.5\pm0.8$ [58.3] \\
    H1413+117       & $0.131\pm0.007$ [0.135] & $0.331\pm0.010$ [0.333] & $0.640\pm0.070$ [0.640] & $0.45\pm0.05$ [0.46] & $16\pm5$ [16] & $0.051\pm0.013$ [0.054] & $51\pm13$ [54] \\
    WFI~J2026$-$4536 & $\equiv0.0125$ & $\equiv0.385$ & $0.667\pm0.019$[0.668] & $0.275\pm0.012$[0.278] & $-73.7\pm1.9$[$-73.8$] & $0.090\pm0.004$[0.090] & $-109\pm1$[$-109$]  \\
    WFI~J2033$-$4723  & $-0.799\pm0.015$ [$-$0.801] & $-1.298\pm0.018$ [$-$0.129] & $0.967\pm0.016$ [0.977] & $0.12\pm0.06$ [0.12] & $25.7\pm9.2$ [29.5] & $0.20\pm0.02$ [0.19] & $12.7\pm1.9$ [12.4] \\
                     & $\equiv$3.301 & $\equiv -$1.243 & $\equiv$0.93 & $\equiv$0.33 & $\equiv$38.5 & - & - \\
                     & $\equiv -$0.868 & $\equiv$0.817 & $\equiv$0.001 & - & - & - & - \\  \hline
    \end{tabular}}
    \label{table:lens_models}
\end{table} 

\begin{table}
\caption{ The properties of the reconstructed continuum emission, corrected for lensing magnification. For H1413+117 we give values after correcting the SED for photometric confusion (see Section~\ref{section:1413_obs}). Note that for all except WFI~J2033$-$4723 and PG~1115+080, the quoted sizes include an additional $\pm10\%$ uncertainty due to empirical scatter in the slope of the mass profile (see Section~\ref{section:reconst} for details).}
    \setlength{\tabcolsep}{9pt}
    \def\arraystretch{1.2}
    \begin{tabular}{ p{2.4cm} c c c c c c c c c c }
    \hline
        & $\mu_{\rm dust}$ & $R_{\rm dust}$ & $n$ & $T_{\rm d}^{\rm cold}$ & $T_{\rm d}^{\rm warm}$ & $\beta$ & $L_{\rm FIR}$ & $SFR$ & $\Sigma_{\rm SFR}$ & $M_{\rm d}$ \\
        & & (kpc) &  & (K) & (K) &  & (L$_{\odot}$) & (M$_{\odot}$~yr$^{-1}$) & (M$_{\odot}$~yr$^{-1}$~kpc$^{-2}$) & (M$_{\odot}$) \\ \hline
    HS~0810+2554 & $25\pm4$ & $0.13\pm0.04$ & $0.57\pm0.05$ & $39^{+4}_{-5}$ & $135^{+25}_{-18}$ & $\equiv2$ & $1.3^{+1.0}_{-0.9}\times10^{11}$ & $40^{+30}_{-30}$ & $400^{+300}_{-300}$ & $1.1^{+0.4}_{-0.5}\times10^{7}$ \\   
    RX~J0911+0551 & $14.8\pm0.2$ & $0.51\pm0.03$ & $0.56\pm0.04$ & $54^{+3}_{-3}$ & -- & $1.9^{+0.1}_{-0.1}$ & $2.3^{+0.1}_{-0.1}\times10^{12}$ & $750^{+50}_{-50}$ & $500^{+100}_{-100}$ & $5.3^{+0.9}_{-0.8}\times10^{6}$  \\
    SDSS~J0924+0219 & $15.1\pm0.7$ & $1.9\pm0.2$ & $1.2\pm0.1$ & $38^{+8}_{-7}$ & -- & $1.5^{+0.6}_{-0.4}$ & $1.9^{+0.7}_{-0.5}\times10^{11}$ & $60^{+20}_{-20}$ & $3^{+1}_{-1}$ & $3.9^{+1.5}_{-1.4}\times10^{7}$ \\
    PG~1115+080 & $21.5\pm0.4$ & $0.14\pm0.04$ & $0.47\pm0.01$ & $66^{+17}_{-21}$ & -- & $1.9^{+0.3}_{-0.3}$ & $3.4^{+1.2}_{-1.0}\times10^{11}$ & $110^{+40}_{-30}$ & $1000^{+400}_{-400}$ & $2.5^{+1.5}_{-1.0}\times10^{6}$ \\
    H1413+117 & $11.2\pm0.2$ & $0.8\pm0.1$ & $0.66\pm0.07$ & $53^{+1}_{-1}$ & $246^{+14}_{-24}$ & $1.88^{+0.04}_{-0.04}$ & $4.9^{+0.2}_{-0.2}\times10^{12}$ & $1600^{+60}_{-60}$ & $370^{+80}_{-80}$ & $1.2^{+0.1}_{-0.1}\times10^{8}$ \\
    WFI~J2026$-$4536 & $19.6\pm0.8$  & $0.19\pm0.02$  & $0.61\pm0.02$ &  $63^{+4}_{-3}$ & -- & $1.8^{+0.1}_{-0.1}$ & $3.1^{+1.7}_{-1.5}\times10^{12}$ & $1010^{+60}_{-50}$ & $4500^{+1700}_{-1700}$ & $3.9^{+0.1}_{0.1}\times10^7$  \\ 
    WFI~J2033$-$4723 & $17.6\pm0.9$ & $2.5\pm0.2$ & $1.1\pm0.2$ & $31^{+7}_{-5}$ & -- & $1.7^{+0.3}_{-0.3}$ & $3.1^{+0.9}_{-0.6}\times10^{11}$ & $100^{+30}_{-20}$ & $2.5^{+0.9}_{-0.7}$ & $4.5^{+0.8}_{-0.7}\times10^{7}$ \\ \hline
    \end{tabular}
    \label{table:intrinsic_dust}
\end{table}

\begin{table}
\caption{ The properties of the reconstructed CO line emission, corrected for lensing magnification. For SDSS~J0924+0219 and WFI~J2033$-$4723, the S\'ersic index, $n$, is fixed to a value of 1 (consistent with the value found for the dust continuum). All parameters are derived from moment maps of the reconstructed source. Note that for all except WFI~J2033$-$4723 and PG~1115+080, the quoted sizes (but not magnifications) include an additional $\pm10\%$ uncertainty due to empirical scatter in the slope of the mass profile (see Section~\ref{section:reconst} for details).}
    \setlength{\tabcolsep}{9pt}
    \begin{tabular}{ p{2.4cm} c c c c c c c c }
    \hline
         & $\bar{\mu}_{\rm CO}$ & $R_{\rm CO}$ & $n$ & $I_{\rm CO}$ & $L_{\rm CO}$ & $L^{'}_{\rm CO}$ & $M_{\rm dyn}$ \\
        & &  (kpc) & & (Jy~km~s$^{-1}$) & (L$_{\odot}$) & (K~km~s$^{-1}$~pc$^{2}$) & (M$_{\odot}$) \\ \hline
    HS~0810+2554 & $20.3\pm0.7$ & $0.27\pm0.02$ & $0.57\pm0.06$ & $0.19\pm0.01$ & $(3.4\pm0.1)\times10^{6}$ & $(2.6\pm0.1)\times10^{9}$ & $(7\pm1)\times10^{9}$ \\
    RX~J0911+0551 & $13.6\pm0.2$ & $0.6\pm0.1$ & $0.56\pm0.01$ & $0.47\pm0.01$ & $(4.2\pm0.1)\times10^{7}$ & $(6.9\pm0.1)\times10^{9}$ & $(1.1\pm0.3)\times10^{9}$ \\
    SDSS~J0924+0219 & $17\pm1$ & $1.7\pm0.4$ & $\equiv1.0$ & $0.13\pm0.01$ & $(6.4\pm0.4)\times10^{6}$ & $(2.5\pm0.2)\times10^{8}$ & $(6\pm4)\times10^{9}$  \\
    H1413+117 & $11.1\pm0.2$ & $0.9\pm0.1$ & $0.54\pm0.05$ & $3.4\pm0.1$ & $(4.8\pm0.1)\times10^{8}$ & $(1.34\pm0.02)\times10^{10}$ & $(2.2\pm0.3)\times10^{10}$ \\
    WFI~J2026$-$4536 & $19.6\pm0.8$  & $0.23\pm0.02$  & $0.52\pm0.03$ &  $0.54\pm0.02$ & $(6.6\pm0.3)\times10^7$ & $(1.4\pm0.1)\times10^9$ & $(3.2\pm0.3)\times10^9$  \\ 
    WFI~J2033$-$4723 & $17\pm2$ & $2.1\pm0.3$ & $\equiv1.0$ & $0.026\pm0.003$ & $(1.5\pm0.2)\times10^{6}$ & $(6.1\pm0.7)\times10^{7}$ & $(3\pm2)\times10^{9}$ \\ \hline
    \end{tabular}
    \label{table:intrinsic_CO}
\end{table}
\end{landscape}

\section{Discussion}
\label{section:discussion}

\subsection{Evidence for extreme star formation in quasar host galaxies}

The seven objects investigated here were selected from a larger survey of 104 lensed quasars with {\it Herschel}/SPIRE that was reported by \cite{Stacey:2018a}, who performed SED fitting to constrain the level of dust-obscured star formation in the sample. In all but three cases (including H1413+117), a single dust component was used for the fitting due to the lack of data in the MIR to FIR. From this analysis, \citet{Stacey:2018a} found a median dust temperature of $38^{+12}_{-5}$~K (based on optically thin dust models) and a median star formation rate of $120^{+160}_{-80}$~M$_{\odot}$~yr$^{-1}$, both of which are typical of DSFGs. A concern from such a simple SED model is that AGN-heated dust emission could contribute to the FIR and result in a higher dust temperature and, hence, a higher effective dust temperature (e.g. \citealt{Kirkpatrick:2012}). Indeed, in some cases, effective dust temperatures were found that are higher than typical for star formation. However, we find that the quasar hosts are generally consistent with the Stefan-Boltzmann relation between temperature and luminosity when considering an optically thick dust model (see Fig.~\ref{fig:size_temp}), suggesting that the dust is uniformly heated to higher temperatures by an obscured starburst, as opposed to the central AGN. This does not exclude the possibility that the AGN contributes to the cold dust emission, but in such a scenario it is unlikely be a significant fraction without a conspiracy between the effective radius, temperature and luminosity. The higher dust temperatures found for some quasar hosts relative to DSFGs can be explained by the more compact physical size of their dust emission.

We estimate that five compact quasar hosts in our sample have high star formation rate surface densities which may be Eddington-limited. This finding is consistent with previous investigations of H1413+117, which have independently verified an Eddington-limited starburst with radiative transfer modelling of spectral line ratios \citep{Bradford:2009,Riechers:2011a,Uzgil:2016}. In general, while an X-ray radiation field may penetrate large column densities of molecular gas, models suggest it will not efficiently heat dust grains over large ($\sim$kpc) distances \citep{Meijerink:2007}. These factors further suggest that the AGN are unlikely to be responsible for a significant fraction of the sub-mm dust emission, but is more ambiguous for more compact sources with an effective radius of $\sim100$~pc and super-Eddington starbursts. We cannot determine the origin of the molecular gas heating from single CO lines, but a comparison between multiple transitions will help constrain the contribution from the AGN to the emission reported here. Radiative transfer models of quasar hosts find that these are very likely to be enhanced by X-rays from the AGN \citep{Vallini:2019}, so could probe the effect of radiative feedback on the host galaxy ISM.

Another potential source of uncertainty in the analysis by \cite{Stacey:2018a} was the assumption on the unknown lensing magnification of the FIR emission, which was set to be $10_{-5}^{+10}$ where the FIR magnification is unknown. If the lensing magnifications were significantly higher than was conservatively assumed, then the level of the inferred star formation would be much lower. From our analysis of the seven lensed quasars investigated here, we find dust magnifications factors in the range 11--25, roughly consistent with the general estimate of 5--20 by \citet{Stacey:2018a}. While on the higher side of the estimate, these are four-image lens systems so high magnifications are typical. Accounting for the inferred magnifications, we find that the star formation rates for the sample here range from 40 to 1600~M$_{\odot}$~yr$^{-1}$.

\subsection{Evidence for compact quasar hosts}

The quasar hosts in our study can be divided into two types: DSFGs characterised by clumpy dust distributions, with sizes and star formation rate densities similar to typical sub-mm-selected DSFGs (i.e. WFI~J2033$-$4723 and SDSS~J0924+0219), and DSFGs characterised by compact ($R_{\rm eff}<1$~kpc) sizes and generally high star formation densities (see Fig.~\ref{fig:size_sigma}) with no evidence of clumpy features on kpc-scales\footnote{For HS~0810+2554 there may be extended dust emission that is undetected due to the low surface brightness sensitivity of the data.}. 

The sizes of the compact quasar hosts in this work are similar to the sub-mm sizes of compact star-forming galaxies \citep{Ikarashi:2015,Barro:2016,Barro:2017}. The lensed DSFGs (which we expect to represent more compact DSFGs due to their selection bias, e.g. \citealt{Serjeant:2012}) show a large range of sizes, yet the quasar hosts of similar luminosity are generally more compact and have higher dust temperatures. This may indicate that quasars are preferentially hosted in more compact systems. Even considering that two of these quasar hosts are likely to be selected on the basis of their FIR properties, the remaining optically selected systems are distinct from the remaining sample.

In this work, by selecting hosts of Type~1 quasars, we have explicitly chosen galaxies whose AGN are rapidly accreting and generating relativistic winds that have exposed their accretion discs. It is predicted that a significant fraction of DSFGs host AGN (e.g. \citealt{Hickox:2014}). While it has been found that at least one in five DSFGs have an X-ray luminous AGN \citep{Wang:2013}, this is true for at least half of compact star-forming galaxies \citep{Barro:2014,Kocevski:2017}\footnote{We assume these are lower limits as observations are limited by sensitivity, and simulations suggest that he actual AGN contribution to the SED may be much higher \citep{Roebuck:2016}.}. The relative prevalence of X-ray luminous AGN may point towards elevated black hole growth in compact DSFGs. The small size of compact galaxies requires that there has been a rapid infall of gas such that the bulk of the star formation occurs within the central region of the galaxy: this could allow the AGN to accrete more efficiently from the dense ISM. 

The sub-mm sizes of both the compact DSFGs and quasar hosts are similar to the optical/infrared sizes of compact quiescent galaxies at similar redshifts ($\tilde{R}_{\rm eff}=0.9$~kpc; \citealt{vanDokkum:2008}). Assuming the sub-mm dust emission traces the bulk of the stellar component, this is consistent with the hypothesis that compact starbursts are progenitors of compact quiescent galaxies (and ultimately massive ellipticals). The still-high star formation rates and molecular gas densities of these hosts could mean that we catch them at the moment of compaction, at the onset of gas depletion and quenching. This would require that at least some compact quiescent galaxies form rapidly, rather than through a slow evolution \citep{Wellons:2015}.

Overall, the sizes of $<$\,1 to 3~kpc we find for our sample are similar to those of \cite{Silverman:2019}, who found the optical/infrared sizes of quasar hosts at $z\sim1.5$ are in the range $<1$ to 6~kpc ($\bar{R}_{\rm eff}=2.2$~kpc), intermediate between main sequence and quiescent galaxies. Additionally, \cite{Ikarashi:2017} also found that $z\sim1$--3 quasar-starburst composites have smaller sub-mm dust sizes than both starburst-dominated DSFGs and quasars whose SEDs have a more dominant AGN fraction in the MIR. \cite{DAmato:2020} also found similar sizes for the dust components of X-ray-selected quasars at $2<z<5$. These results and ours are inconsistent with models that predict the size of quasar hosts to be larger than those of normal star-forming galaxies, due to adiabatic expansion that results from negative AGN feedback \citep{Fan:2008} or positive AGN feedback on kpc-scales \citep{Ishibashi:2013}.  

\subsection{Mechanism of formation}

In the context of the ongoing debate about how compact quiescent galaxies form their very high stellar densities, our finding that the host galaxies of quasars are compact is in agreement with a model whereby they form rapidly in a period of dissipative contraction. Simulations suggest that star-forming galaxies are able to maintain high star formation rates in quasi-stable discs, fed by smooth accretion from the cosmic web (e.g. \citealt{Keres:2005}). For a galaxy to compactify, dynamical instabilities must be induced by mergers or intense inflows that can drive gas into the centre of the galaxy on timescales shorter than the star formation rate \citep{Dekel:2014,Zolotov:2015}. The high star formation rates and the existence of clumpy, non-axisymmetric dust features that we detect strongly suggest that quasar hosts are globally unstable due to gravitational fragmentation, as has been frequently observed for DSFGs \citep{Iono:2016,Oteo:2017,Hodge:2019}.

Optical imaging of DSFGs has often found that their molecular gas and obscured star formation is significantly more compact than pre-existing stellar distributions \citep{Simpson:2015,Ikarashi:2015,Kaasinen:2020}. The optical/UV-luminous stellar emission observed for PG~1115+080 (see \citealt{Peng:2006}) must also be more extended or offset from the dust, as it is lensed into an Einstein ring, whereas the dust is not. These unobscured stellar features may exist for the other quasar systems we observe here, but are too faint to be seen in optical imaging or cannot be distinguished from the bright quasar emission. This UV-luminous emission likely contributes only a small fraction of the star formation, which is largely obscured, but these features may hint at the formation histories of the galaxies. For example, the unobscured stellar emission may be quenched as a result of gas compaction and stellar feedback \citep{Maiolino:2015}. However, this may be difficult to interpret due to the much higher sensitivity to star formation achieved by optical/UV imaging compared to sub-mm imaging.

\subsection{Mechanism of quenching}

Simulations find that compaction can naturally lead to quenching, through rapid gas consumption and stellar feedback, coupled with increasing dynamical stability (e.g. \citealt{Zolotov:2015}). Observations of dense starbursts suggest that radiation pressure from stars could be a feasible mechanism to suppress further star formation \citep{Murray:2005,Andrews:2011}. If all the dust emission we observe here is associated with star formation, the five most compact sources in our work are in the regime of Eddington-limited maximum starbursts. This is consistent with observations of compact quiescent galaxies, whose star formation histories suggest a short starburst of $\sim50$~Myr before quiescence \citep{Valentino:2019}. In the absence of reliable stellar mass estimates, we cannot determine whether star formation in the compact quasar hosts has begun to quench by direct comparison with DSFGs. However, the higher implied star formation rate surface densities imply that stellar feedback could play the primary role.

For the one system where we can trace the extent of the gas reservoir, we find a very low gas fraction in comparison to its dynamical mass. This suggests that this galaxy may be transitioning into quiescence following the depletion and removal of gas, as observed for compact star-forming galaxies at $z\sim2$ \citep{Spilker:2016,Spilker:2019}. The implied depletion timescale of $50\pm40$~Myr is comparable to the finding of $\sim$\,100~Myr for compact star-forming galaxies by \cite{Spilker:2016} and consistent with $\sim50$~Myr found by \cite{Valentino:2019}. 

The compact starbursts and implied short depletion timescales could suggest that AGN feedback does not play an important role in the immediate quenching of star formation, contrary to simulations. However, long-term maintenance of quenching requires that the supply of fresh cold gas into the galaxy is halted. As the quasars in this work are likely hosted in massive haloes \citep{Kormendy:2013}, virial shock-heating alone could help suppress accretion of cold gas from the cosmic web \citep{Dekel:2006}. It is likely that AGN feedback becomes important at some point to prevent re-accretion of ejected gas \citep{Croton:2006}, as the energy input from star formation falls far short of the level required to completely unbound the gas from the galaxy and maintain a quenched state. Indeed, we expect that compact galaxies quench abruptly \citep{Valentino:2019} and that `maintenance mode' (or `jet mode') AGN feedback tends to be observed after galaxies have already begun to quench \citep{Hardcastle:2007}. 

Observations of CO~(1--0) in high-redshift lensed AGN-starbursts have revealed disturbed morphologies, suggestive of feedback (AGN or stellar) or ongoing major mergers \citep{Thomson:2012,Spingola:2020}, whilst others do not appear to show such features \citep{Riechers:2011b,Sharon:2016}. Such observations for this sample will be useful to understand to what extent the CO lines investigated here probe the mass and kinematics of the gas reservoir, particularly in comparison to the population of DSFGs that do not appear to host rapidly accreting AGN.

\subsection{Selection effects and confusion}

A source of caution in the interpretation of our findings is the combination of systematic biases that stem from the comparison of samples with different selection effects. Ideally, we would like to compare coeval samples of DSFGs and quasar hosts. However, the detection of DSFGs is strongly influenced by selection effects, such that systems with lower dust temperatures are preferentially detected at lower redshifts due to the shape of the modified black body spectrum. The selection of strongly lensed DSFGs at FIR wavelengths is much less dependent on dust temperature and expected to be more strongly dependent on source size \citep{Serjeant:2012,Hezaveh:2012}. Together, these represent a diverse selection of DSFGs. The quasar systems, on the other hand, were optically selected, based on the emission from their unobscured accretion discs, and their selection was in most cases based on their lens configuration (RX~J0911+0551 and H1413+117 being exceptions). Therefore, we do not implicitly select quasar hosts that have high temperatures or are more compact. However, analyses with larger samples, including less strongly magnified systems, will provide a better statistical comparison between coeval galaxy populations selected with different methods.

Notably, we identify a nearby source of one object (H1413+117) that likely causes photometric confusion at shorter wavelengths. This has the important implication that source blending may contribute to the extreme infrared luminosities found for some quasar-starbursts. Observations with ALMA of (unlensed) optically selected quasars have revealed that 30~percent of FIR identifications can be resolved into secondary counterparts that contribute at least 25~percent to the measured infrared luminosity \citep{Hatziminaoglou:2018}. This may be unsurprising as massive galaxies seem to be formed in over-dense regions \citep{Zeballos:2018}. The sources investigated here are gravitationally lensed, so the relative contribution from companions is expected to be lower on average than for field sources. Nevertheless, our findings suggest care should be taken to account for field sources when compiling an SED with photometric measurements obtained with low spatial resolution (e.g. {\it Herschel}/SPIRE).

\section{Conclusions}
\label{section:conclusion}

The properties of cold dust and molecular gas in quasar hosts at cosmologically important redshifts can give insights into the mechanism responsible for the transformation of these galaxies from gas-rich, dusty starbursts into passive, quiescent systems. We have presented high angular resolution imaging with ALMA of the host galaxies of seven optically selected quasars at redshifts between 1.5 and 3. Using pixellated lens modelling, we reconstructed the sources and their intrinsic properties.

In comparison to unlensed DSFGs with similar redshifts and infrared luminosities (i.e. ALESS DSFGs and FIR-selected lensed DSFGs) and DSFGs at higher redshifts with similar dust temperatures, the quasar hosts in this work are generally more compact. The observed luminosities are broadly consistent with the expectations for more compact star formation, disputing the case for a significant contribution from black hole accretion to the global dust emission at sub-mm wavelengths.

We find that two of the quasar hosts are characterised by extended, clumpy dust distributions, but the remainder are compact starbursts. These differences may represent quasar hosts at different stages of morphological change into compact galaxies. The more compact systems have sizes that are already similar to quiescent galaxies at $z\sim2$, with extreme star formation rate densities that imply a rapid consumption of gas. This is consistent with a picture in which the inflow of gas resulting from mergers and/or dynamical instabilities triggers both the intense starburst and a period of efficient black hole accretion \citep{Kocevski:2017}. These intense starbursts could be responsible for both the formation of the high stellar densities and rapid quenching of compact galaxies.

If quasar hosts are in the process of forming stellar bulges, we should find that these systems are globally characterised by unstable gas discs and rapidly depleting gas reservoirs \citep{Zolotov:2015,Tacchella:2016}. Matched, high angular resolution observations of low $J$-level CO emission would probe the size of the cold gas reservoir and determine whether the star formation rate efficiency and gas depletion timescales of quasar hosts are significantly different relative to those of DSFGs. Furthermore, cold gas diagnostics will be useful to determine whether outflows are prevalent and whether AGN or stellar feedback is ongoing, as expected from simulations and semi-analytic models (e.g. \citealt{Hopkins:2008}). 

Our conclusions here are based on the data for just seven objects from the sample of lensed quasars. In the current observing cycle of ALMA, we will obtain data for at least 27 lensed quasars in Band~6 and 7 which will further probe the size and structure of the heated dust emission from this class of objects, and probe the structure and kinematics of the molecular gas that is feeding both the star formation and AGN activity.

\section*{Acknowledgements}

JPM acknowledges support from the Netherlands Organization for Scientific Research (NWO, project number 629.001.023) and the Chinese Academy of Sciences (CAS, project number 114A11KYSB20170054). SV has received funding from the European Research Council (ERC) under the European Union's Horizon 2020 research and innovation programme (grant agreement No 758853). CS is grateful for support from the National Research Council of Science and Technology, Korea (EU-16-001).

This research made use of Astropy and Matplotlib packages for Python \citep{Astropy:2018,Hunter:2007}.

This paper makes use of the following ALMA data: \\ ADS/JAO.ALMA\#2012.1.00175.S, ADS/JAO.ALMA\#2015.1.01309.S, ADS/JAO.ALMA\#2017.1.01368.S, ADS/JAO.ALMA\#2017.1.01081.S, ADS/JAO.ALMA\#2018.1.01519.S. ALMA is a partnership of ESO (representing its member states), NSF (USA) and NINS (Japan), together with NRC (Canada), MOST and ASIAA (Taiwan), and KASI (Republic of Korea), in cooperation with the Republic of Chile. The Joint ALMA Observatory is operated by ESO, AUI/NRAO and NAOJ.

\section*{Data availability}
The data underlying this article will be shared on reasonable request to the corresponding author.

\bibliographystyle{mnras}
\bibliography{references}
\bsp 

\appendix

\section{Supplementary figures and tables}

\begin{table}
    \centering
    \caption{List of DSFGs used in this study and their properties. ($\dagger$) notes where the dust temperature is obtained from the literature (from an optically thick dust model); in all other cases, the SED fitting was performed in this work using the methodology described in Section~\ref{section:sed_fitting}. For the unlensed DSFGs, the sizes are taken from the literature; for the lensed DSFGs, the sizes and ALMA photometry are derived with the same methodology used for the lensed quasar systems in this work.}
    \def\arraystretch{1.2}
    \begin{tabular}{l | c c c c c c} \hline
                   & $z$ & $T_{\rm d}$ & $L_{\rm FIR}$ & $R_{\rm eff}$ & Ref \\
                   &  & (K) & ($10^{12}$ L$_{\odot}$) & (kpc) & \\ \hline
        \hfill  $z\sim2$ DSFGs          &     \\
    ALESS~3.1   & 3.374 & $43^{+5}_{-4}$ & $6.7^{+0.8}_{-0.8}$ & $1.6\pm0.1$ & [1,2] \\
    ALESS~5.1   & 2.86 & $33^{+4}_{-3}$ & $3.4^{+0.5}_{-0.4}$ & $2.1\pm0.1$ & [3,2] \\
    ALESS~10.1  & 2.02 & $36^{+3}_{-3}$ & $2.1^{+0.4}_{-0.4}$ & $2.9\pm0.4$ & [3,2] \\
    ALESS~15.1  & 2.67 & $34^{+3}_{-4}$ & $3.1^{+0.4}_{-0.4}$ & $2.7\pm0.2$ & [1,2] \\
    ALESS~17.1  & 1.540 & $25^{+4}_{-4}$ & $1.0^{+0.2}_{-0.2}$ & $1.8\pm0.1$ & [1,2] \\
    ALESS~29.1  & 1.439 & $25^{+5}_{-4}$ & $0.7^{+0.2}_{-0.2}$ & $1.4\pm0.1$ & [3,2] \\
    ALESS~39.1  & 2.44 & $31^{+5}_{-3}$ & $1.7^{+0.3}_{-0.3}$ & $2.2\pm0.1$ & [3,2] \\
    ALESS~45.1  & 2.34 & $32^{+5}_{-4}$ & $2.1^{+0.4}_{-0.3}$ & $2.2\pm0.2$ & [3,2] \\
    ALESS~67.1  & 2.123 & $39^{+9}_{-5}$ & $4.1^{+0.8}_{-0.6}$ & $2.1\pm0.2$ & [3,2] \\
    ALESS~112.1 & 2.315 & $35^{+4}_{-4}$ & $2.6^{+0.5}_{-0.4}$ & $1.9\pm0.1$ & [1,2] \\ \hline
        \hfill  $z>4$ DSFGs        &     \\
    ALESS~9.1   &  4.867 & $47^{+6}_{-7}$ & $10^{+1}_{-1}$ & $1.6\pm0.1$ & [1,2] \\
    ADFS~27 (1)   &  5.655 & $55^{+8}_{-8}$ $^{\dagger}$ & $9.1^{+1.5}_{-1.5}$ & $0.8\pm0.1$ & [4] \\
    ADFS~27 (2)   &        &                             & $7.2^{+1.5}_{-1.5}$ & $0.8\pm0.1$  & [4] \\
    HFLS~3       &  6.334 & $56^{+9}_{-12}$ $^{\dagger}$ & $29^{+3}_{-3}$ & $1.3\pm0.1$ & [5] \\
    AzTEC~3     &  5.299 & $53^{+5}_{-5}$ $^{\dagger}$ & $11^{+2}_{-2}$ & $0.9\pm0.1$  & [6] \\
    GN20        &  4.055 & $52^{+5}_{-5}$ $^{\dagger}$ & $16^{+1}_{-1}$ & $1.9\pm0.1$  & [7] \\
    SGP~196076(1)  &  4.425 & $55^{+3}_{-3}$ $^{\dagger}$ & $8^{+2}_{-2}$ & $1.1\pm0.1$ & [8] \\
    SGP~196076(2)  &        &                & $4^{+1}_{-1}$ & $0.9\pm0.1$ & [8] \\ \hline
        \hfill  Lensed DSFGs        &     \\
    HELMS~5   &  3.503 &  $48^{+1}_{-1}$ & $2.5^{+0.3}_{-0.3}$ & $0.54\pm0.05$ & [9,10] \\    
    HELMS~8   &  1.195 &  $36^{+1}_{-1}$ & $1.2^{+0.2}_{-0.2}$ & $0.43\pm0.05$ & [10,11] \\    
    HELMS~9   &  1.441 &  $36^{+3}_{-2}$ & $0.9^{+0.1}_{-0.1}$ & $1.6\pm0.3$ & [10,11] \\    
    HELMS~13  &  2.765 &  $44^{+2}_{-1}$ & $2.7^{+0.5}_{-0.5}$ & $3.8\pm1.1$  & [10,11] \\
    HELMS~22  &  2.509 &  $36^{+1}_{-1}$ & $16.5^{+1.7}_{-1.7}$ & $1.0\pm0.1$ & [10,11] \\
    G15v2.779 &  4.243 &  $48^{+1}_{-1}$ & $12.8^{+1.3}_{-1.3}$ & $1.7\pm0.2$ & [10,11] \\   
    G15v2.19  &  1.027 &  $39^{+2}_{-2}$ & $1.4^{+0.3}_{-0.2}$ & $0.38\pm0.04$  & [10,11] \\ \hline
    \end{tabular} 
\raggedright References: [1] \cite{Hodge:2019}; [2] \cite{Swinbank:2010}; [3] \cite{Hodge:2016}; [4] \cite{Riechers:2017}; [5] \cite{Riechers:2013}; [6] \cite{Riechers:2014}; [7] \cite{Cortzen:2020}; [8] \cite{Oteo:2016}; [9] \cite{Nayyeri:2016}; [10] Stacey et al. in prep; [11] \cite{Dye:2018}
    \label{table:dsfgs}
\end{table}

\begin{figure*}
\centering
    \includegraphics[width=0.38\textwidth]{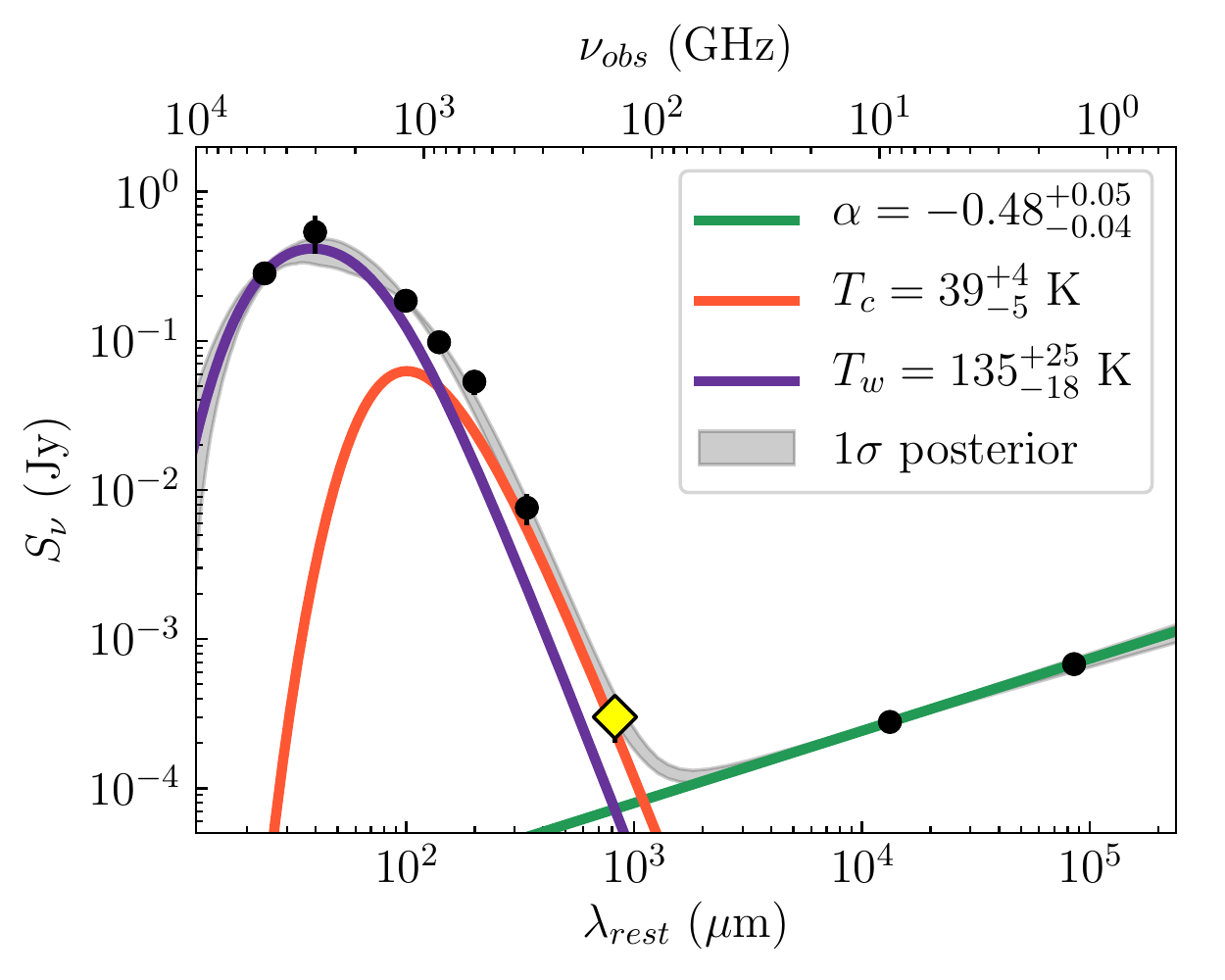}
    \includegraphics[width=0.38\textwidth]{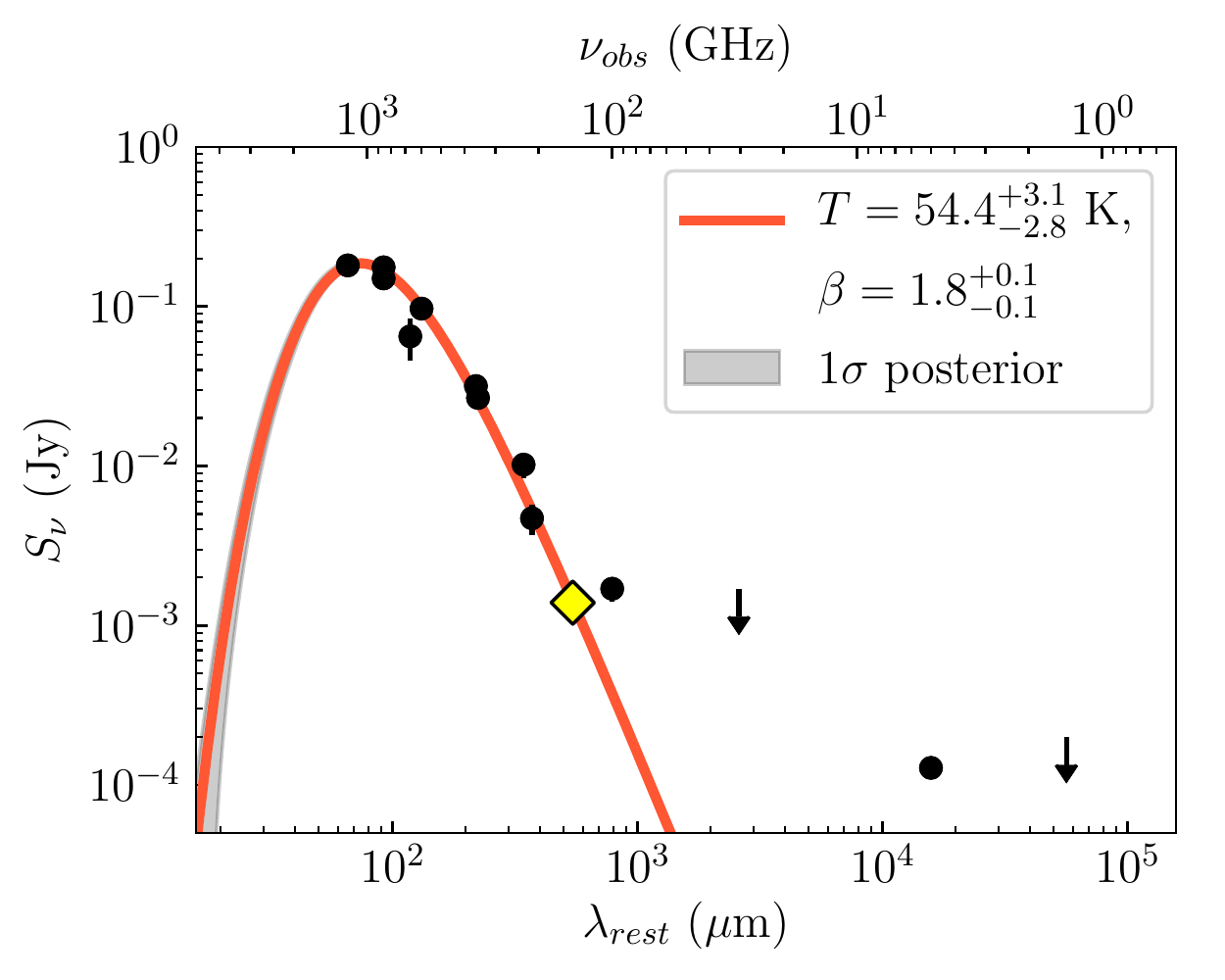}
    \includegraphics[width=0.38\textwidth]{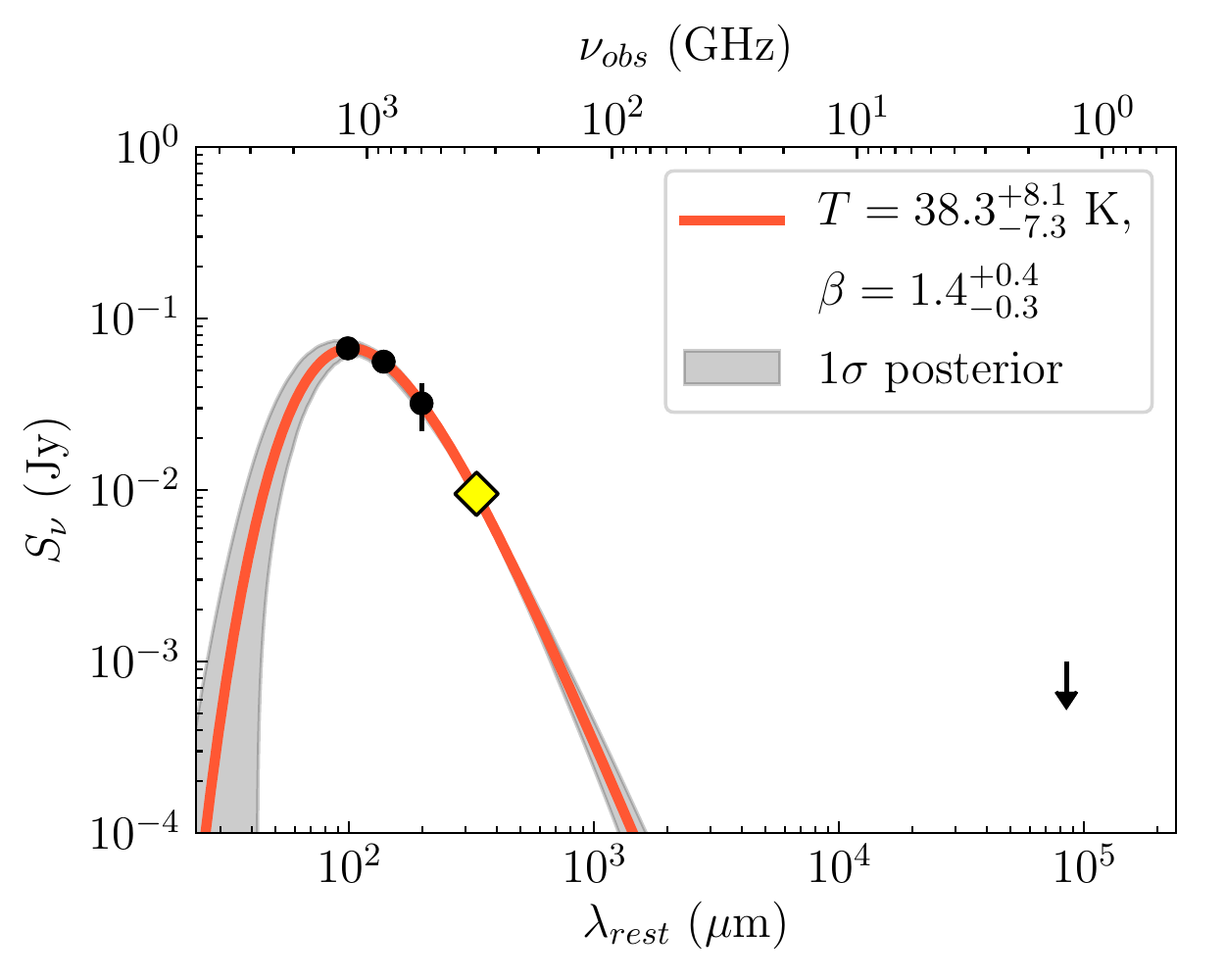}
    \includegraphics[width=0.38\textwidth]{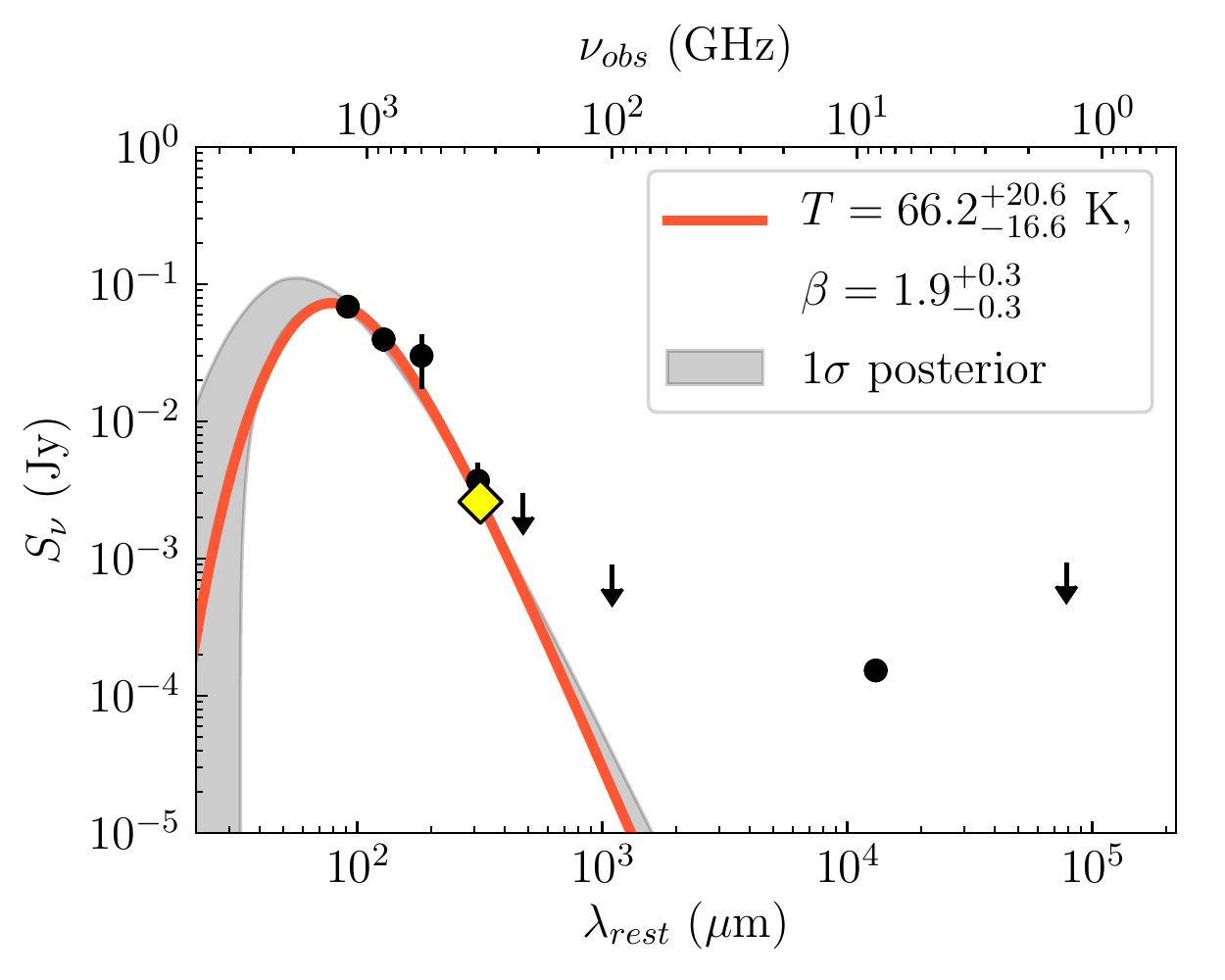}
    \includegraphics[width=0.38\textwidth]{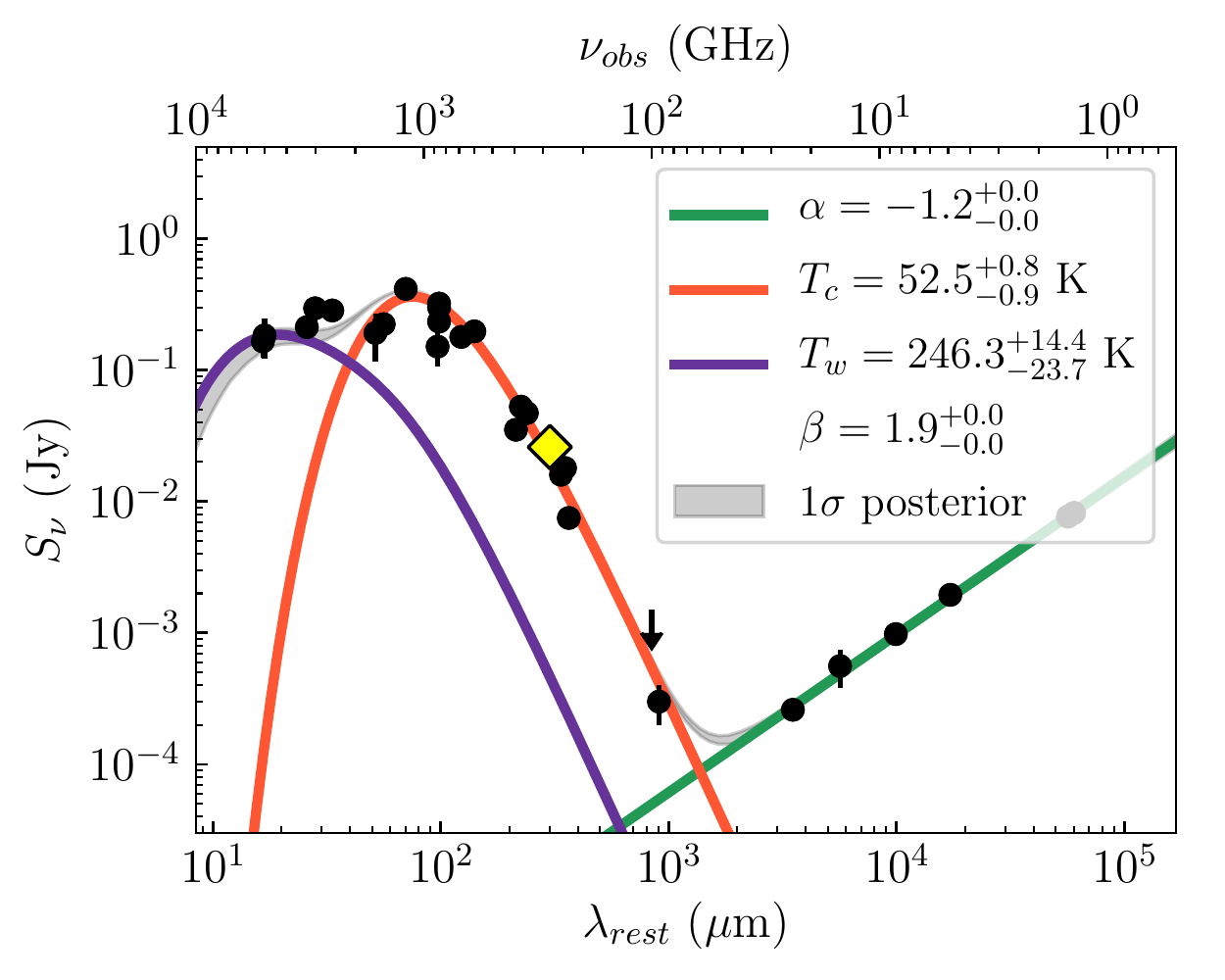}
    \includegraphics[width=0.38\textwidth]{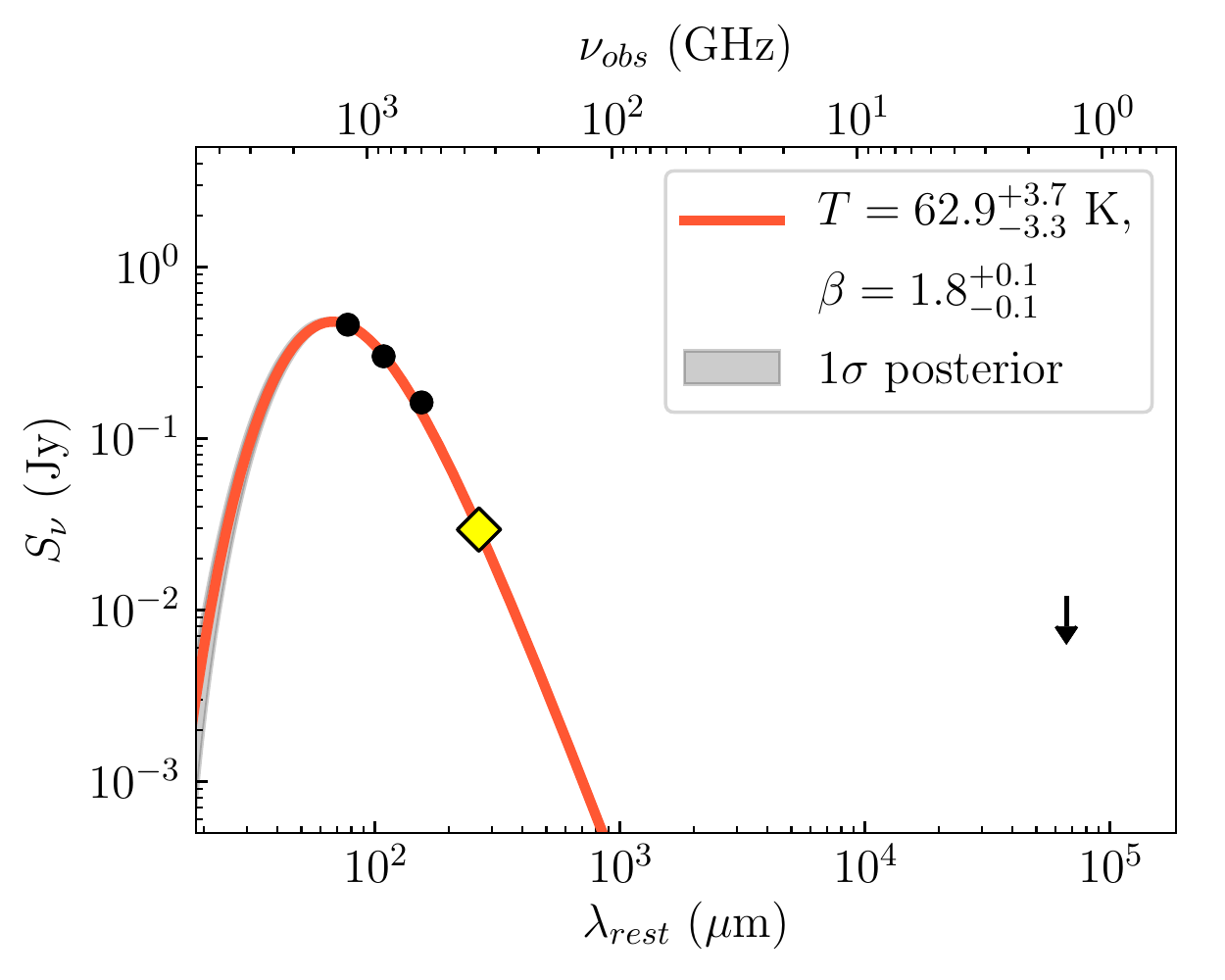}
    \includegraphics[width=0.38\textwidth]{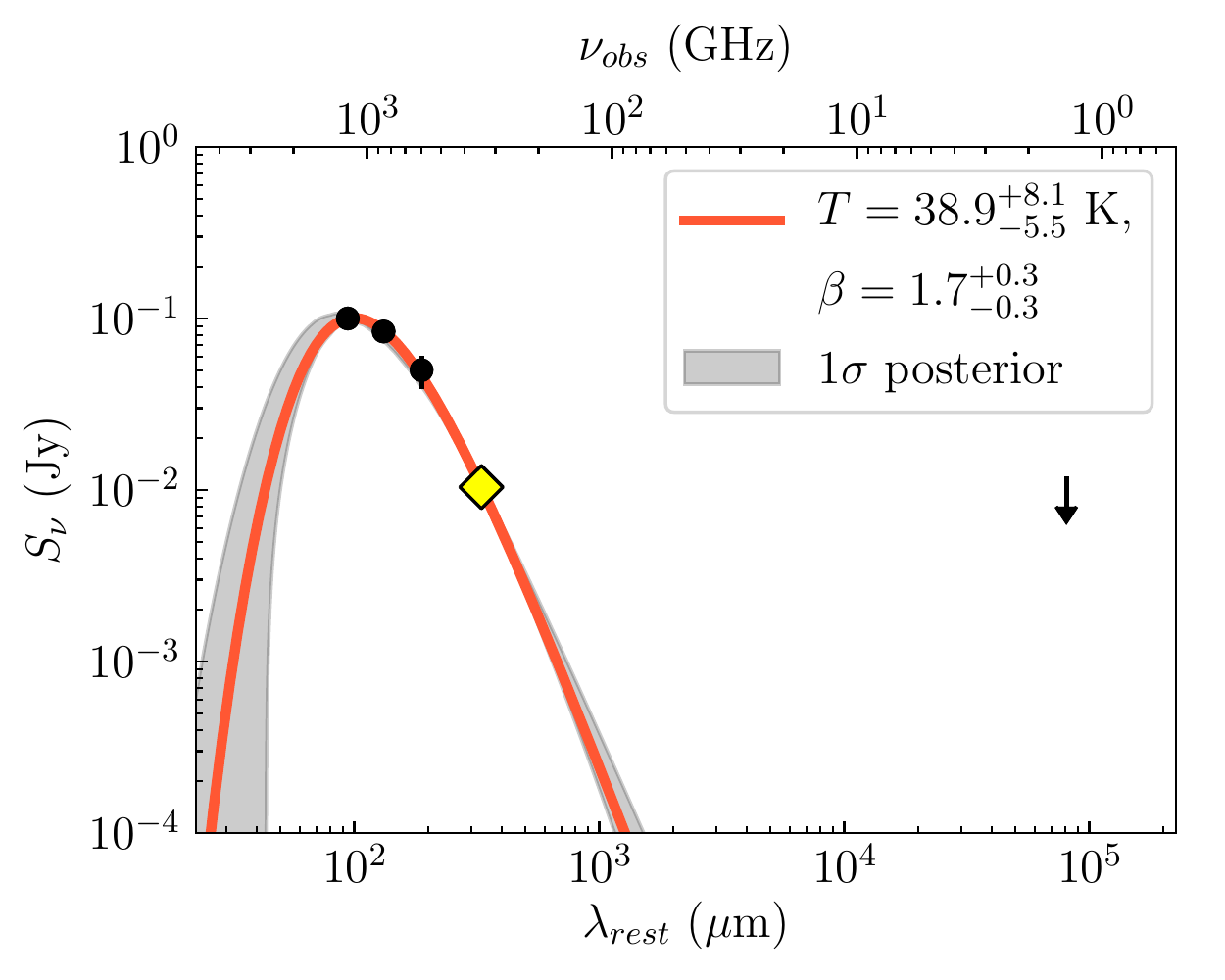}
    \caption{SED fits from FIR to radio wavelengths. Top row, HS~0810+2554 and RX~J0911+0551; second row, SDSS~J0924+0219 and PG~1115+080; third row, H1413+117 and WFI~J2026$-$4536; bottom, WFI~J2033$-$4723. The data are fit with one or two modified black body component for thermal dust emission (orange and purple), and in some cases power-law for radio synchrotron emission (green). The grey shaded region shows the 1$\sigma$ distribution of the posterior from the MCMC sampling. The flux density measurement from this work is shown by a yellow diamond; ancillary data points are listed and referenced in \protect\cite{Stacey:2018a}.}
    \label{fig:seds1}
\end{figure*}

\begin{figure*}
\centering
    \includegraphics[width=0.7\textwidth]{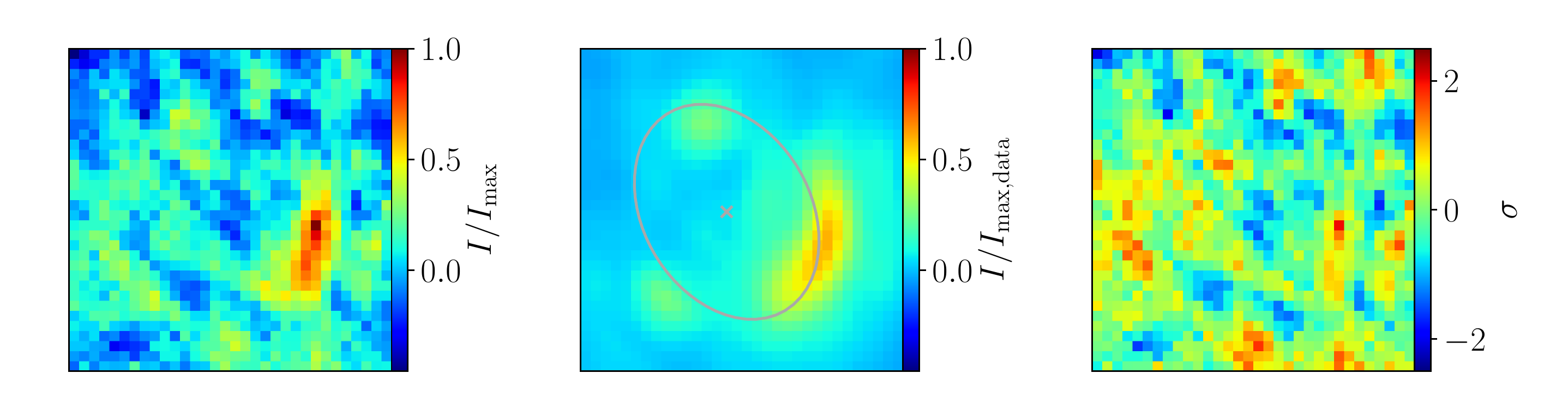}
    \vspace*{-0.2cm}
    \includegraphics[width=0.7\textwidth]{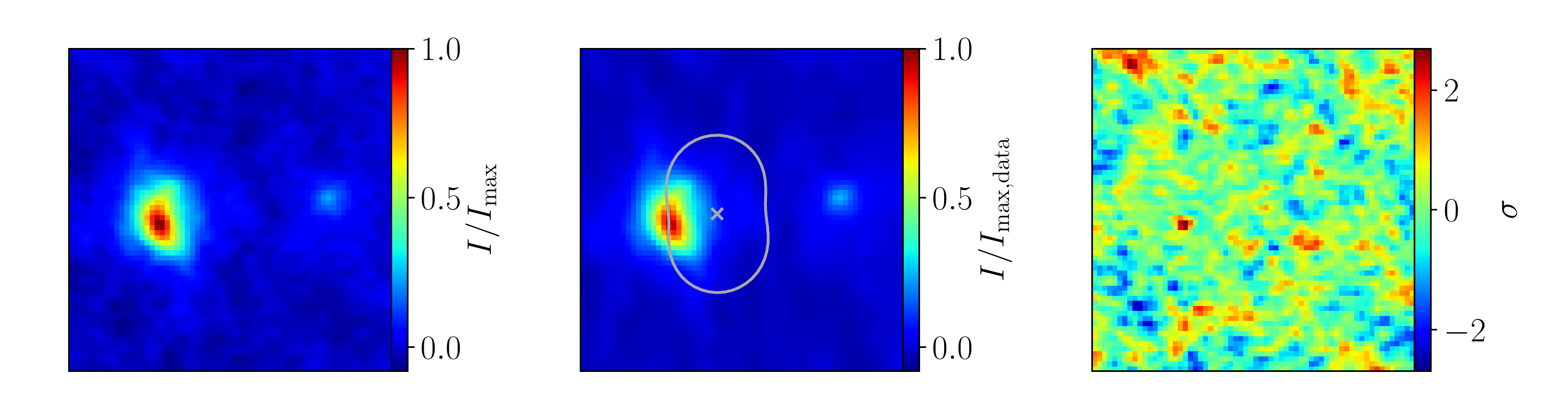}
    \vspace*{-0.2cm}
    \includegraphics[width=0.7\textwidth]{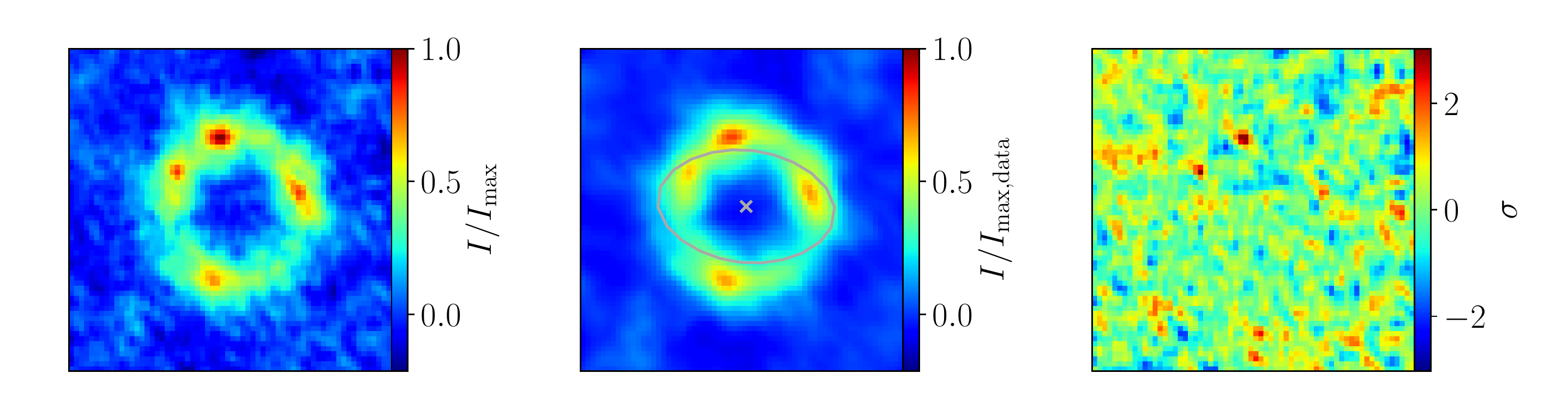}
    \vspace*{-0.2cm}
    \includegraphics[width=0.7\textwidth]{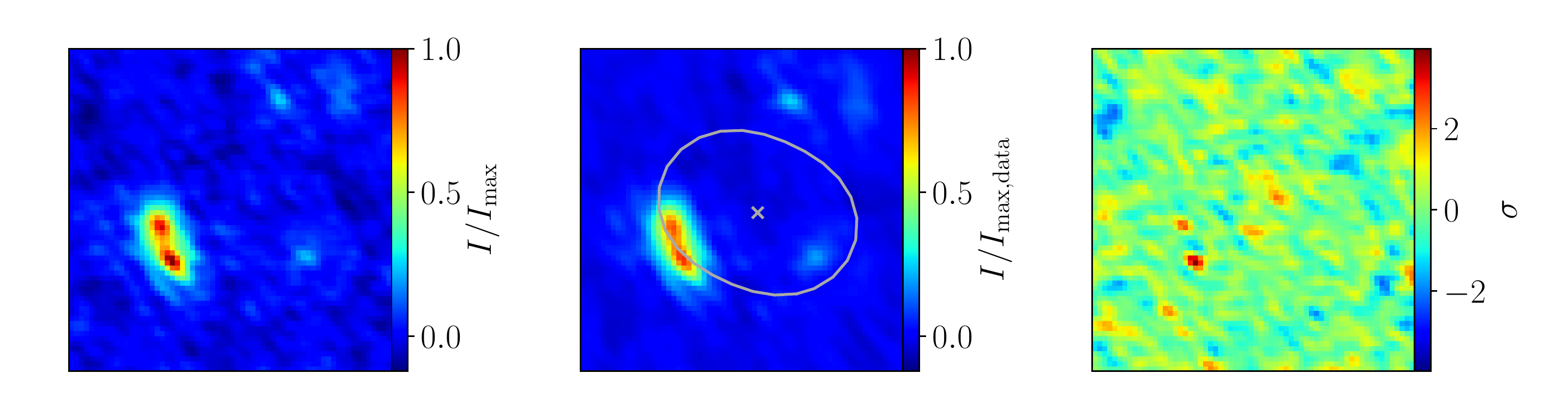}
    \vspace*{-0.2cm}
    \includegraphics[width=0.7\textwidth]{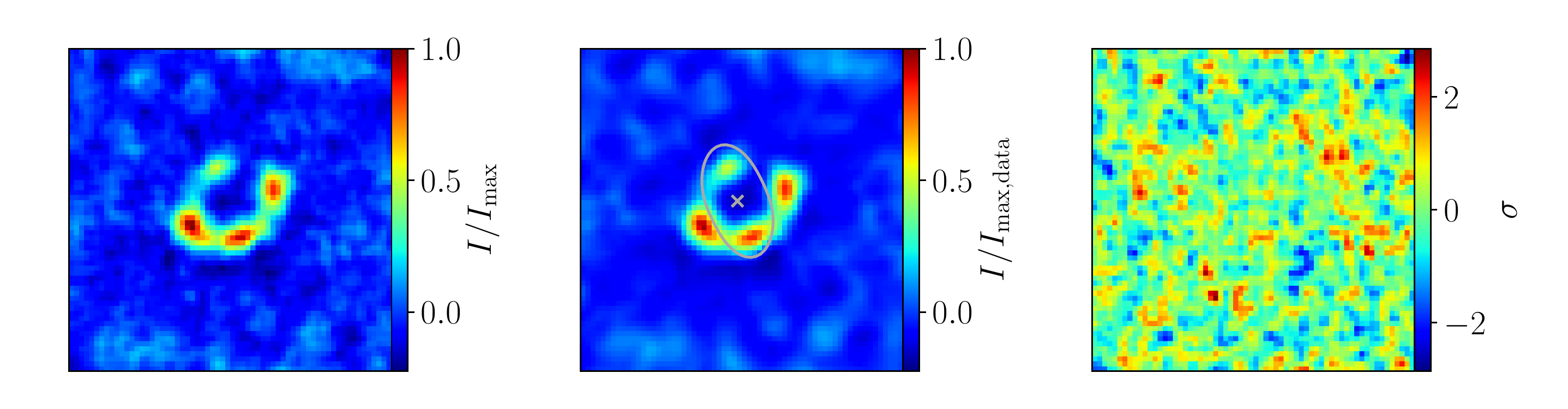}
    \vspace*{-0.2cm}
    \includegraphics[width=0.7\textwidth]{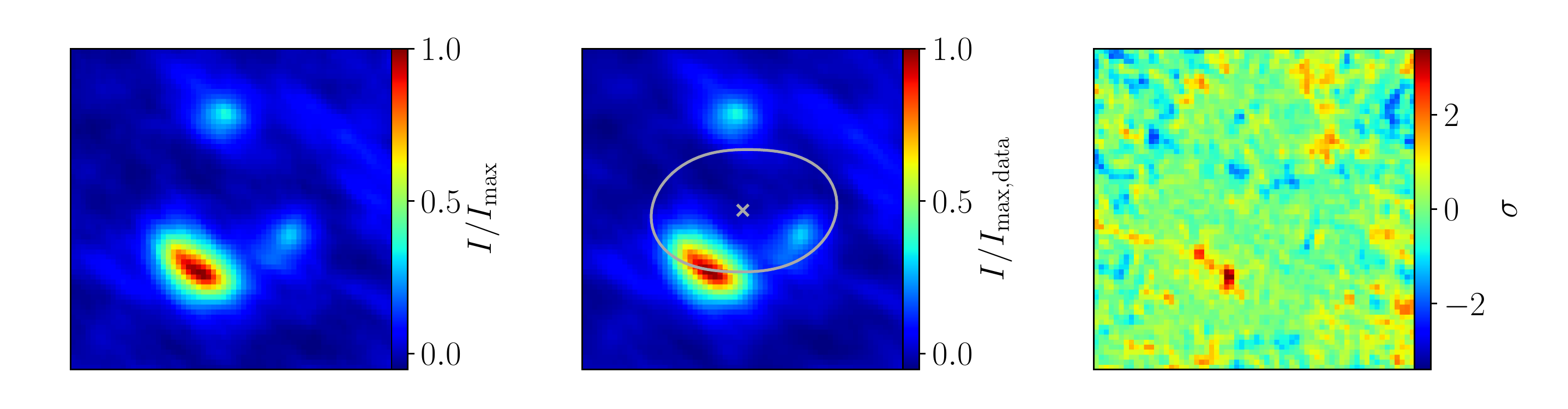}
    \vspace*{-0.2cm}
    \includegraphics[width=0.7\textwidth]{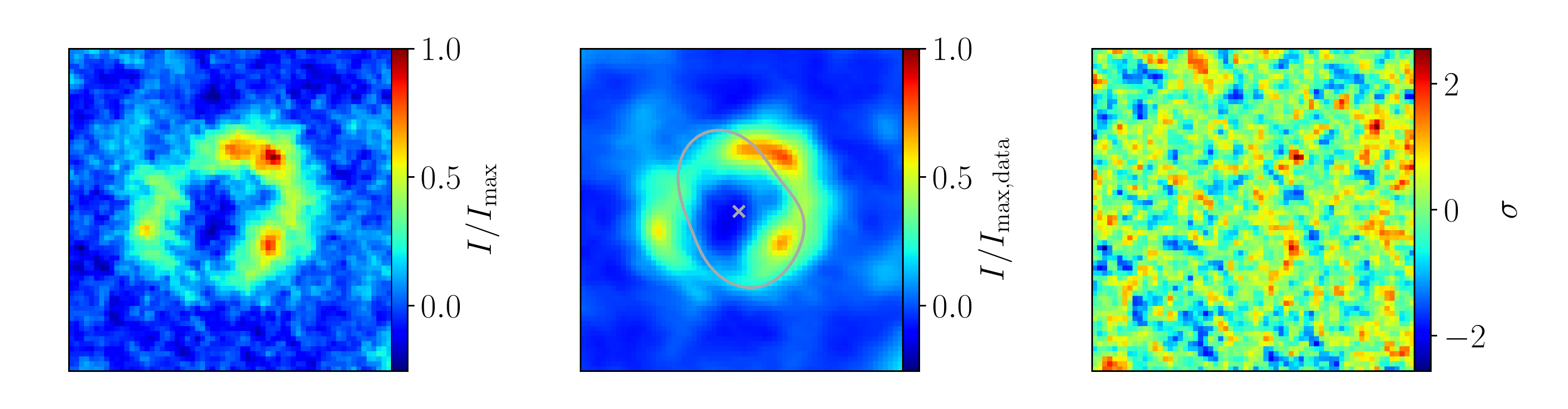}
    \caption{Grid-based lens models of the continuum data. The rows, top to bottom, show HS~0810+2554, RX~J0911+0551, SDSS~J0924+0219, PG~1115+080, WFI~J2026$-$4536 and WFI~J2033$-$4723. Panels, left to right, show the dirty image of the data in arbitrary flux units, the dirty image of the model on the same scale as the data, and the residuals (data$-$model) in units of $\sigma$, where $\sigma$ is the rms noise of the visibilities. The lens position and critical curve is shown in grey.}
    \label{fig:reconst_cont}
\end{figure*}

\begin{figure*}
    \centering
    \includegraphics[width=0.7\textwidth]{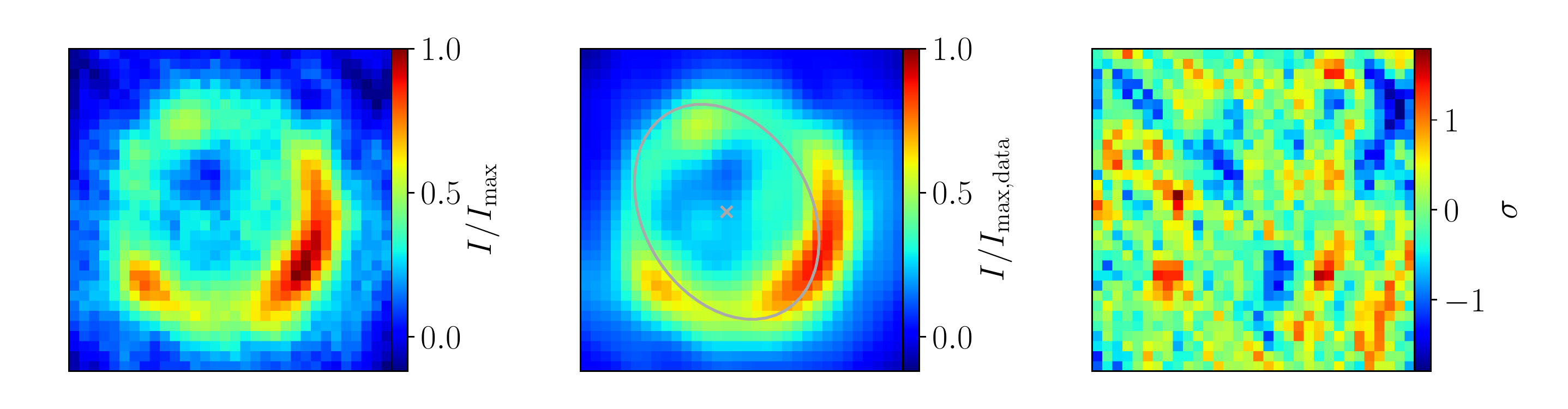}
    \vspace*{-0.2cm}
    \includegraphics[width=0.7\textwidth]{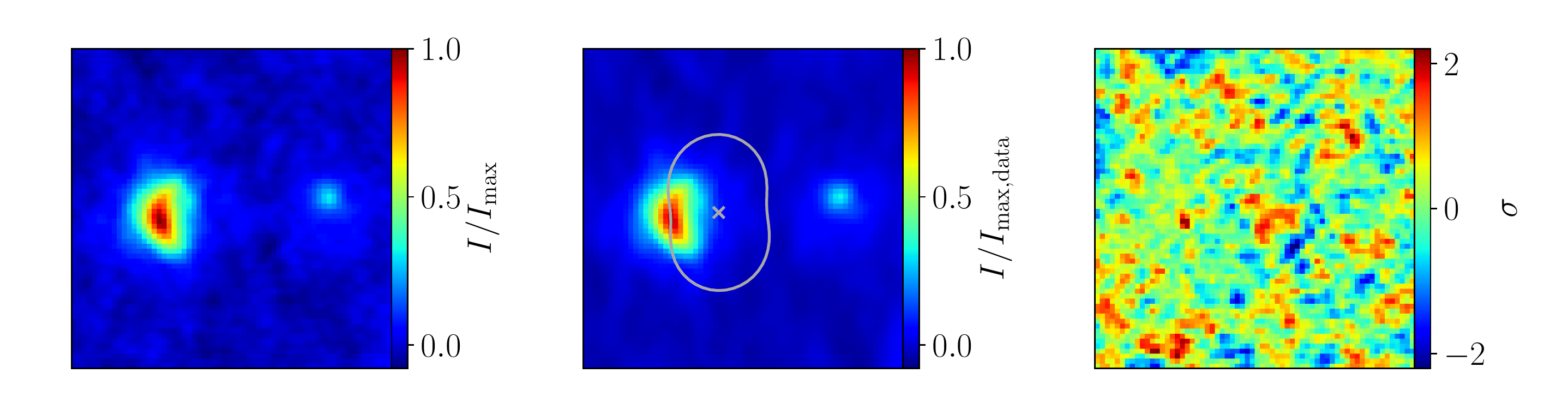}
    \vspace*{-0.2cm}
    \includegraphics[width=0.7\textwidth]{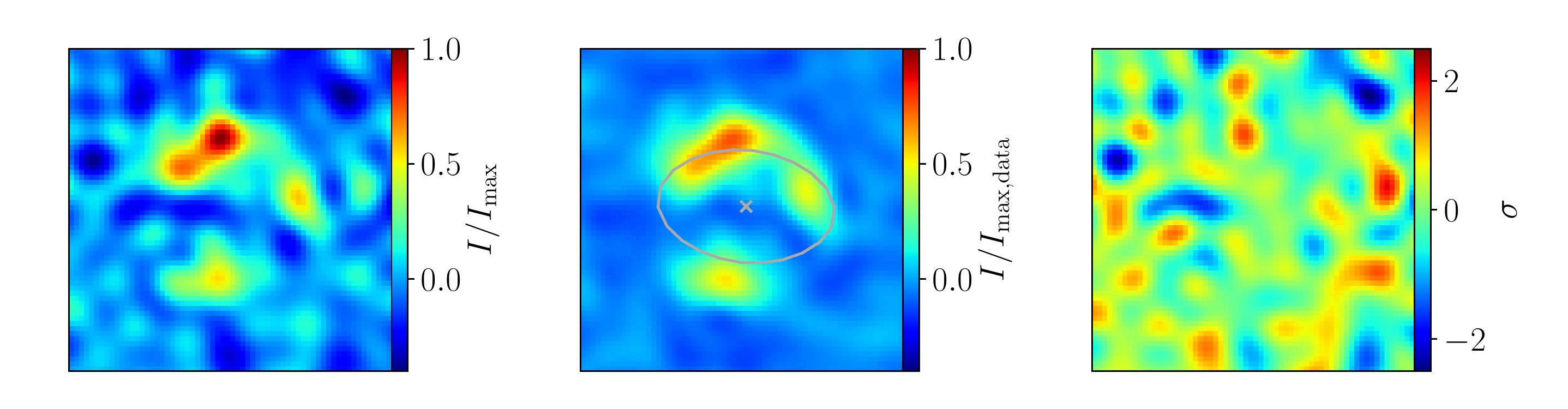}
    \vspace*{-0.2cm}
    \includegraphics[width=0.7\textwidth]{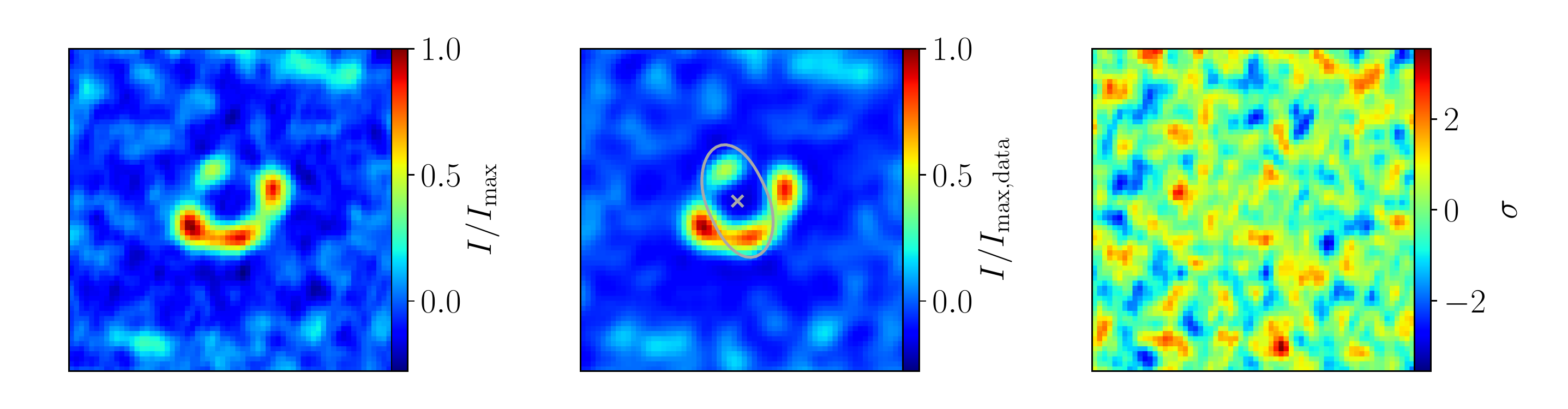}
    \vspace*{-0.2cm}
    \includegraphics[width=0.7\textwidth]{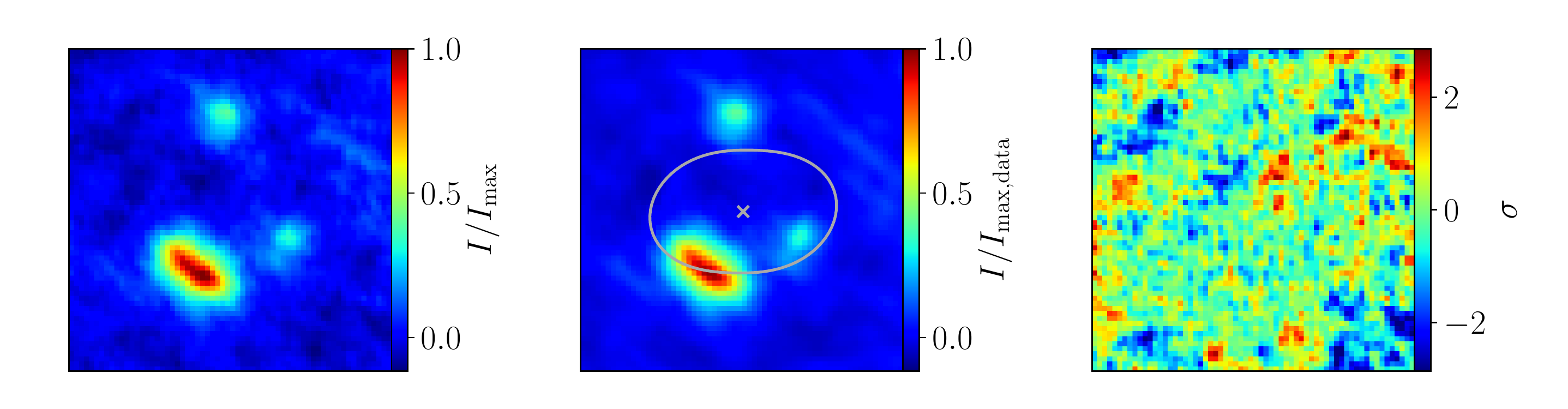}
    \vspace*{-0.2cm}
    \includegraphics[width=0.7\textwidth]{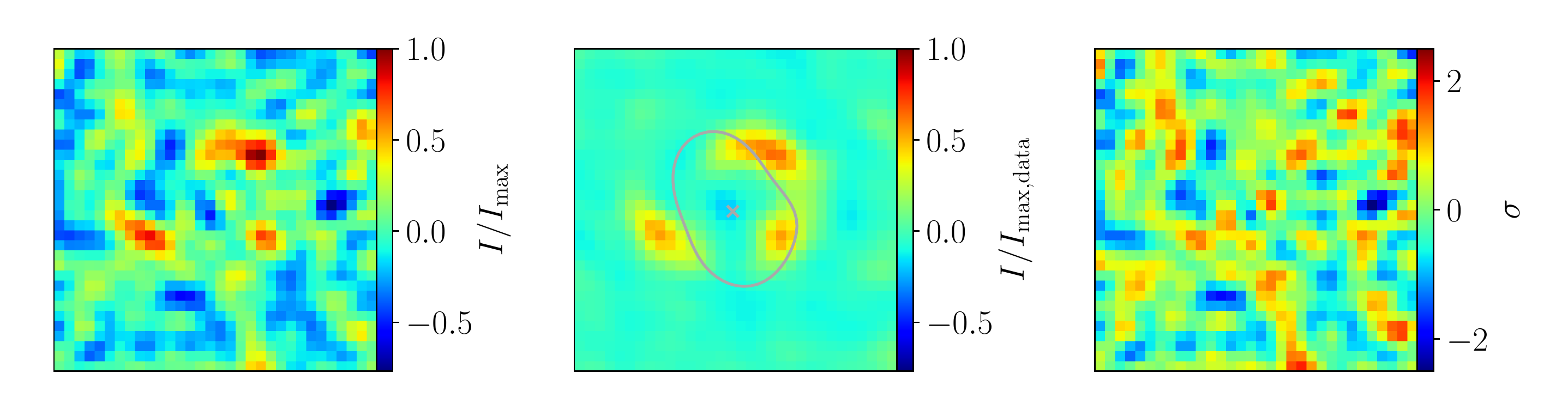}
    \caption{Grid-based lens modelling of the integrated CO line data. Rows, top to bottom, show HS~0810+2554, RX~J0911+0551, SDSS~J0924+0219, H1413+117, WFI~J2026$-$4536 and WFI~J2033$-$4723. Panels, left to right, show the dirty image of the data in arbitrary flux units, the dirty image of the model (on the same scale as the data), and the residuals (data$-$model) in units of $\sigma$, where $\sigma$ is the rms noise of the visibilities. The lens position and critical curve is shown in grey.}
    \label{fig:reconst_line}
\end{figure*}
\label{lastpage}
\end{document}